\documentclass[preprint,nofootinbib,floats,superscriptaddress,tightenlines,amsfonts,amsmath,amssymb]{revtex4}
\usepackage[colorlinks=true,
            linkcolor=blue,
            urlcolor=blue,
            citecolor=green,          
            bookmarks=true,
            bookmarksnumbered=true,
            breaklinks=true,
            pdfpagemode=Fullscreen,
            pdfstartview=FitBH]{hyperref}
\usepackage{subcaption}
\usepackage{graphicx}
\usepackage{bm}
\usepackage{xcolor}
\usepackage{multirow}
\usepackage{slashed}
\allowdisplaybreaks
\usepackage{siunitx}
\usepackage{diagbox}
\usepackage{ulem}
\usepackage{tablefootnote}

\newcommand{\calO}{{\cal O}}
\newcommand{\calB}{{\cal B}}
\newcommand{\calM}{{\cal M}}
\newcommand{\Tr}{\rm Tr}

\newcommand{\hc}{{\rm h.c.}}
\newcommand{\C}{\tt C}

\begin{document}
\baselineskip=17pt \parskip=5pt

\hspace*{\fill}

\title{ FCNC $B$ and $K$ Meson Decays with Light Bosonic Dark Matter}

\author{Xiao-Gang He}
\email{hexg@phys.ntu.edu.tw}
\affiliation{Tsung-Dao Lee Institute (TDLI) \& School of Physics and Astronomy (SPA),
Shanghai Jiao Tong University (SJTU), Shanghai 200240, China}
\affiliation{Department of Physics, National Taiwan University, Taipei 10617}

\author{Xiao-Dong Ma}
\email{maxid@sjtu.edu.cn}
\affiliation{Tsung-Dao Lee Institute (TDLI) \& School of Physics and Astronomy (SPA),
Shanghai Jiao Tong University (SJTU), Shanghai 200240, China}

\author{German Valencia}
\email{german.valencia@monash.edu}
\affiliation{School of Physics and Astronomy, Monash University, Wellington Road, Clayton, VIC-3800, Australia\bigskip}

\begin{abstract}

We consider decays of $B$ and $K$ mesons into a pseudo-scalar or vector meson plus missing energy. Within the SM, these modes originate from flavor changing neutral current (FCNC) processes with two neutrinos in the final state. In this paper we consider the experimental upper bounds on these modes and interpret the difference between these bounds and the SM prediction as a window into new light invisible particles. In particular we consider the case where some new symmetry requires the new particles to be produced in pairs. We first construct the general low energy effective Lagrangian coupling an FCNC with two dark sector particles of spin zero, one-half and one. We then present numerical estimates for the constraints that can be placed on these interactions, finding that an effective new physics scale from $\calO(10)$-$\calO(10^{11})$ GeV can be probed, with the exact value strongly depending on the interaction structure as well as the mass of the invisible particle. For $K^+\to \pi^+ \slashed E$  we incorporate into our constraints the effect of using only the signal regions of NA62, and for $B^+\to K^+ \slashed E$ the $q^2$-dependent efficiency of Belle II.

\end{abstract}

\maketitle

{\small\hypersetup{linkcolor=black}\tableofcontents}

\newpage

\section{Introduction}

It is well known that flavor changing neutral current (FCNC) processes are severely suppressed within the standard model (SM) due to both their loop origin and the GIM mechanism  \cite{Glashow:1970gm}. For this reason, these processes are very sensitive to new physics (NP) beyond the SM. We consider $B$ and $K$  decay modes with one meson and a neutrino pair in the final state because  experimental upper bounds are available for them. As the neutrinos are not detected, these modes become decays with missing energy ($\slashed{E}$) in the final state from the experimental perspective. At the quark level, any  transition of the form $b\to s(d)+\slashed{E}$ and $s\to d +\slashed{E}$, would contribute to these modes and  $\slashed{E}$ can originate from any  sufficiently light invisible particle. In addition to the SM neutrinos, the invisible particles can be any hypothetical  neutral particle that escapes the current experimental detection. One well-motivated choice is to relate these light invisible particles to dark matter (DM) or other dark sectors. 
These rare meson decays can then be used to constrain light DM and are particularly  important  in the face of current stringent experimental constraints on heavy DM from direct detection experiments \cite{Roszkowski:2017nbc,Bottaro:2021snn}. 
\begin{table}[!h]
\centering
\resizebox{\linewidth}{!}{
\renewcommand{\arraystretch}{1.1}
\begin{tabular}{|c | l | l | l |}
\hline
\diagbox[width=7em]{Topology}{Particle}  &\, Fermion  $(\chi)$ &\, Scalar  $(\phi)$ &\, Vector  $(X)$
\\\hline 
\multirow{2}*{2-body decay} 
 &\, $b \to s(d) +\chi$ \cite{Dib:2022ppx} 
 &\,  $b \to s(d) +\phi $ \cite{Kamenik:2011vy,Li:2021sqe}  
 &\,  $b \to s(d) + X$ \cite{Kamenik:2011vy,Li:2021sqe}
\\\cline{2-4} %
 & \quad\quad\quad\quad --- 
 &\,  $s \to d +\phi $ \cite{Kamenik:2011vy,He:2020jly}
  &\,  $s \to d + X$ \cite{Kamenik:2011vy}
\\\hline
\multirow{2}*{3-body decay} 
&\,  $b \to s(d) +\chi\chi$\, \cite{Kamenik:2011vy, Altmannshofer:2009ma, Li:2020dpc, Felkl:2021uxi} \quad  
&\,  $b \to s(d) +\phi\phi $ \, \cite{Bird:2004ts,Altmannshofer:2009ma} ({\color{red}\checkmark })  \quad 
&\,  $b \to s(d) + XX$ \, ({\color{red}\checkmark\checkmark }) \quad
\\\cline{2-4} %
&\,  $s \to d +\chi\chi$\, \cite{Kamenik:2011vy, Tandean:2019tkm, Su:2019tjn, Li:2019fhz,Deppisch:2020oyx, He:2021yoz}
&\,  $s \to d +\phi\phi $ \cite{Li:2019cbk,Geng:2020seh}\, ({\color{red}\checkmark})  \quad
&\,  $s \to d + XX$\, ({\color{red}\checkmark\checkmark}) \quad 
\\
\hline
\end{tabular}
}
\caption{List of possible $b\to s (d) + \slashed{E}$ and $s\to d + \slashed{E}$ modes related to $B$ and $K$ meson decays:  a ``\checkmark\checkmark'' means that they have not been studied before while a ``\checkmark'' indicates they have been partially studied, both are the subject of this paper.
 }
\label{tab:b2smisE1}
\end{table} 

Without knowledge of the fundamental interactions in dark sectors, it is suitable to study these FCNC processes in a model independent manner by incorporating new light, neutral, degrees of freedom with an effective field theory (EFT) approach. 
The new particles can be either scalar, fermion, or vector in nature if we limit our study to spin less than or equal to one. Well known examples for these invisible particles are the axion \cite{Weinberg:1977ma, Wilczek:1977pj}, the sterile neutrino \cite{Dasgupta:2021ies}, and the dark photon \cite{Fabbrichesi:2020wbt}. 
Model independent studies of the FCNC $b\to s(d) + \slashed{E}$ and $s\to d + \slashed{E}$ transitions  within the EFT framework have been carried out in the past for these three types of invisible particles for $B$ and $K$ meson decays. A summary of what has been done is shown in Tab.\,\ref{tab:b2smisE1}.  These processes are further divided into 2-body decays with a single new particle and 3-body decays with a pair of new particles. All the 2-body channels plus the 3-body decay with a pair of new fermions have been extensively studied before \cite{Dib:2022ppx, Kamenik:2011vy,Altmannshofer:2009ma, Li:2021sqe, He:2020jly, Li:2020dpc, Felkl:2021uxi, Tandean:2019tkm,Su:2019tjn, Li:2019fhz, Deppisch:2020oyx, He:2021yoz}.
\footnote{Note that the 2-body modes with a single new fermion are only possible when the meson decays into a baryon and $\chi$ \cite{Dib:2022ppx}. For the 3-body decay with a pair of fermions, $\chi$ can be either the neutrino or some DM particle, the former is studied in \cite{Altmannshofer:2009ma,Felkl:2021uxi,Li:2019fhz, Deppisch:2020oyx, He:2021yoz}, while the latter in \cite{Li:2020dpc}. }
The table indicates with a double-checkmark processes that have not received much attention, namely three body modes with a pair of  vector particles. We are only aware of a classification of relevant SMEFT operators  in \cite{Kamenik:2011vy}.
The 3-body decays with a pair of scalar particles indicated with a checkmark in the table, have been partially studied before. 
For the $B\to K$ and $K\to \pi$ transitions: \cite{Altmannshofer:2009ma} has considered only two of the four possible operators we enumerate in Eq.\,\ref{eq:Oqphi} (the scalar ones), whereas  \cite{Bird:2004ts} has only considered one of them. For $s\to d+\phi\phi$ transitions, \cite{Li:2019cbk} has considered both kaon and hyperon decays and \cite{Geng:2020seh} has considered additional kaon decay modes. Our study of the kaon modes includes the new NA62 results for the relevant signal window.
 In this paper,  we systematically investigate these channels using a general  low energy effective theory  (LEFT), and find constraints on all the relevant effective operators with the help of the most recent experimental results. 

The observables we use to set the bounds are summarized in Tab.\,\ref{tab:B2KmisE2}.  
In the second column, we list the SM background with a pair of neutrinos. For most modes in the table, we obtain the SM prediction using the {\tt flavio} package \cite{Straub:2018kue}.
In some cases, the SM calculation of the form factors is not trivial and we use instead the predictions for $B\to \pi\bar\nu\nu$ from \cite{Hambrock:2015wka}.  The kaon decay modes are very clean theoretically, but their parametric uncertainty due to CKM angles can be large, with central values changing by up to 20\% \cite{Buras:2015qea}, we quote the values from the PDG in this case \cite{Workman:2022ynf}.

The new physics we discuss, will always add incoherently to the SM di-neutrino background in $B(K)\to M \slashed E$ processes (with $M$ representing the final state mesons shown in 
Tab.\,\ref{tab:B2KmisE2}). In view of this we define the ``room for new physics'' in these modes as the difference between the experimental upper bound and the standard model prediction. We adopt the simple prescription of subtracting from the 90\,\% experimental upper bound the lower limit of the $90\,\%$ C.L. SM range and show this number in the last column. The constraints can be easily adapted for more sophisticated subtractions if desired. The mode $K^+\to \pi^+ \bar\nu\nu$ has been measured by both BNL787/949 \cite{E949:2008btt,BNL-E949:2009dza} and NA62 \cite{NA62:2020fhy,NA62:2021zjw}, so in this case we use the upper limit of the $90\,\%$ range quoted by PDG  \cite{Workman:2022ynf} as the experimental upper limit.

\begin{table}
\centering
\resizebox{\linewidth}{!}{
\renewcommand{\arraystretch}{1.1}
\begin{tabular}{| c | c  | c | c | }
\hline
\quad\quad   Observable \quad\quad   
& \quad  SM prediction ($\slashed{E}=\nu\bar\nu$) \quad  
& \quad 90\,\% C.L. upper bound 
\tablefootnote{Except for $K^+\to \pi^+\slashed{E}$ channel, for which the $1\,\sigma$ measured value quoted by PDG  \cite{Workman:2022ynf} is used. }
& New physics bound $ {\cal B}^{\rm UL}$ 
\\\hline 
$\calB(B^+ \to K^+ \slashed{E}) $ 
& $(4.4\pm 0.6)\times 10^{-6}$ 
& $1.6\times 10^{-5} $ \cite{Workman:2022ynf}
&  $1.3\times 10^{-5} $
\\\hline 
$\calB(B^0 \to K^0 \slashed{E})$ 
& $(4.1\pm 0.6)\times 10^{-6}$  
& $2.6\times 10^{-5} $ \cite{Belle:2017oht} 
& $2.3\times 10^{-5} $
\\\hline 
$ \calB(B^+ \to K^{*+}  \slashed{E})  $ 
&  $ (1.0\pm 0.1)\times 10^{-5}$
&  $4.0\times 10^{-5} $ \cite{Belle:2013tnz}
& $3.1\times 10^{-5} $
\\\hline 
   $ \calB(B^0 \to K^{*0} \slashed{E} ) $ 
& $(9.5\pm 1.0)\times 10^{-6}$
& $1.8\times 10^{-5} $ \cite{Belle:2017oht} 
& $1.0\times 10^{-5} $
\\\hline
  $\calB(B^+ \to \pi^+ \slashed{E}) $ 
& $(2.39_{-0.28}^{+0.30})\times 10^{-7}$ \cite{Hambrock:2015wka}
& $1.4\times 10^{-5} $ \cite{Belle:2017oht} 
& $1.4\times 10^{-5} $
\\\hline 
   $\calB(B^0 \to \pi^0 \slashed{E}) $ 
& $(1.2_{-0.14}^{+0.15})\times 10^{-7}$ \cite{Hambrock:2015wka}
& $9.0\times 10^{-6} $ \cite{Belle:2017oht} 
& $8.9\times 10^{-6} $
\\\hline 
   $ \calB(B^+ \to \rho^{+}  \slashed{E})  $ 
  & $(4.5\pm 1.0)\times 10^{-7}$ 
& $3.0\times 10^{-5} $ \cite{Belle:2017oht} 
& $3.0\times 10^{-5} $
\\\hline 
   $\calB(B^0 \to \rho^{0} \slashed{E} )$ 
  & $(2.0\pm 0.4)\times 10^{-7}$   
& $4.0\times 10^{-5} $ \cite{Belle:2017oht} 
& $4.0\times 10^{-5} $
\\\hline
   $\calB(K^+ \to \pi^+ \slashed{E} )$ 
&  $(8.1\pm 0.4)\times 10^{-11}$   \cite{Workman:2022ynf}
& $(1.14_{-0.33}^{+0.40})\times 10^{-10}$   \cite{Workman:2022ynf}
& $1.1\times 10^{-10} $
\\\hline
   $\calB(K_L \to \pi^0 \slashed{E} )$   
&  $(2.8\pm 0.2 )\times 10^{-11}$  \cite{Workman:2022ynf}
& $4.9\times 10^{-9} $  \cite{Workman:2022ynf}
& $4.9\times 10^{-9} $
\\\hline
\end{tabular}
}
\caption{Summary on the status of FCNC $B \to (K, K^*, \pi, \rho) \slashed{E}$ and $K\to \pi \slashed{E}$ decays with missing energy. }
\label{tab:B2KmisE2}
\end{table}

The paper is organized as follows.  In section \ref{sec:LEFT_ope} we classify all the relevant effective interactions in the framework of low energy effective field theory. 
In section \ref{sec:Bdecay} we consider the FCNC $B$ meson decay $B\to (K,K^*, \pi, \rho)$+DM+DM with a pair of scalar or vector DM,  and use the current experimental bounds to  constrain  the relevant effective new physics scale involving $(bs)$ and $(bd)$ quark flavors. In section \ref{sec:kaondecay}, we use chiral perturbation theory to analyze the FCNC kaon decay $K\to \pi$+DM+DM to constrain the effective scale involving $(ds)$ quark flavors. In all cases we will refer to the invisible light particles as ``DM'' regardless of their origin. 
In section \ref{sec:conclusion}, we draw our conclusions. Supplementary material presented in the Appendix includes: the phase space integration in \ref{sec:phasespace}; the reduction of operators with vector DM fields in  \ref{sec:reductionofVope}; operators for  lepton-DM interactions in \ref{sec:leptonDMope}; a collection of form factors involving $B$ meson decays in \ref{sec:formfactor}, and  specific renormalizable model realizations for some scalar and vector DM operators as an illustration in appendix \ref{sec:models}.  

\section{Quark-DM interaction in LEFT}
\label{sec:LEFT_ope}

In this section we list the most general local quark-DM interactions in the framework of low energy effective field theory, with the FCNC interactions being a subset of these operators with appropriate flavor indices. In LEFT, only the unbroken $SU(3)_{\rm c}\times U(1)_{\rm em}$ symmetry of the standard model is imposed. Operators at the weak scale in SMEFT have been listed before \cite{Grzadkowski:2010es, Lehman:2014jma, Liao:2016hru} along with their contributions to the LEFT operators \cite{Jenkins:2017jig,Liao:2020zyx}. Since we are considering cases with light new particles it is more general to start from LEFT \cite{Jenkins:2017jig}. In this way we cover scenarios that contain both light DM as well as weak scale new mediators that can be integrated out to reach the LEFT description containing only the relevant degrees of freedom \cite{Lehmann:2020lcv}.      

The scenario in which modes with two DM particles becomes relevant is that in which  they  must be pair produced due to  underlying symmetries in the dark sector which can also guarantee that the DM is stable. In this case the relevant quark-DM interactions have to involve a pair of quark fields and a pair of  DM fields. These interactions can be classified in terms of the spin of the DM particle, and we consider the cases of spin zero, one-half and one  with corresponding scalar, fermion and  vector DM fields. 
We denote the light SM quarks as $q\in\{ d,s,b, u,c\}$, the fermionic DM particle as  $\chi$ (Dirac or Majorana fermion), the scalar DM as $\phi$ (complex or real scalar), and the vector DM as $X$ (complex or real vector), respectively. For each type of DM particle, the relevant quark-DM interactions are enumerated below. 

\noindent
{\bf Fermion case}: The leading dimension-6 (dim-6) operators for this case have been considered before \cite{Kumar:2013iva,Badin:2010uh} and we list them (with slightly different convention) here for completeness. They are,
\begin{subequations}
\begin{align}
\calO_{q\chi1}^{S} &= (\overline{q} q)(\overline{\chi}\chi),
&
\calO_{q\chi2}^{S}  &= (\overline{q} q)(\overline{\chi}i \gamma_5\chi), 
\\
\calO_{q\chi1}^{P} &=  (\overline{q} i \gamma_5 q)(\overline{\chi}\chi),
&
\calO_{q\chi2}^{P} & = (\overline{q} \gamma_5 q)(\overline{\chi} \gamma_5\chi), 
\\
\calO_{q\chi1}^{V} &=  (\overline{q}\gamma^\mu  q)(\overline{\chi}\gamma_\mu  \chi),
\, (\times)
&
\calO_{q\chi2}^{V} &= (\overline{q}\gamma^\mu q)(\overline{\chi}\gamma_\mu  \gamma_5\chi), 
\\
\calO_{q\chi1}^{A} &=  (\overline{q}\gamma^\mu\gamma_5  q)(\overline{\chi}\gamma_\mu  \chi),
\, (\times)
&
\calO_{q\chi2}^{A} & = (\overline{q}\gamma^\mu\gamma_5 q)(\overline{\chi}\gamma_\mu  \gamma_5\chi), 
\\
\calO_{q\chi1}^{T} &=  (\overline{q}\sigma^{\mu\nu}  q)(\overline{\chi}\sigma_{\mu\nu}   \chi),
\, (\times)
&
\calO_{q\chi2}^{T} &= (\overline{q}\sigma^{\mu\nu} q)(\overline{\chi}\sigma_{\mu\nu}  \gamma_5\chi), 
\, (\times)
\end{align}
\end{subequations}
where the quark flavor indices have been omitted for notational simplicity, but are understood to be those required for the specific FCNC process throughout the paper. The ``$(\times) $'' indicates the accompanying operator vanishes for the Majorana DM case due to the fermion bilinear identity, $\overline{\chi}\Gamma\chi =- \overline{\chi^{\C}}\Gamma\chi^{\C}$ for $\Gamma\in \{\gamma_\mu, \sigma_{\mu\nu},\sigma_{\mu\nu}\gamma_5\}$, where $\chi^{\C}\equiv - i\gamma^2 \chi^*$ is the charge conjugation of $\chi$. 

\noindent
{\bf Scalar case}: The leading order operators for the scalar DM appear at dimension 5 and 6,\footnote{They would all arise at dimension 6 in the SMEFT framework \cite{Kamenik:2011vy}. A specific renormalizable model realization for the operators $\calO_{q\phi}^S$ and $\calO_{q\phi}^P$ is given in appendix \ref{sec:models}. The realization for the operators $\calO_{q X}^S$ and $\calO_{q X}^P$ involving vector DM is also given there.} they can be parametrized in the following manner, 
\begin{subequations}
\begin{eqnarray}
\calO_{q\phi}^S &= & (\overline{q} q)(\phi^\dagger \phi), 
\\
\calO_{q\phi}^P &= & (\overline{q} i \gamma_5 q)(\phi^\dagger \phi), 
\\
\calO_{q\phi}^V &= & (\overline{q}\gamma^\mu q) (\phi^\dagger i \overleftrightarrow{\partial_\mu} \phi), \, (\times) 
\\
\calO_{q\phi}^A &= & (\overline{q}\gamma^\mu\gamma_5 q) (\phi^\dagger i \overleftrightarrow{\partial_\mu} \phi),  \, (\times).  
\end{eqnarray}
\label{eq:Oqphi}
\end{subequations}
Once again the implicit quark flavor indices should be understood. The symbol ``$(\times) $'' indicates the related operator vanishes for real scalar DM, and the double arrow derivative is defined as $A\overleftrightarrow{\partial_\mu}  B \equiv A (\partial_\mu B) - (\partial_\mu A)  B  $. 

\noindent 
{\bf Vector case A}: For the vector DM, we consider separately two cases: when the DM field is represented by the four-vector potential $X_\mu$ (scenario A) or by the field strength tensor $X_{\mu\nu}\equiv \partial_\mu X_\nu - \partial_\nu X_\mu$ (scenario B). 
For scenario A, we find there are 4 independent dim-5 operators with the field content $\bar q q X^\dagger X$ and 12 dim-6 operators with the content $\bar qq X^\dagger XD_\mu$. 
By requiring the flavor diagonal operators to be self-conjugate, we can parametrize those operators in the following way, 
\begin{subequations}
\label{eq:OqX}
\begin{eqnarray}
\calO_{q X}^S &= & (\overline{q} q)(X_\mu^\dagger X^\mu), 
\\
\calO_{q X}^P &= & (\overline{q}i \gamma_5 q)(X_\mu^\dagger X^\mu), 
\\
\calO_{q X1}^T &= &  {i \over 2} (\overline{q}  \sigma^{\mu\nu} q) (X_\mu^\dagger X_\nu - X_\nu^\dagger X_\mu),  \, (\times) 
\\
\calO_{q X2}^T &= & {1\over 2} (\overline{q}\sigma^{\mu\nu}\gamma_5 q) (X_\mu^\dagger X_\nu - X_\nu^\dagger X_\mu),  \, (\times) 
 \\
 \calO_{q X1}^V &= &{1\over 2} [ \overline{q}\gamma_{(\mu} i \overleftrightarrow{D_{\nu)} } q] (X^{\mu \dagger} X^\nu + X^{\nu \dagger} X^\mu  ), 
\\
\calO_{q X2}^V &= & (\overline{q}\gamma_\mu q)\partial_\nu (X^{\mu \dagger} X^\nu + X^{\nu \dagger} X^\mu  ), 
\\
\calO_{q X3}^V &= & (\overline{q}\gamma_\mu q)( X_\rho^\dagger \overleftrightarrow{\partial_\nu} X_\sigma )\epsilon^{\mu\nu\rho\sigma}, 
\\
\calO_{q X4}^V &= & (\overline{q}\gamma^\mu q)(X_\nu^\dagger  i \overleftrightarrow{\partial_\mu} X^\nu), 
 \, (\times)
 \\
\calO_{q X5}^V &= & (\overline{q}\gamma_\mu q)i\partial_\nu (X^{\mu \dagger} X^\nu - X^{\nu \dagger} X^\mu  ),  \, (\times)
 \\
\calO_{q X6}^V &= & (\overline{q}\gamma_\mu q) i \partial_\nu ( X^\dagger_\rho X_\sigma )\epsilon^{\mu\nu\rho\sigma},
\, (\times)  
 \\
\calO_{q X1}^A &= &{1\over 2} [\overline{q}\gamma_{(\mu} \gamma_5 i \overleftrightarrow{D_{\nu)} }  q](X^{\mu \dagger} X^\nu + X^{\nu \dagger} X^\mu  ), 
\\
\calO_{q X2}^A &= & (\overline{q}\gamma_\mu \gamma_5 q)\partial_\nu (X^{\mu \dagger} X^\nu + X^{\nu \dagger} X^\mu  ), 
\\ 
\calO_{q X3}^A &= & (\overline{q}\gamma_\mu\gamma_5 q) (X_\rho^\dagger \overleftrightarrow{ \partial_\nu} X_\sigma )\epsilon^{\mu\nu\rho\sigma}, 
\\
\calO_{q X4}^A &= & (\overline{q}\gamma^\mu\gamma_5 q)(X_\nu^\dagger  i \overleftrightarrow{\partial_\mu} X^\nu), 
 \, (\times)
  \\
\calO_{q X5}^A &= &  (\overline{q}\gamma_\mu \gamma_5 q)i \partial_\nu (X^{\mu \dagger} X^\nu - X^{\nu \dagger} X^\mu  ),  \, (\times)
 \\
\calO_{q X 6}^A &= & (\overline{q}\gamma_\mu\gamma_5 q)i \partial_\nu (  X^\dagger_\rho X_\sigma)\epsilon^{\mu\nu\rho\sigma},
\, (\times)  
\end{eqnarray}
\end{subequations}
where the current $\overline{q}\gamma_{(\mu} i \overleftrightarrow{D_{\nu)} } q \equiv \overline{q}\gamma_{\mu} i \overleftrightarrow{D_{\nu} } q + \mu \leftrightarrow \nu$, and similarly for the current $\overline{q}\gamma_{(\mu}\gamma_5 i \overleftrightarrow{D_{\nu)} } q$.
The symbol ``$(\times)$'' indicates the corresponding operator vanishes for real vector DM.
Other operators with different Lorentz contractions or derivatives are not independent and can always be reduced to those given above by using the Dirac gamma identities (DI), integration by parts (IBP), and equation of motions (EoM). 
The construction of the above dim-6 operators with a derivative is expanded upon in appendix \ref{sec:reductionofVope}. 
We have also checked our result by using the Hilbert series method with a modified conformal representation for the vector DM field \cite{Henning:2017fpj}. 

Using the above operators to calculate the amplitudes for physical processes results in rates that are divergent in the limit of massless DM particles. This divergence originates from the longitudinal part in the polarization sum and is a well known problem. One way to deal with this problem is to assume that the vectors are gauge bosons of some dark symmetry and gauge invariance under that symmetry forbids the direct appearance of the field $X_\mu$. These would appear instead in covariant derivatives acting on other dark matter fields which are not present in our effective Lagrangian. The net effect of such a scenario is that $X_\mu$ acquires mass by some Higgs mechanism \cite{Kamenik:2011vy,Williams:2011qb} and the effective operators inherit a coefficient that vanishes in the limit of massless $X_\mu$. An example of how this could work in a specific model is given in appendix  \ref{sec:models}. 
Operationally, for our numerical analysis, we require the Wilson coefficients for the above operators to contain an explicit factor of the DM mass to the minimum power necessary to cancel  potential divergences as $m\to 0$.   

\noindent
{\bf Vector case B}: With the vector DM entering through field strength tensors the minimal dimensionality of the operators is 7.\footnote{If working with SMEFT assumption, the resulting operators would be at dimension 8.}  In this case two explicit DM fields are required again for processes with two DM particles because a single field strength tensor referring to a dark non-abelian symmetry could not couple to a quark current that is not charged under the dark group.
The operators in this scenario produce amplitudes that are well behaved as $m\to 0$ and we find there are 6 operators as follows,
\begin{subequations}
\label{eq:OtildeqX}
\begin{eqnarray}
\tilde \calO_{qX1}^S& = & (\overline{q}q)X_{\mu\nu}^\dagger  X^{\mu\nu},
\\
\tilde \calO_{qX2}^S& = & (\overline{q}q)X_{\mu\nu}^\dagger \tilde X^{ \mu\nu},
\\
\tilde \calO_{qX1}^P& = & (\overline{q}i \gamma_5q)X_{\mu\nu}^\dagger X^{ \mu\nu},
\\
\tilde \calO_{qX2}^P& = & (\overline{q}i \gamma_5q)X_{\mu\nu}^\dagger \tilde X^{ \mu\nu},
\\
\tilde \calO_{qX1}^T& = &{i \over 2} (\overline{q}\sigma^{\mu\nu} q)(X^{\dagger}_{ \mu\rho} X^{\rho}_{\,\nu}-X^{\dagger}_{ \nu\rho} X^{\rho}_{\,\mu}), \, (\times) 
\\
\tilde \calO_{qX2}^T& = & {1\over 2}  (\overline{q} \sigma^{\mu\nu}\gamma_5 q)(X^{\dagger}_{ \mu\rho} X^{\rho}_{\,\nu}-X^{\dagger}_{ \nu\rho} X^{\rho}_{\,\mu}). \,  (\times) 
\end{eqnarray}
\end{subequations}
The dual field strength is defined as $\tilde X^{\mu\nu} =(1/2)\epsilon^{\mu\nu\rho\sigma}X_{\alpha\beta}$, and the symbol ``$(\times)$'' denotes an operator that vanishes for real vector fields. 

The method used to construct the LEFT quark-DM interactions given above can also be used to  obtain the corresponding lepton-DM interactions. We list those in appendix \ref{sec:leptonDMope} for completeness. 

In the literature, Ref.\,\cite{Kumar:2013iva} provides a list of LEFT operators for the fermion, scalar, and vector DM (scenario A) cases coupled to flavor diagonal currents. The fermion and scalar cases agree with our list, but the vector case does not.\footnote{The list of dim-6 operators with a derivative in \cite{Kumar:2013iva} is not complete. For example, it does not contain operators corresponding to $\calO_{qX3}^{V(A)}$.}  Here we discuss the more general case with flavor non-diagonal operators relevant to the FCNC processes we study. 

As mentioned above, we only impose the unbroken $SU(3)_{\rm c}\times U(1)_{\rm em}$ symmetry to obtain the LEFT operators. If we instead assume that all mediators are far beyond the weak scale, we can start from a SMEFT with the full SM gauge symmetry $SU(3)_{\rm c}\times SU(2)_L\times U(1)_{Y}$ and obtain the corresponding LEFT in the Higgs phase. The relevant DM EFT operators in this picture can be found in \cite{Kamenik:2011vy, Brod:2017bsw,Criado:2021trs, Arina:2021nqi, Aebischer:2022wnl}.
\footnote{ For the vector DM case, we find that the DM-quark (lepton) operators given in Tab.\,8 in \cite{Criado:2021trs} are neither independent nor complete. For example, for vector DM coupled to the down-type right-handed quark current, that paper provides 2 operators: $(\rho^\dagger_\mu D_\nu \rho^\mu)(\bar d\gamma^\nu d)$ and $(\rho^\dagger_\mu D_\nu \rho^\nu)(\bar d\gamma^\mu d)  $. The first one is equivalent to $\calO_{qX1}^{V+A}$ in our list, while the second one vanishes from the on-shell condition $\partial_\mu \rho^\mu=0$.   
} 
The FCNC $B$ and $K$ meson decays into fermion DM (or similar invisible particles like sterile neutrinos) have been studied in \cite{Kamenik:2011vy, Li:2020dpc, Felkl:2021uxi,Li:2019fhz, Deppisch:2020oyx, He:2021yoz}. For this reason, in the following we restrict ourselves to the scalar and vector DM cases and investigate the experimental sensitivity to the interactions in Eqs.\,(\ref{eq:Oqphi}-\ref{eq:OtildeqX}) in detail. We will first consider  $B$ meson decay in the next section, followed by the $K$ meson decay after that.

\section{$B\to (K,K^*, \pi, \rho)$+DM+DM}
\label{sec:Bdecay}

To calculate the decay rate for the $B\to M$ transition (where $M$ denotes either a pseudo-scalar meson $P=K,\pi$ or a vector meson $V=K^*, \rho$) from the effective interactions in Eqs.\,(\ref{eq:Oqphi}-\ref{eq:OtildeqX}), we first need to know the hadronic transition matrix elements $\langle M| \bar q \Gamma b|B\rangle$. These are usually parametrized in terms of  scalar form factors associated with each possible allowed Lorentz structure. The Lorentz structures can be organized according to  parity and charge conjugation. While some of the form factors can be determined from experimental data, others require theoretical models for the non-perturbative aspects of QCD. In the following subsections, we first collect the relevant form factors and their determination using light-cone sum rules \cite{Gubernari:2018wyi, Ball:2004ye, Bharucha:2015bzk, Lu:2018cfc, Gao:2019lta}. We then consider the decay rates for both scalar and vector DM scenarios and the implications for the parameter space. 

\subsection{Form factors}

We follow the parametrization of $B$ meson form factors in \cite{Gubernari:2018wyi}. 
For the $B \to P(J^P=0^-)$ transition with a final state pseudo-scalar $P = \pi,\, K$, the non-vanishing hadronic matrix elements from the scalar, vector, and tensor quark currents are parametrized by the form factors  $f_0, \ f_+$, and $f_T$, 
\begin{subequations}
\label{eq:formfacB2P}
\begin{eqnarray}
\langle P(k) | {\bar q} b | B(p) \rangle & = & 
{m_B^2 - m_P^2 \over m_b - m_{q} } f_0(q^2), 
\\
\langle P(k) | {\bar q} \gamma^\mu b | B(p) \rangle & = &
\left[ \left( p+k \right)^\mu - \frac{m_B^2-m_P^2}{q^2} q^\mu \right] f_{+}(q^2) 
+ \frac{m_B^2-m_P^2}{q^2} q^\mu f_{0}(q^2),
\\
\langle P(k) | {\bar q} \sigma^{\mu \nu} b | B(p) \rangle & = &
\frac{2 i }{m_B + m_P} ( p^\mu q^\nu - p^\nu q^\mu )  f_T(q^2) ,
\end{eqnarray}
\end{subequations}
where $k$ and $p$ are the 4-momenta of  $P$ and $B$ respectively,  $q^\mu = p^\mu - k^\mu $,  $m_B$ and $m_P$ are the masses of the initial $B$ meson and final state $P$ meson, and $m_b$ and $m_q$ are the masses of the quarks appearing in the currents. 
In the $q^2\to 0$ limit, $f_+(0) = f_0(0)$, and we follow the light-cone sum rule (LCSR) methods \cite{Ball:2004ye} to parametrize the dependence on the momentum transfer $s\equiv q^2$ as, 
\begin{eqnarray}
f_0(s) =  {r_2 \over 1-  s/m_{\rm fit}^2  }, 
\,
f_{+(T)}^\pi(s) = {r_1 \over 1- s/m_R^2 } +  {r_2 \over  1- s/m_{\rm fit}^2 }, 
\,
f_{+(T)}^K(s) = {r_1 \over 1-  s/m_R^2 } +  {r_2 \over \left( 1- s/m_R^2 \right)^2}.
\label{eq:FFpara1}
\end{eqnarray}
Above, $r_{1,2}$, $m_R^2$, and $m_{\rm fit}^2$, are parameters  with the preferred values given in \cite{Ball:2004ye} and collected in appendix \ref{sec:formfactor} for reference. 
\footnote{After submitting the manuscript, we became aware of a recent lattice calculation of the $B\to K$ form factors \cite{Parrott:2022rgu}. We find that using these new lattice results has no significant impact on the sensitivity curves we obtained with the LCSR calculation of the form factors from \cite{Ball:2004ye}.}

For the transition into a vector meson $V$, $B \to V(J^P=1^-)$ with $V= \rho,\, K^*$, the non-vanishing form factors $V_0, \ A_{0,1,2,3}, \ T_{1,2,3}$ are defined as
\begin{subequations}
\label{eq:formfacB2V}
\begin{eqnarray}
\langle V(k) | {\bar q} \gamma_5b | B(p) \rangle & = & 
- i \epsilon_{V,\nu}^* q^\nu {2 m_V \over m_b + m_{q}} A_0,  
\\
\langle V(k) | {\bar q} \gamma^\mu b | B(p) \rangle & = & 
 \epsilon^{\mu \nu \rho \sigma} \epsilon_{V,\nu}^* p_\rho k_\sigma \frac{2}{m_B+m_V}V_0,
 \\
\langle V(k) | {\bar q} \gamma^\mu \gamma_5 b | B(p) \rangle & = &  
i \epsilon_{V,\nu}^* \left[ g^{\mu \nu}(m_B+m_V)A_1 - \frac{(p+k)^\mu q^\nu}{m_B+m_V} A_2 - q^\mu q^\nu \frac{2m_V}{q^2} (A_3 - A_0) \right],
\\
\langle V(k) | {\bar q} \sigma_{\mu \nu}  b | B(p) \rangle & = & 
  i \epsilon_{\mu \nu  \rho  \sigma} \epsilon_{V,\alpha}^* 
 \left\{ 
g^{\alpha \rho}(p+k)^\sigma T_1 - g^{\alpha \rho }q^\sigma {m_B^2 - m_V^2 \over q^2} (T_1 - T_2)
\right.
\nonumber
\\
& + & \left. 2 q^\alpha p^\rho k^\sigma 
 \left[ {1\over m_B^2 - m_V^2} T_3 - {1 \over q^2 }(T_1 - T_2) \right]
 \right\},
\end{eqnarray}
\end{subequations}
where $\epsilon_V$ is the polarization vector of the spin-one meson and $m_V$ its mass.\footnote{In literature, the tensor current is usually parametrized by multiplying by the four-momentum $q_\nu$. From the tensor current in Eq.\,\eqref{eq:formfacB2V}, they can be directly calculated to take the form,   
\begin{subequations}
\begin{eqnarray}
\langle V(k) | {\bar q}  i\sigma^{\mu \nu} q_\nu b | B(p) \rangle & = & 
2 \epsilon^{\mu \nu \rho \sigma} \epsilon_{V,\nu}^* p_\rho k_\sigma T_1,
\\
 \langle V(k) | {\bar q} i \sigma^{\mu\nu}\gamma_5  q_\nu  b | B(p) \rangle & = & 
i \epsilon_{V,\nu}^* \left\{ \left[ g^{\mu \nu}(m_B^2-m_V^2)-(p+k)^\mu q^\nu \right] T_2
+ q^\nu \left[ q^\mu - \frac{q^2 (p+k)^\mu }{m_B^2-m_V^2} \right] T_3 \right\}. 
\end{eqnarray}
\end{subequations}
}
The axial-vector and pseudo-scalar current matrix elements can be related by EoM  via the relation $i\partial_\mu  \langle P|\bar q \gamma^\mu\gamma_5 b| B\rangle =
\langle P|\bar q i(\slashed{D} + \overleftarrow{ \slashed{D}})\gamma_5 b| B\rangle$.
Equivalently in momentum space, $q_\mu \langle P(k)|\bar q \gamma^\mu\gamma_5 b| B(p)\rangle =
- (m_b + m _q) \langle P(k)|\bar q\gamma_5  b| B(p)\rangle$, which implies
that the form factor $A_3$ is a redundant  and can be expressed in terms of $A_1$ and $A_2$ as, 
\begin{eqnarray}
A_3 \equiv \frac{m_B+m_V}{2 m_V}A_1 - \frac{m_B - m_V}{2 m_V} A_2.
\label{eq:A3}
\end{eqnarray}
It is common practice to replace $A_2$ and $T_3$ by
\begin{subequations}
\begin{eqnarray}
A_{12} & \equiv& \frac{(m_B + m_V)^2 (m_B^2 - m_V^2 - q^2)A_1 - \lambda(m_B^2, m_V^2,q^2) A_2}{16 m_B m_V^2 (m_B + m_V)},\\
T_{23} & \equiv& \frac{(m_B^2 - m_V^2) (m_B^2 + 3 m_V^2 - q^2)T_2 - \lambda(m_B^2, m_V^2,q^2) T_3}{8 m_B m_V^2 (m_B - m_V)}.
\label{eq:A12T23}
\end{eqnarray}
\end{subequations}
where the K$\ddot{a}$llen function $\lambda(x,y,x)$ is the usual,
\begin{eqnarray}
\lambda(x,y,z) \equiv x^2 +y^2 +z^2 - 2(xy+yz+zx). 
\label{eq:Kallen}
\end{eqnarray} 
In the $q^2\to 0$ limit, some of form factors are related with as follows, 
\begin{eqnarray}
A_0(0) = A_3(0),
\quad 
T_1(0) =T_2(0),
\quad
A_{12}(0) = \frac{m_B^2 - m_V^2}{8 m_B m_V} A_0(0). 
\end{eqnarray}
Relabeling the form factors $\{A_0, A_1, A_{12}, V_0, T_1, T_2, T_{23}\}$ as  
$F_{1,2,3,4,5,6,7}$, their momentum transfer dependence can be  parametrized  as  \cite{Bharucha:2015bzk}, 
\begin{eqnarray}
F_i(s) = {1 \over 1 - s / m_{R,i}^2 } \sum_k \alpha_k^i [ z(s) - z(0) ]^k,
\quad
z(s) \equiv {\sqrt{s_+ -s} -\sqrt{s_+ -s_0} \over  \sqrt{s_+ -s} + \sqrt{s_+ -s_0} }, 
\label{eq:FFpara2}
\end{eqnarray}
where $s_\pm \equiv (m_B\pm m_V)^2$ and $s_0 \equiv s_+(1 -\sqrt{1 - s_-/s_+})$. $m_{R,i}$ are the resonance masses associated with the transition modes and  are taken from Tab.\,3 in \cite{Bharucha:2015bzk}. The parameters $\alpha_k^i$ are truncated at quadratic order in $z$, $k_{\rm max}=2$, so that three fit parameters $\alpha_0^i,\alpha_1^i,\alpha_3^i$ are needed for each form factor $i$. They are given in Tab.\,14 of \cite{Bharucha:2015bzk} and we  collect them  in appendix \ref{sec:formfactor} for convenience.

\subsection{ $B  \to M + \phi \phi$ with scalar DM $\phi$}
\label{subsec:scalarDM}

For the quark-scalar DM interactions in Eq.\,\eqref{eq:Oqphi}, with the hadronic matrix elements given in Eq.\,\eqref{eq:formfacB2P} and Eq.\,\eqref{eq:formfacB2V}, the non-vanishing amplitudes for the processes $B(p) \to P(k) \phi( k_1) \phi^*(k_2)$ and $B(p) \to V(k) \phi( k_1) \phi^*(k_2)$ take the following general form, 
\begin{subequations}
\begin{eqnarray}
i\calM_{B\to P\phi\phi} &= & C_{q\phi}^{S,x b}\langle P(k)| \overline{q}_x b|B(p)\rangle
+C_{q\phi}^{V,x b} (k_1^\mu - k_2^\mu ) \langle P(k)| \overline{q}_x \gamma_\mu b |B(p)\rangle,
\\
i\calM_{B\to V\phi\phi} &= &  C_{q\phi}^{P,x b} \langle V(k)| \overline{q}_x i\gamma_5 b |B(p)\rangle
+ C_{q\phi}^{V,x b} (k_1^\mu - k_2^\mu ) \langle V(k)| \overline{q}_x \gamma_\mu b |B(p)\rangle
\nonumber
\\
& + & C_{q\phi}^{A,x b} (k_1^\mu - k_2^\mu) \langle V(k)| \overline{q}_x \gamma_\mu\gamma_5 b |B(p)\rangle,
\end{eqnarray}
\end{subequations}
where $x=d,s$ is a quark flavor label characterizing the final state meson $P(V)=\pi,K (\rho, K^*)$. Using the hadronic matrix elements in Eq.\,\eqref{eq:formfacB2P} and Eq.\,\eqref{eq:formfacB2V}, with the help of {\tt Feyncalc} \cite{Shtabovenko:2016sxi}, the differential decay widths  take the following compact form 
\begin{eqnarray}
{d\Gamma_{B\to P\phi\phi} \over d q^2} 
& = & 
{(m_B^2 - m_P^2)^2 \over 256 \pi^3 m_B^3 (m_b - m_{q_x})^2} 
\lambda^{1\over2}(m_B^2, m_P^2, s)  \kappa^{1\over2}(m^2,s)  f_0^2 \left|C_{q\phi}^{S,xb}\right|^2
\nonumber
\\
& + & 
{1 \over 768 \pi^3 m_B^3} \lambda^{3\over2}(m_B^2, m_P^2, s)  \kappa^{3\over2}(m^2,s) f_+^2 \left|C_{q\phi}^{V,xb}\right|^2, 
\\%
{d\Gamma_{B\to V\phi\phi} \over d q^2} 
& = & {1 \over 256 \pi^3 m_B^3 (m_b + m_{q_x})^2}  \lambda^{3\over2}(m_B^2, m_V^2, s)  \kappa^{1\over2}(m^2,s) A_0^2 \left|C_{q\phi}^{P,xb}\right|^2
\nonumber
\\
& + & { s \over 384\pi^3 m_B^3 (m_B + m_V)^2}  \lambda^{3\over2}(m_B^2, m_V^2, s)  \kappa^{3\over2}(m^2,s) V_0^2 \left|C_{q\phi}^{V,xb}\right|^2
\nonumber
\\
& + & {1 \over 384 \pi^3 m_B^3} \lambda^{1\over2}(m_B^2, m_V^2, s) \kappa^{3\over2}(m^2,s)
\left[ (m_B + m_V)^2 s A_1^2 +  32 m_B^2 m_V^2 A_{12}^2 \right] 
\left|C_{q\phi}^{A,xb}\right|^2, \quad
\end{eqnarray}
where $\kappa(m^2,s)$ is a kinetic ``endpoint'' function defined as 
\begin{eqnarray}
\kappa(m^2,s) \equiv 1 - {4m^2 \over s},
\end{eqnarray}
and the $q^2$ dependence of the form factors is left implicit for notational simplicity. 
In the above results, there are no interference effects between any pair of operators because the relevant hadronic matrix elements have different parity and/or charge conjugation properties and thus cannot mix. The dependence on the DM mass enters only through kinematics, unlike the vector DM case discussed below. For the case of real scalar DM, only the scalar and pseudo-scalar quark currents appear (due to some operators vanishing as noted in Eq.\,\eqref{eq:Oqphi}); there is also an additional factor of two in the decay width. 

\begin{figure}
\centering
\includegraphics[width=7cm]{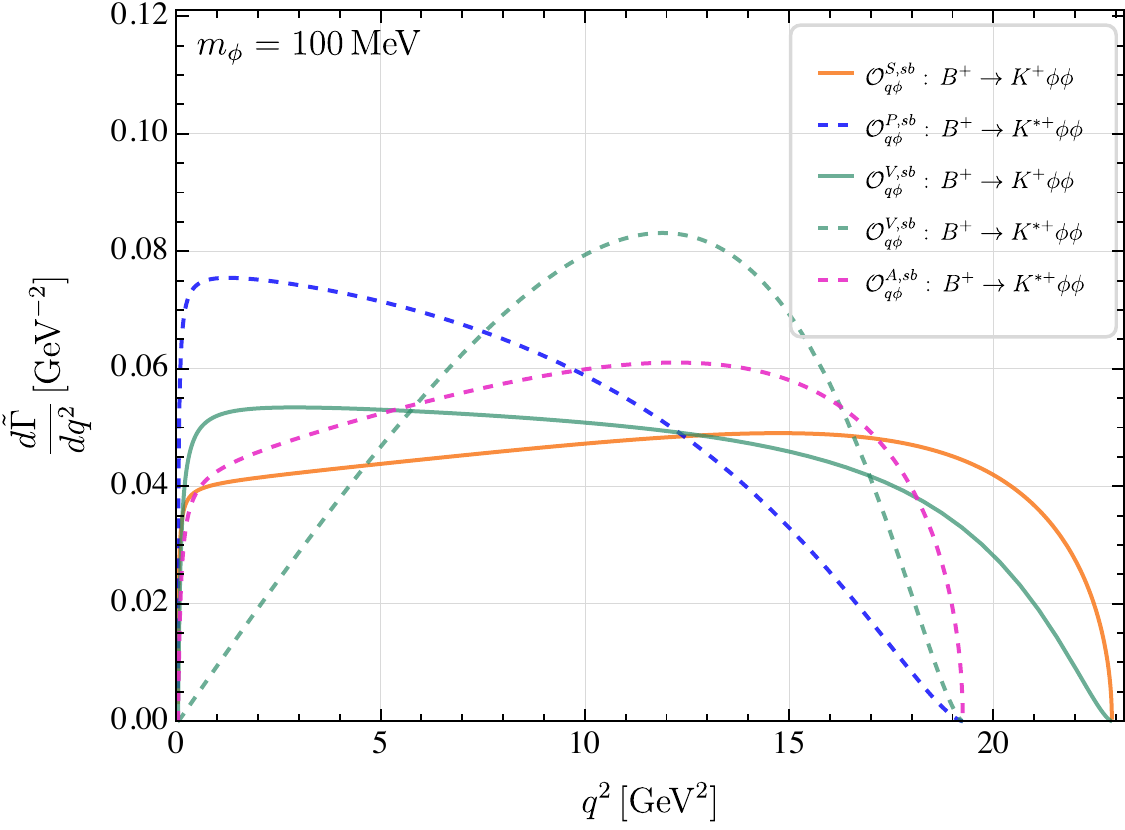}\qquad\quad
\includegraphics[width=7cm]{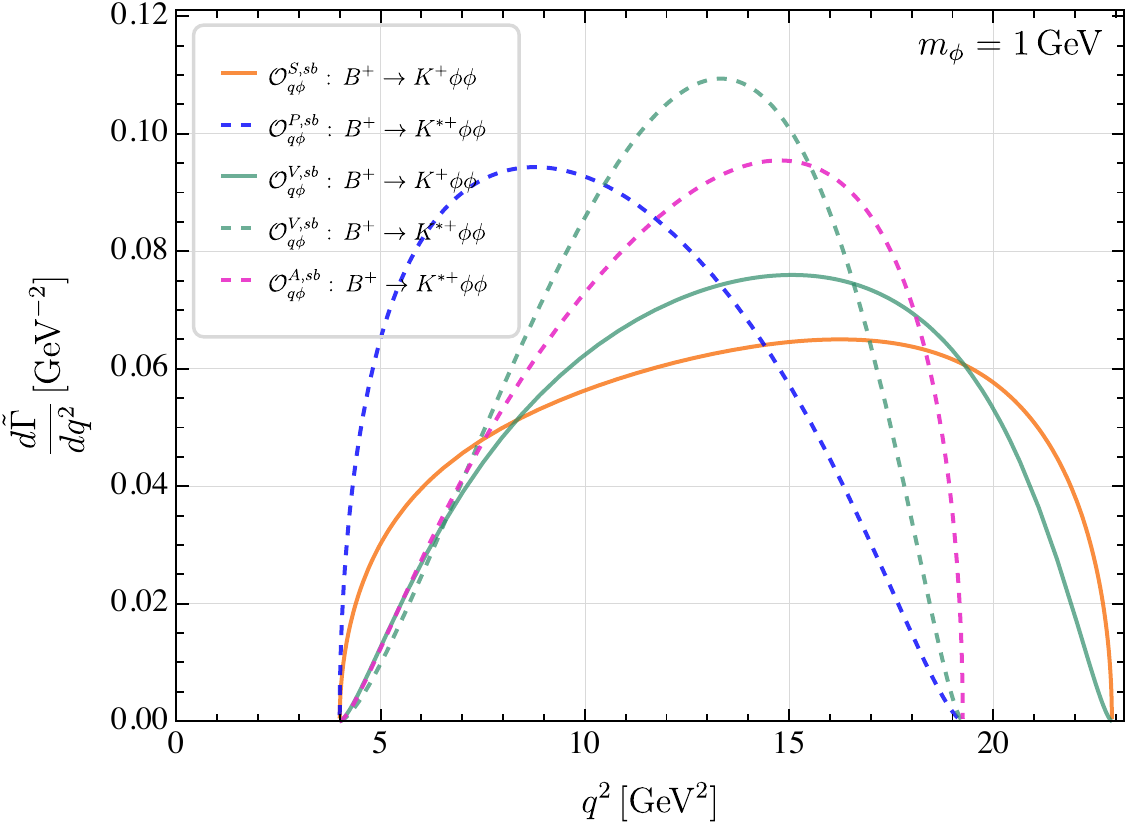}
\caption{Normalized differential decay width for $B\to K^{(*)+}\phi\phi$ decay from different types of operators. {\it Left panel}: $m=100\,\rm MeV$; {\it Right panel}: $m=1\,\rm GeV$. }
\label{fig:dBdq2scalar}
\end{figure}

The different operators result in different $q^2$ distributions in the $B\to M\slashed{E}$ decay. For example, we illustrate the  normalized distributions in the $b\to s$ transition for different operators in Fig.\,\ref{fig:dBdq2scalar}.  
\footnote{In the figures we add a subscript to differentiate the type of DM in question, but in the text we refer generically to the mass of any DM particle as $m$.}
In the left (right) panel, we consider scalar DM mass $m=100\, \rm MeV (1\, GeV)$ and in both cases a solid (dashed) line is used for the $B\to K^+\phi\phi$ ($K\to K^{*+}\phi\phi$) channels. One can clearly see that the distribution varies significantly between operators and as a function of the DM mass. This feature could be exploited to differentiate the various cases in the upcoming experimental search from Belle II. 

\begin{figure}
\centering
\includegraphics[width=7cm]{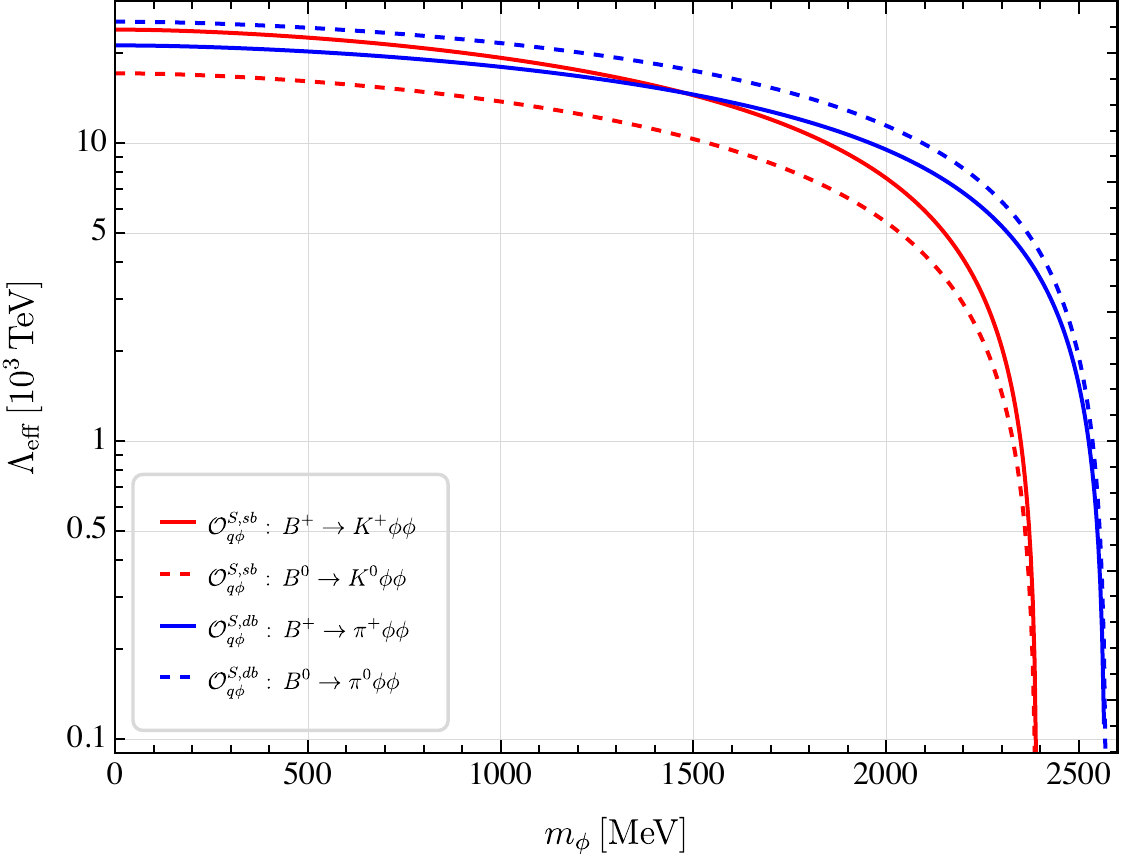}\qquad\quad
\includegraphics[width=7cm]{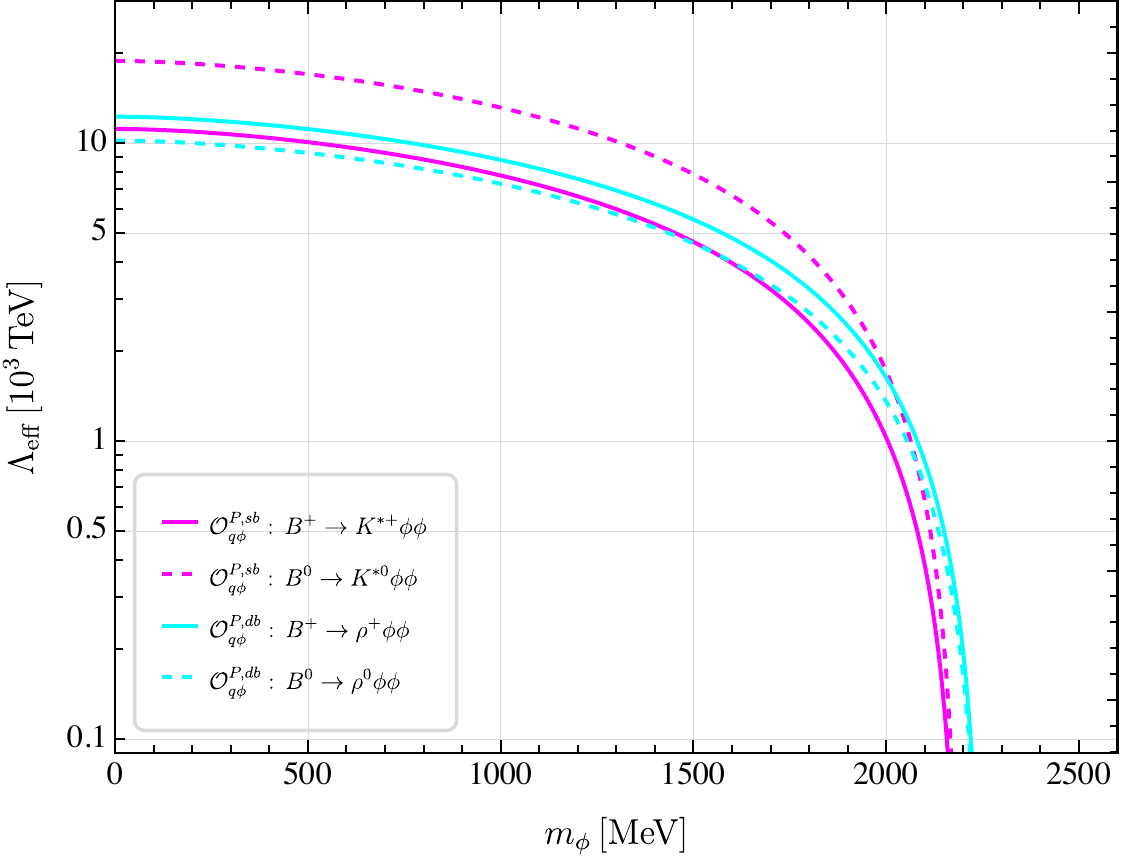}

 \bigskip
 
\includegraphics[width=7cm]{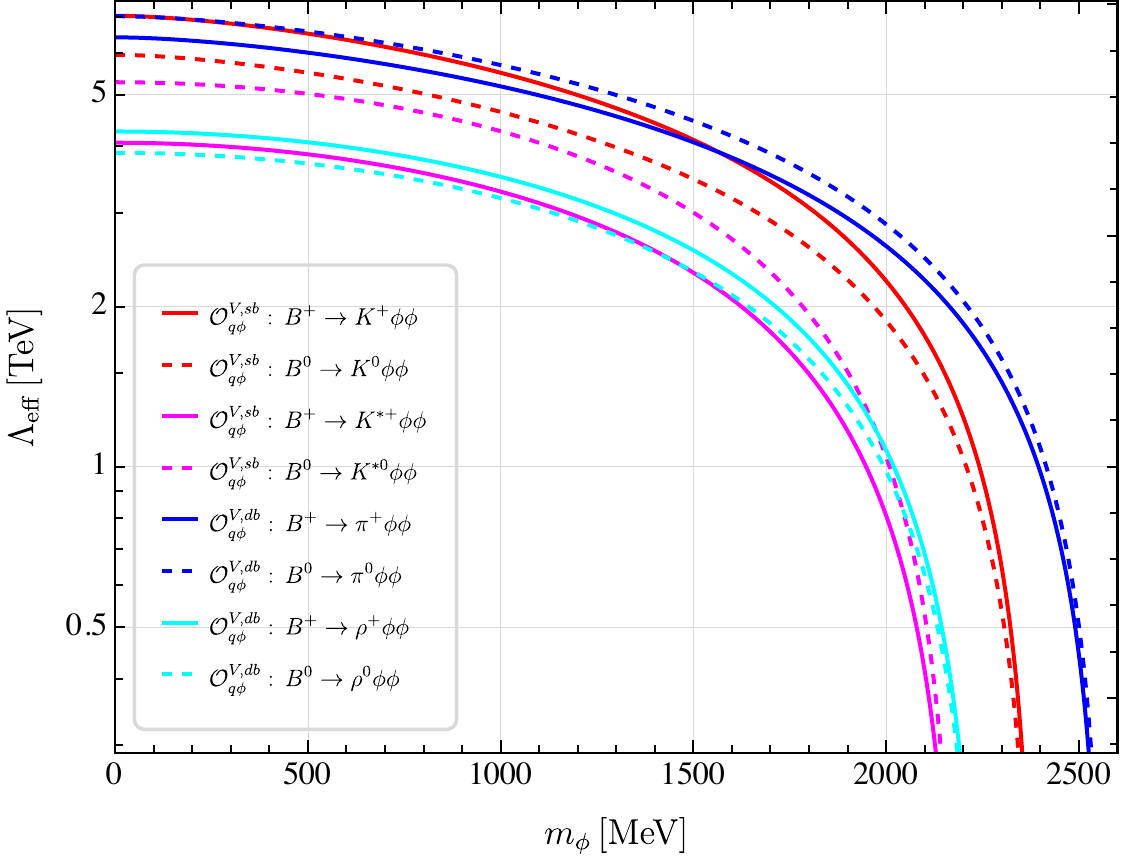}\qquad\quad
\includegraphics[width=7cm]{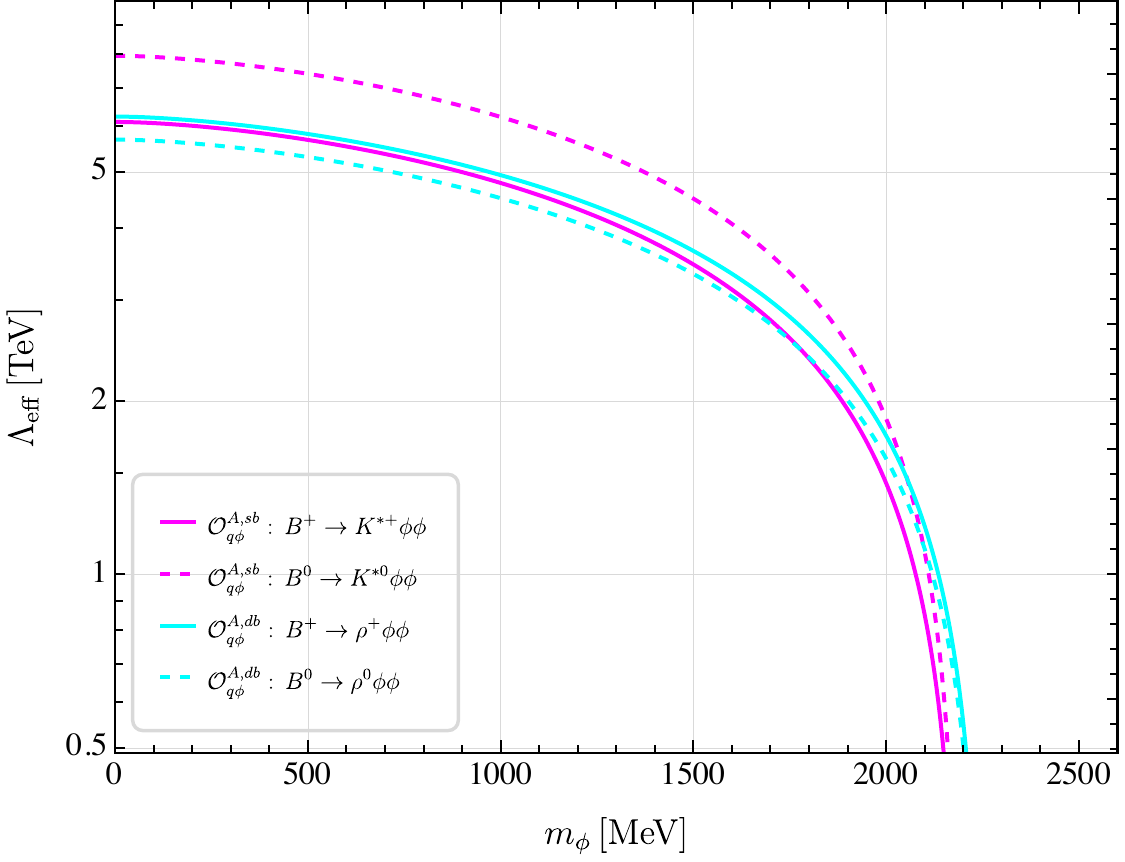}
\caption{Constraints on the effective new physics scale for each operator involving a $b$ quark as a function of the DM mass $m$ from all possible relevant $B$ meson decay channels. }
\label{fig:scalar_constraint}
\end{figure}
\begin{figure}
\centering
\includegraphics[width=7cm]{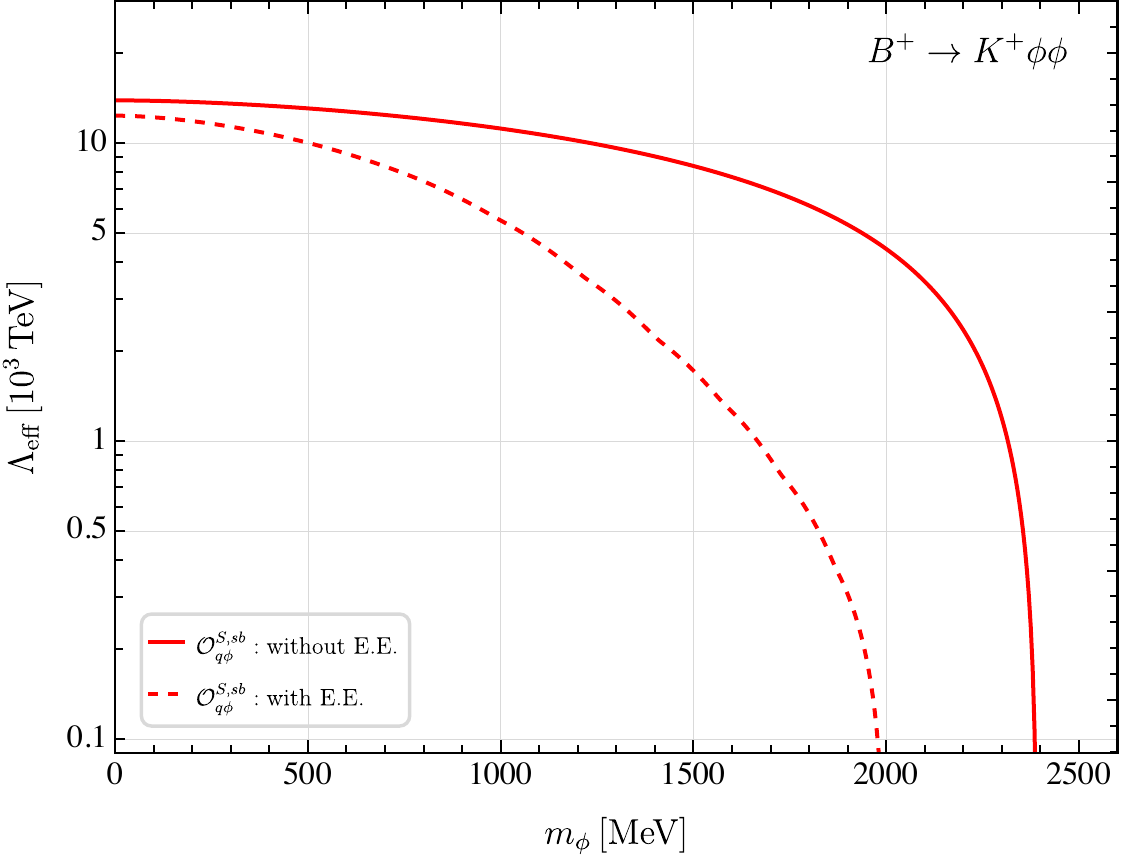}\qquad\quad
\includegraphics[width=7cm]{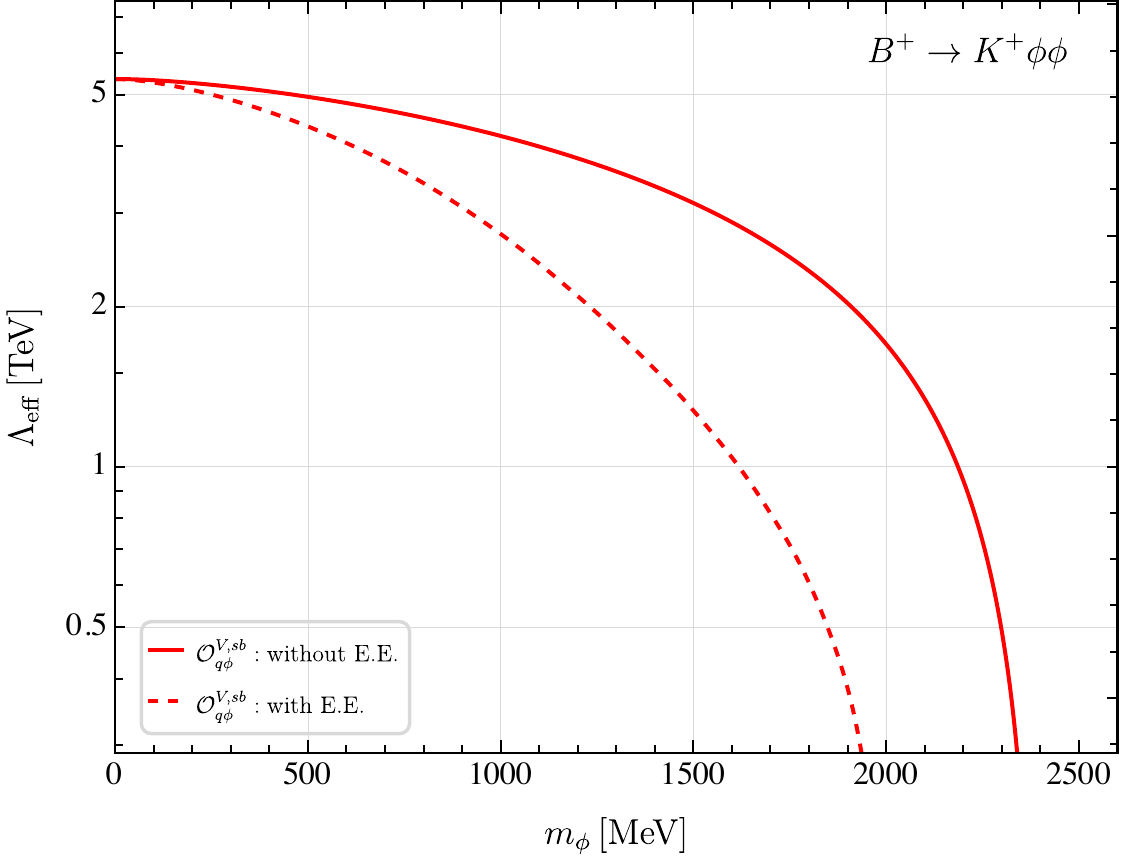}
\caption{Constraints on the effective new physics scale as a function of the DM mass $m$ from the inclusive tag Belle II $B^+\to K^+ \nu\bar\nu$  search without (solid lines) and with (dashed lines) experimental efficiency (E.E.) included. }
\label{fig:scalarBelleeff}
\end{figure}

To quantify the constraints set on the parameter by the current experimental bounds listed in Tab.\,\ref{tab:B2KmisE2}, we use an effective scale  $\Lambda_{\rm eff}$ associated with each operator from dimensional analysis:  $ C_{q\phi}^{S(P)} \equiv \Lambda_{\rm eff}^{-1}$ and  $C_{q\phi}^{V(A)}\equiv\Lambda_{\rm eff}^{-2}$. Fig.\,\ref{fig:scalar_constraint} shows the current experimental sensitivity in the $m$-$\Lambda$ plane for each operator with $(bs)$ and $(bd)$ flavor changing quark combinations. 
The solid (dashed) lines correspond to the constraints from charged (neutral) decay modes for both pseudo-scalar and vector final state mesons. The largest possible DM mass is restricted by the kinematic relation $m \leq (m_B-m_M)/2$ as reflected in each panel. Generally, for the scalar and vector current operators on the left  two panels, the charged mode $B^+\to K^+$ sets the stronger constraints for $(bs)$ transitions and the neutral mode $B^0\to \pi^0$ for the $(bd)$ transitions. For the pseudo-scalar and axial-vector quark current operators shown on the right two panels, the stronger bounds for the $(bs)$ and $(bd)$ transitions are set respectively by the neutral mode $B^0\to K^{*0}$ and the charged mode $B^+\to \rho^{+}$. This feature is just a reflection of the current experimental bounds as can be seen in Tab.\,\ref{tab:B2KmisE2}.

The situation depicted in Fig.\,\ref{fig:scalar_constraint} will, of course, be modified by experimental considerations. For example, Belle II has reported with its current measurement of $B^+\to K^+  \slashed{E}$ a signal efficiency that varies with $q^2$, peaking at low values and becoming very small for $q^2\gtrsim 12~{\rm GeV}^2$ \cite{Belle-II:2021rof}. Fig.\,\ref{fig:dBdq2scalar} then suggests that searches for very light DM will be more sensitive than searches for heavier DM. 
Inclusion of this experimental sensitivity changes the corresponding constraint, and we illustrate this in Fig.\,\ref{fig:scalarBelleeff}, where the left panel is for the scalar operator ${\cal O}_{q\phi}^{S,sb}$ while the right panel for the vector operator ${\cal O}_{q\phi}^{V,sb}$. The signal efficiency reported in \cite{Belle-II:2021rof} applies only to the search for $B^+\to K^+ \nu\bar\nu$ with an inclusive tag, and results in the 90\% confidence level limit ${\cal B}(B^+\to K^+ \nu\bar\nu) \leq 4.1\times 10^{-5}$, a few times weaker than the PDG value we quote in Table\,\ref{tab:B2KmisE2}. The solid lines in Fig.\,\ref{fig:scalarBelleeff} show the constraints from 
Fig.\,\ref{fig:scalar_constraint} for the $B^+\to K^+  \phi\phi$ mode, but using the weaker upper limit from the inclusive tag Belle II search. When the experimental efficiency is included, these limits turn into the ones depicted by dashed lines. To estimate these corrections, we scale the Belle II upper limit by a ratio of normalized rates weighted by the reported efficiency,
\begin{eqnarray}
\omega (m)
= {\sum_i   \tilde \Gamma_{i, \rm SM} \epsilon_i \over \sum_i  \tilde \Gamma_{i, \rm NP}(m) \epsilon_i }, 
\end{eqnarray}
where $\tilde \Gamma_{i, \rm NP}(m)$ is the normalized width from NP contribution in $i$-th bin, i.e., 
\begin{eqnarray}
\tilde \Gamma_{i, \rm NP}(m) = {1 \over \Gamma_{\rm NP}(m) } \int_{{\rm bin}_i} dq^2  {d \Gamma_{\rm NP}(m) \over d q^2},
\end{eqnarray}
and similarly for $\tilde \Gamma_{i, \rm SM}$.

\subsection{$B \to M + XX$ with vector DM $X$: scenario A }

For spin one DM, we first consider scenario A in which the operators were constructed using the vector field as given in Eq.\,\eqref{eq:OqX}.  Two of these operators involve a covariant derivative acting on the quark current, $\calO_{qX1}^{V}$ and $\calO_{qX1}^{A}$. To evaluate their contribution one needs form factors that have not been studied before. Within specific quark models for mesons one could estimate these form factors by replacing the derivatives with the corresponding quark momentum. As this is beyond the scope of this paper, we simply ignore these two operators in the following numerical study. The non-vanishing amplitudes for the processes $B(p) \to P(k) X( k_1) X^*(k_2)$ and $B(p) \to V(k) X( k_1) X^*(k_2)$ from the remaining operators in Eq.\,\eqref{eq:OqX} take the following form, 
\begin{subequations}
\begin{eqnarray}
i\calM_{B\to PXX}^A  &= & \epsilon^{*}_\rho (k_1) \epsilon^{*}_\sigma (k_2)
\left\{  g^{\rho\sigma} C_{qX}^{S,xb}  \langle P(k)|\bar q_x b| B(p)\rangle 
\right.
\nonumber
\\
& + & {i \over 2}  \left( 2 g^{\mu\rho} g^{\nu\sigma} C_{qX1}^{T,xb} + \epsilon^{ \mu\nu \rho\sigma}  C_{qX2}^{T,xb}\right)
 \langle P(k)| \bar q_x \sigma_{\mu\nu} b |B(p)\rangle
\nonumber
\\
& + & \left.
\left[ 
 i(g^{\mu\rho}k_{1\sigma}+g^{\mu\sigma}k_{2\rho})C_{qX2}^{V,xb} 
-i \epsilon^{\mu\nu\rho\sigma}(k_1-k_2)_\nu C_{qX3}^{V,xb} 
+ g^{\rho\sigma}(k_1- k_2)^\mu C_{qX4}^{V,xb}
\right.
\right.
\nonumber
\\
& - & \left.\left.
 (g^{\mu\rho}k_{1\sigma} - g^{\mu\sigma}k_{2\rho})C_{qX5}^{V,xb}
- \epsilon^{\mu\nu\rho\sigma} (k_1+ k_2)_\nu C_{qX6}^{V,xb} 
\right]
\langle P(k)|\bar q_x \gamma_\mu b| B(p)\rangle 
\right\},
\\%
i\calM_{B\to VXX}^A &= & \epsilon^{*}_\rho (k_1) \epsilon^{*}_\sigma (k_2)
\left\{  g^{\rho\sigma} C_{qX}^{P,xb}  \langle V(k)|\bar q_x i \gamma_5 b| B(p)\rangle 
\right.
\nonumber
\\
& + & {i \over 2}  \left(2  g^{\mu\rho} g^{\nu\sigma} C_{qX1}^{T,xb} +  \epsilon^{ \mu\nu \rho\sigma}  C_{qX2}^{T,xb} \right)
 \langle V(k)| \bar q_x \sigma_{\mu\nu} b |B(p)\rangle
\nonumber
\\
& + & \left.
 \left[ 
i(g^{\mu\rho}k_{1\sigma}+g^{\mu\sigma}k_{2\rho})C_{qX2}^{V,xb} 
-i \epsilon^{\mu\nu\rho\sigma}(k_1-k_2)_\nu C_{qX3}^{V,xb}  
+ g^{\rho\sigma}(k_1- k_2)^\mu C_{qX4}^{V,xb}
\right.
\right.
\nonumber
\\
& -& \left.
 (g^{\mu\rho}k_{1\sigma} - g^{\mu\sigma}k_{2\rho})C_{qX5}^{V,xb}
- \epsilon^{\mu\nu\rho\sigma} (k_1+ k_2)_\nu C_{qX6}^{V,xb} 
\right]
\langle V(k)|\bar q_x \gamma_\mu b| B(p)\rangle 
\nonumber
\\
& + & \left.
 \left[ 
 i(g^{\mu\rho}k_{1\sigma}+g^{\mu\sigma}k_{2\rho})C_{qX2}^{A,xb}
-i \epsilon^{\mu\nu\rho\sigma}(k_1-k_2)_\nu C_{qX3}^{A,xb}  
+ g^{\rho\sigma}(k_1- k_2)^\mu C_{qX4}^{A,xb}
\right.
\right.
\nonumber
\\
& -  & \left.\left.
(g^{\mu\rho}k_{1\sigma} - g^{\mu\sigma}k_{2\rho})C_{qX5}^{A,xb} 
- \epsilon^{\mu\nu\rho\sigma} (k_1+ k_2)_\nu C_{qX6}^{A,xb} 
\right]
\langle V(k)|\bar q_x \gamma_\mu\gamma_5 b| B(p)\rangle 
\right\}, 
\end{eqnarray}
\end{subequations}
For the final state with a pseudo-scalar meson $P$, using the hadronic matrix elements in Eq.\,\eqref{eq:formfacB2P}, leads to the differential decay width 
\begin{eqnarray}
{d\Gamma_{B\to PXX}^A \over d q^2} 
& = & 
{(m_B^2 - m_P^2)^2(s^2 - 4 m^2  s +12 m^4) \over 1024 \pi^3 m_B^3(m_b - m_{q_x})^2 m^4 } 
\lambda^{1\over2}(m_B^2, m_P^2, s)  \kappa^{1\over2}(m^2,s)  f_0^2 \left|C_{qX}^{S,xb}\right|^2
\nonumber
\\
& + & 
{s(s + 4 m^2) \over 3072 \pi^3 m_B^3 (m_B + m_P)^2 m^4} 
\lambda^{3\over2}(m_B^2, m_P^2, s)  \kappa^{3\over2}(m^2,s)  f_T^2 \left|C_{qX1}^{T,xb}\right|^2
\nonumber
\\
& + & 
{s + 2 m^2 \over 768 \pi^3 m_B^3(m_B + m_P)^2  m^2 } 
\lambda^{3\over2}(m_B^2, m_P^2, s)  \kappa^{1\over2}(m^2,s)  f_T^2 \left|C_{qX2}^{T,xb}\right|^2
\nonumber
\\
& + & 
{ s \over 3072 \pi^3 m_B^3 m^4}  \lambda^{1\over2}(m_B^2, m_P^2, s)  \kappa^{3\over2}(m^2,s)
\nonumber
\\
& \times & 
\left[ 3(s- 4 m^2) (m_B^2 - m_P^2)^2 f_0^2 +  4 m^2 \lambda(m_B^2, m_P^2, s)f_+^2 \right]
\left|C_{qX2}^{V,xb}\right|^2
\nonumber
\\
& + & 
{1  \over 768 \pi^3 m_B^3 m^2} \lambda^{1\over2}(m_B^2, m_P^2, s)  \kappa^{3\over2}(m^2,s)
\nonumber
\\
& \times & 
\left[ 6 m^2 (m_B^2 - m_P^2)^2 f_0^2 + (s - 4 m^2) \lambda(m_B^2, m_P^2, s)f_+^2 \right]
\left|C_{qX3}^{V,xb}\right|^2
\nonumber
\\
& + & 
{s^2 - 4m^2 s +12 m^4  \over 3072 \pi^3 m_B^3 m^4 } 
\lambda^{3\over2}(m_B^2, m_P^2, s)  \kappa^{3\over2}(m^2,s)  f_+^2 \left|C_{qX4}^{V,xb}\right|^2
\nonumber
\\
& + & 
{s(s + 4 m^2) \over 3072 \pi^3 m_B^3 m^4} 
\lambda^{3\over2}(m_B^2, m_P^2, s)  \kappa^{3\over2}(m^2,s)  f_+^2 \left|C_{qX5}^{V,xb}\right|^2
\nonumber
\\
& + & 
{s + 2 m^2 \over 768 \pi^3 m_B^3 m^2} 
\lambda^{3\over2}(m_B^2, m_P^2, s)  \kappa^{1\over2}(m^2,s)  f_+^2 \left|C_{qX6}^{V,xb}\right|^2
+\cdots. 
\end{eqnarray}
We have dropped interference between different operators (represented by ``$\cdots$'' above) as our numerical study will deal only with one operator at a time.  

For the final state with a vector meson $V$, using the form factors in Eq.\,\eqref{eq:formfacB2V}, we obtain 
\begin{eqnarray}
{d\Gamma_{B\to VXX}^A \over d q^2} 
& = & 
{s^2 - 4 m^2  s +12 m^4 \over 1024 \pi^3 m_B^3 (m_b+ m_{q_x})^2 m^4} 
\lambda^{3\over2}(m_B^2, m_V^2, s)  \kappa^{1\over2}(m^2,s)  A_0^2 \left|C_{qX}^{P,xb}\right|^2
\nonumber
\\
& + &
{1 \over 1536 \pi^3 m_B^3 m^4 s } 
 \lambda^{1\over2}(m_B^2, m_V^2, s)  \kappa^{1\over2}(m^2,s) 
\Big\{
(s^2 - 16 m^4) \lambda(m_B^2, m_V^2, s) T_1^2 
 \nonumber
 \\
 & +  &
\left.
 4m^2 (s + 2 m^2 ) \left[ 
 (m_B^2 - m_V^2)^2  T_2^2 
+ { 8 m_B^2 m_V^2 s \over (m_B+m_V)^2 }T_{23}^2  \right]
\right\}
\left|C_{qX1}^{T,xb}\right|^2
\nonumber
\\
& + &
{1 \over 1536 \pi^3 m_B^3 m^4 s } 
 \lambda^{1\over2}(m_B^2, m_V^2, s)  \kappa^{1\over2}(m^2,s) 
\Big\{
4 m^2 (s + 2 m^2) \lambda(m_B^2, m_V^2, s) T_1^2 
 \nonumber
 \\
 & + &
 \left.
 (s^2 -16 m^4 ) \left[  (m_B^2 - m_V^2)^2 T_2^2 
+ { 8 m_B^2 m_V^2 s \over (m_B+m_V)^2 } T_{23}^2 
\right]
\right\}
\left|C_{qX2}^{T,xb}\right|^2
\nonumber
\\
& + &
 { s^2 \over 384 \pi^3 m_B^3 (m_B+ m_V)^2 m^2} 
 \lambda^{3\over2}(m_B^2, m_V^2, s)  \kappa^{3\over2}(m^2,s)  V_0^2 \left|C_{qX2}^{V,xb}\right|^2
\nonumber
\\
& + &
 { s^2 \over 384 \pi^3 m_B^3 (m_B+ m_V)^2 m^2} 
 \lambda^{3\over2}(m_B^2, m_V^2, s)  \kappa^{5\over 2}(m^2,s)  V_0^2 \left|C_{qX3}^{V,xb}\right|^2
\nonumber
\\
& + & { s(s^2- 4m^2 s + 12 m^4) \over 1536 \pi^3 m_B^3 (m_B+ m_V)^2 m^4  } 
 \lambda^{3\over2}(m_B^2, m_V^2, s)  \kappa^{3\over2}(m^2,s)  V_0^2 \left|C_{qX4}^{V,xb}\right|^2
\nonumber
\\
& + & { s^2(s +4 m^2) \over 1536 \pi^3 m_B^3 (m_B+ m_V)^2 m^4  } 
 \lambda^{3\over2}(m_B^2, m_V^2, s)  \kappa^{3\over2}(m^2,s)  V_0^2 \left|C_{qX5}^{V,xb}\right|^2
\nonumber
\\
& + & { s(s + 2 m^2) \over 384 \pi^3 m_B^3 (m_B+ m_V)^2  m^2 } 
 \lambda^{3\over2}(m_B^2, m_V^2, s)  \kappa^{1\over2}(m^2,s)  V_0^2 \left|C_{qX6}^{V,xb}\right|^2
\nonumber
\\
& + &
  {s \over 3072 \pi^3 m_B^3 m^4 } \lambda^{1\over2}(m_B^2, m_V^2, s) \kappa^{3\over2}(m^2,s) 
\left\{ 3(s-4m^2) \lambda(m_B^2, m_V^2, s) A_0^2
\right.
\nonumber
\\
&+&
\left.
8m^2 \left[ (m_B+m_V)^2 s A_1^2 + 32 m_B^2 m_V^2 A_{12}^2 \right] \right\}
\left|C_{qX2}^{A,xb}\right|^2
\nonumber
\\
& + &
  {1 \over 384 \pi^3 m_B^3 m^2 } \lambda^{1\over2}(m_B^2, m_V^2, s) \kappa^{3\over2}(m^2,s) 
\left\{3 m^2 \lambda(m_B^2, m_V^2, s) A_0^2 
\right.
\nonumber
\\
&+&
\left.
(s-4m^2)\left[ (m_B+m_V)^2 s A_1^2 + 32 m_B^2 m_V^2 A_{12}^2 \right] \right\}
\left|C_{qX3}^{A,xb}\right|^2
\nonumber
\\
& + &
  {s^2 -4 m^2 s + 12 m^4 \over 1536 \pi^3 m_B^3 m^4 } \lambda^{1\over2}(m_B^2, m_V^2, s) \kappa^{3\over2}(m^2,s) 
  \nonumber
\\
&\times&
\left[ (m_B+m_V)^2 s A_1^2 + 32 m_B^2 m_V^2 A_{12}^2 \right]
\left|C_{qX4}^{A,xb}\right|^2
\nonumber
\\
& + &  {s(s + 4m^2) \over 1536 \pi^3 m_B^3 m^4 } \lambda^{1\over2}(m_B^2, m_V^2, s) \kappa^{3\over2}(m^2,s) 
  \nonumber
\\
&\times&
\left[ (m_B+m_V)^2 s A_1^2 + 32 m_B^2 m_V^2 A_{12}^2 \right]
\left|C_{qX5}^{A,xb}\right|^2
\nonumber
\\
& + &  {s + 2m^2 \over 384 \pi^3 m_B^3 m^2 } \lambda^{1\over2}(m_B^2, m_V^2, s) \kappa^{1\over2}(m^2,s) 
  \nonumber
\\
&\times&
\left[ (m_B+m_V)^2 s A_1^2 + 32 m_B^2 m_V^2 A_{12}^2 \right]
\left|C_{qX6}^{A,xb}\right|^2
+\cdots.  
\end{eqnarray}
where again we have dropped interference between different operators.  

It is evident that the above differential decay widths (and also decay widths) diverge in the limit of vanishing DM mass, $m\to 0$. As discussed above, we assume that each relevant Wilson coefficient also depends on the DM mass to some power determined by the number of independent vector four-potentials that cannot be reduced to field strength tensors. Operationally we use effective scales $\Lambda_{\rm eff}$ defined as follows,
\begin{eqnarray}
C_{qX}^{S,P} \equiv {m^2 \over \Lambda_{\rm eff}^3}, 
\quad
C_{qX1,2}^{T} \equiv {m^2 \over \Lambda_{\rm eff}^3}, 
\quad
C_{qX2,4,5}^{V,A} \equiv {m^2 \over \Lambda_{\rm eff}^4},
\quad
C_{qX3,6}^{V,A} \equiv  {m \over \Lambda_{\rm eff}^3}.
\label{eq:CqX2mLam}
\end{eqnarray}
In Eq.\,\eqref{eq:CqX2mLam}, $C_{qX3,6}^{V,A}$ depends linearly on $m$ because one of two vector fields in the operators $\calO_{qX3,6}^{V,A}$ can be rewritten as a field strength tensor. 

\begin{figure}
\centering
\includegraphics[width=7cm]{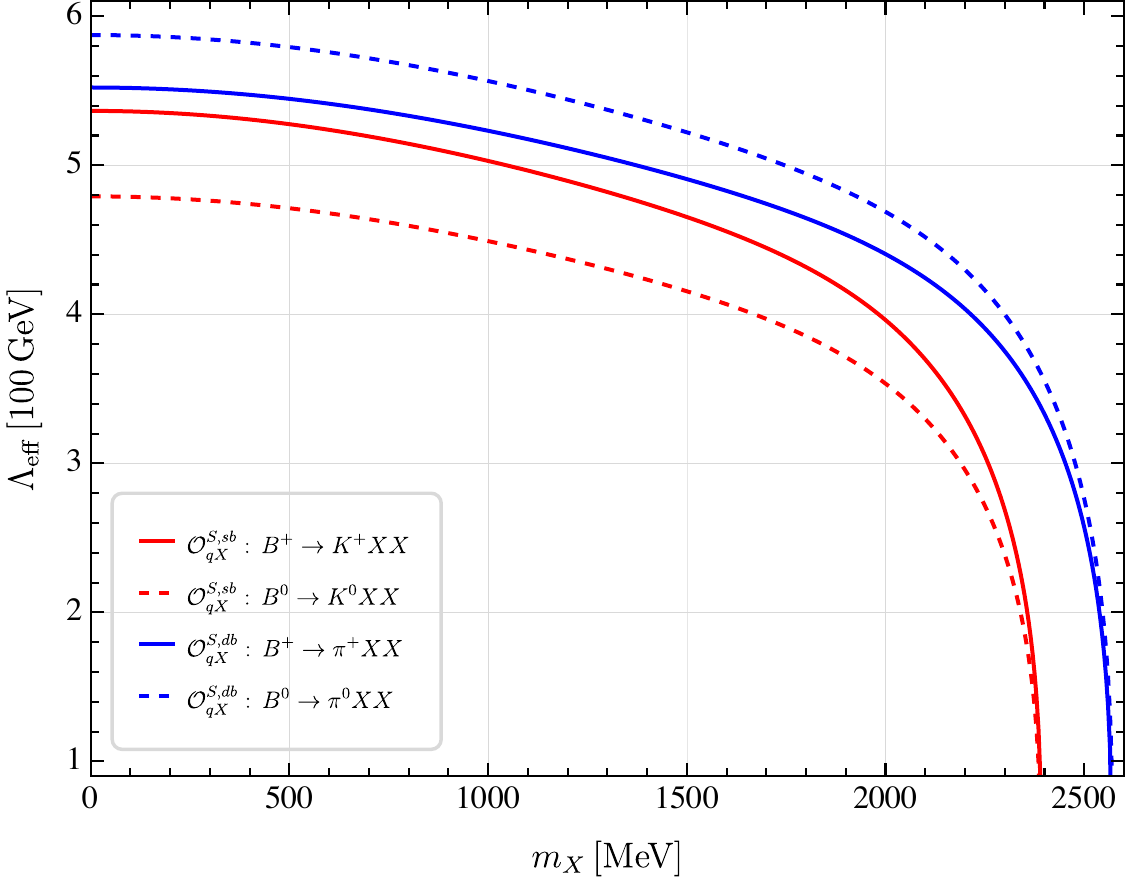}\qquad\quad
\includegraphics[width=7cm]{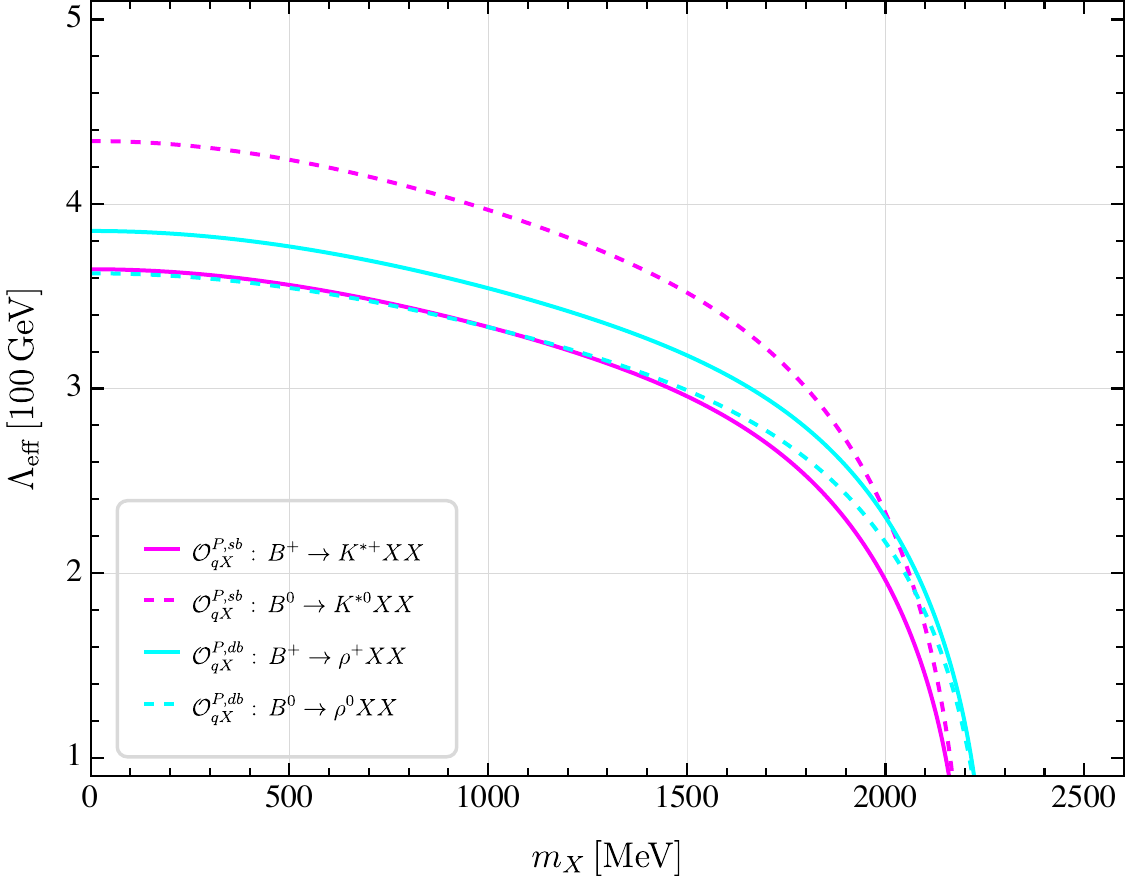}

 \bigskip  
\includegraphics[width=7cm]{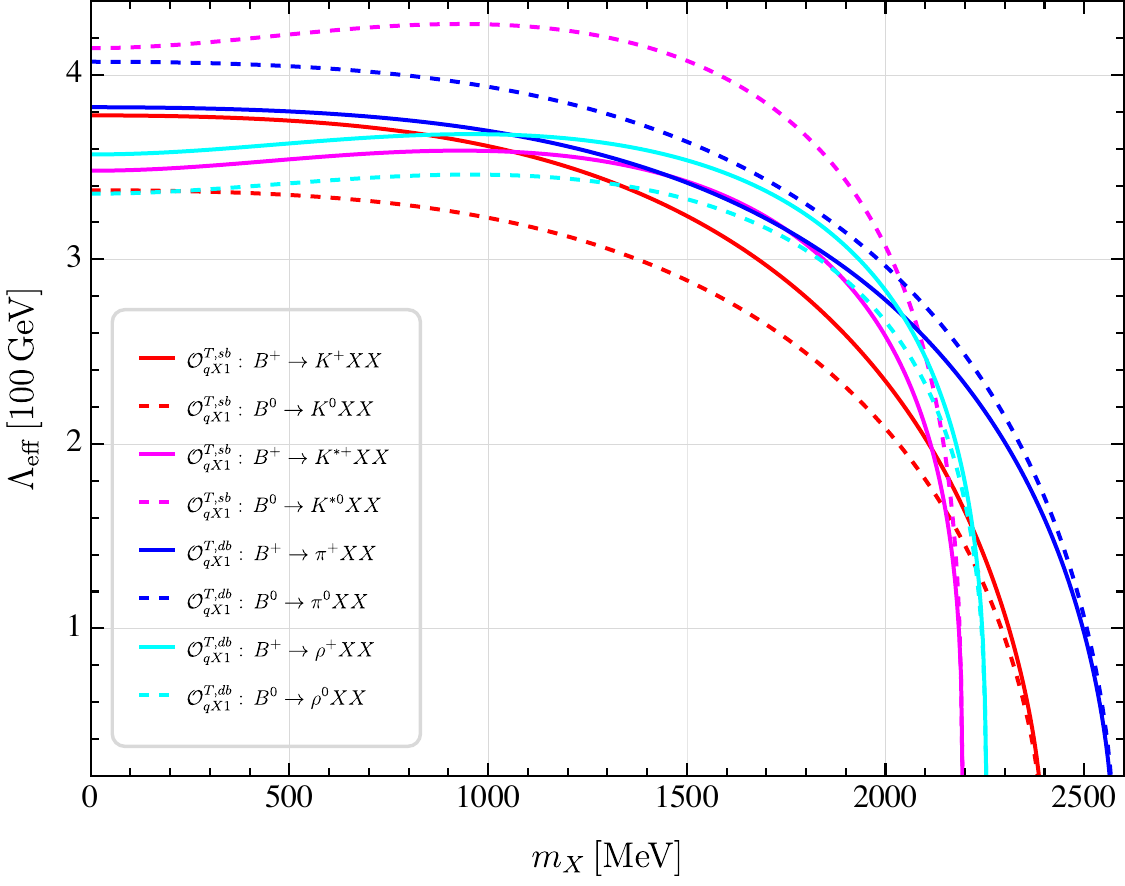}\qquad\quad
\includegraphics[width=7cm]{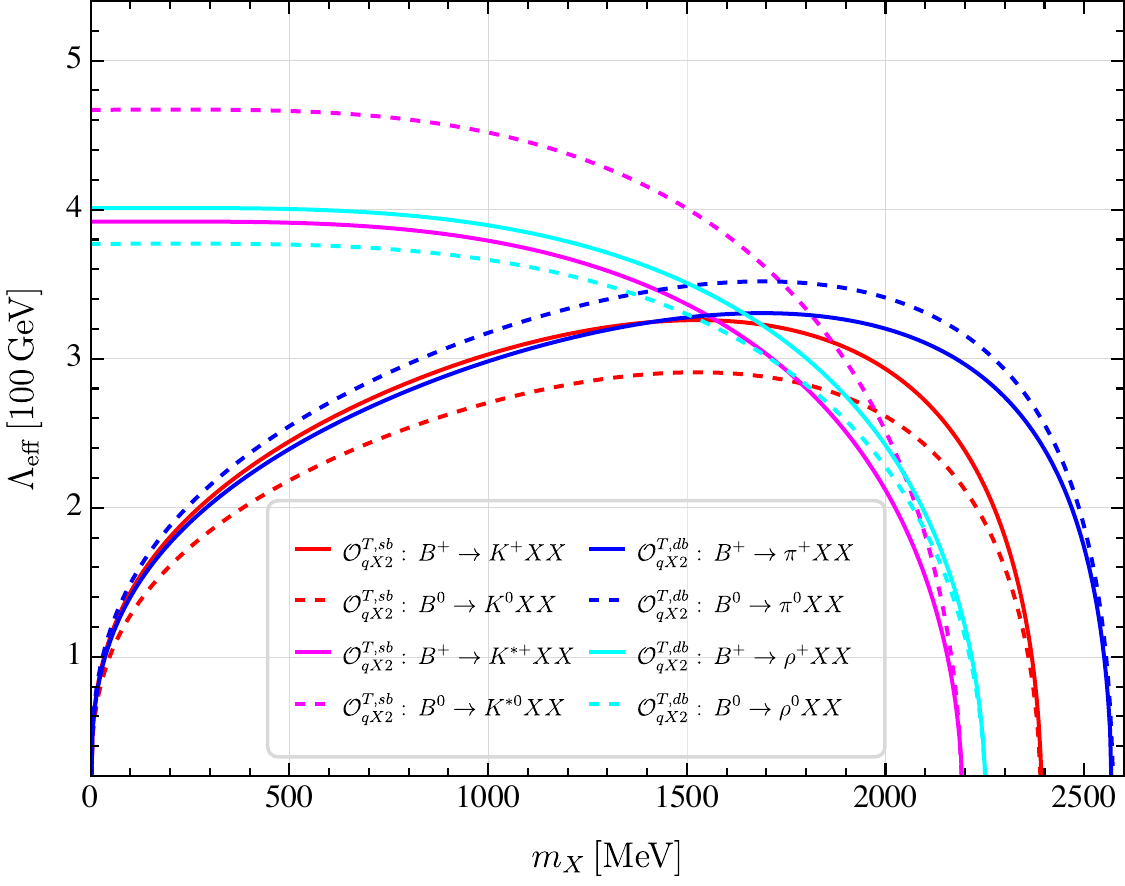}
\caption{Constraints on the effective new physics scale for the 4 dim-5 operators $\calO_{qX}^{S,P}$ and $\calO_{qX1,2}^{T}$ as a function of the DM mass $m$ from $B\to K(\pi) \slashed{E}$ channels.  }
\label{fig:OqXSPT}
\end{figure}

The possible origin of these mass factors is illustrated with an example in appendix \ref{sec:models}.  The divergence as $m\to 0$ also affects the kaon decay mode, $K\to \pi XX$, where we use a  parametrization similar to Eq.\,\eqref{eq:CqX2mLam}. As already mentioned, the problem with the $m\to 0$ limit can be  avoided by assuming these vector particles are gauge bosons of a dark gauge symmetry and requiring them to enter the LEFT as field strength tensors as  in Eq.\,\eqref{eq:OtildeqX}. We elaborate on this second scenario for  vector DM  in the next subsection. 

Fig.\,\ref{fig:OqXSPT} shows the current experimental sensitivity in the $m$-$\Lambda_{\rm eff}$ plane for the 4 dim-5 operators $\calO_{qX}^{S,P}$ and $\calO_{qX1,2}^{T}$ following Eq.\,\eqref{eq:CqX2mLam}. For the quark scalar current operator $\calO_{qX}^{S}$ (left upper panel), the charged mode ($B^+\to K^+XX$) gives the strongest constraint for $(sb)$  flavor indices in the whole DM mass range. Similarly, for $(db)$ flavor indices, the strongest constraint arises from the neutral mode $B^0\to \pi^0XX$.   For the quark pseudo-scalar current operator $\calO_{qX}^{P}$ (right upper panel), the constraints also exhibit a similar behavior to those for the pseudo-scalar DM operator $\calO_{q\phi}^{P}$. However,  due to the different dimensionality of the quark (pseudo-)scalar current operators in the two cases, the numerical results are very different for the two cases, as clearly seen in Fig.\,\ref{fig:scalar_constraint} and Fig.\,\ref{fig:OqXSPT}.

\begin{figure}
\centering
 
\includegraphics[width=7cm]{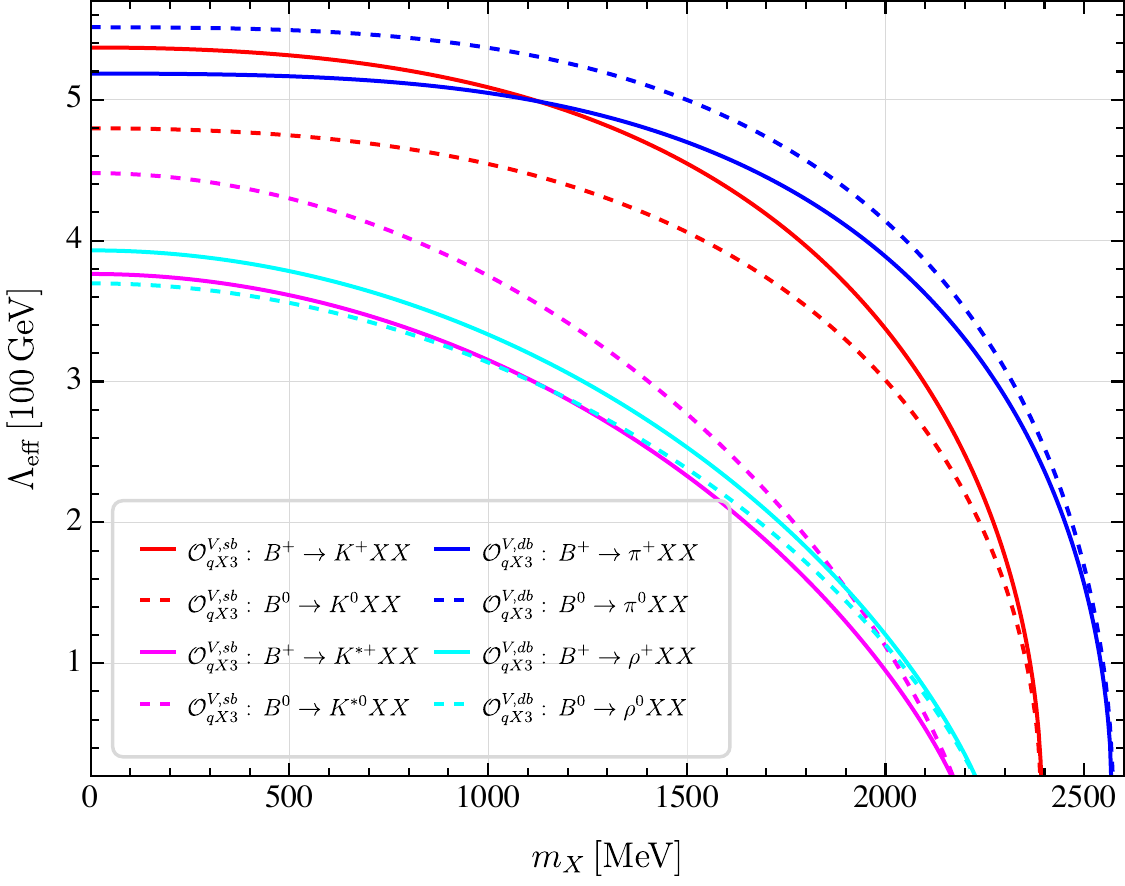}\qquad\quad
\includegraphics[width=7cm]{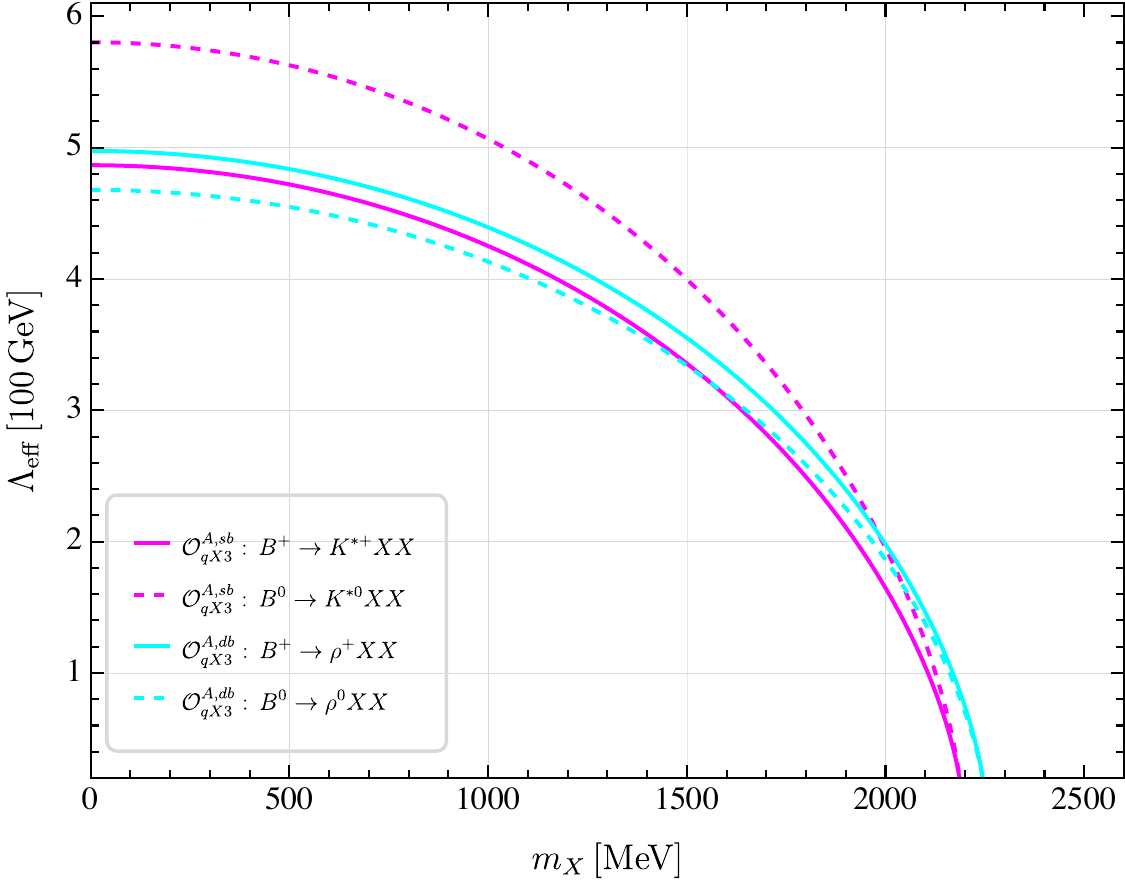}

\bigskip
\includegraphics[width=7cm]{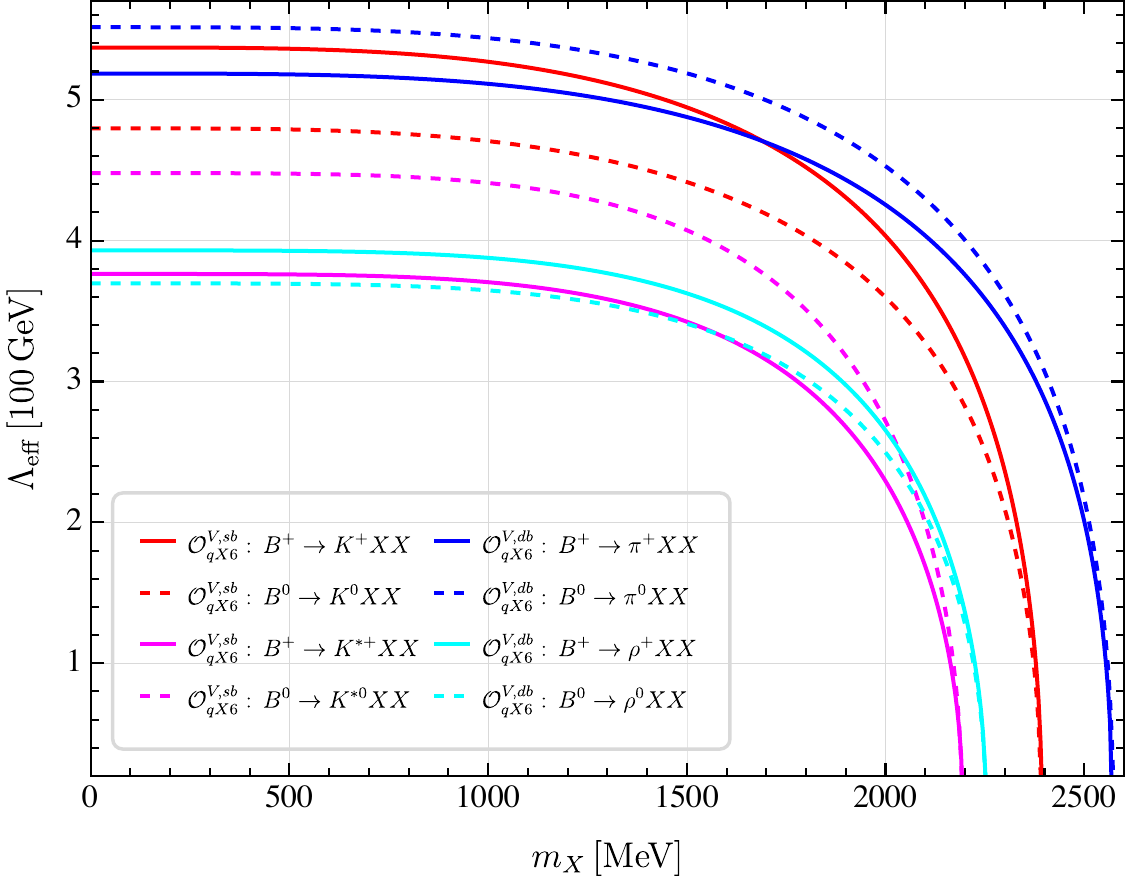}\qquad\quad
\includegraphics[width=7cm]{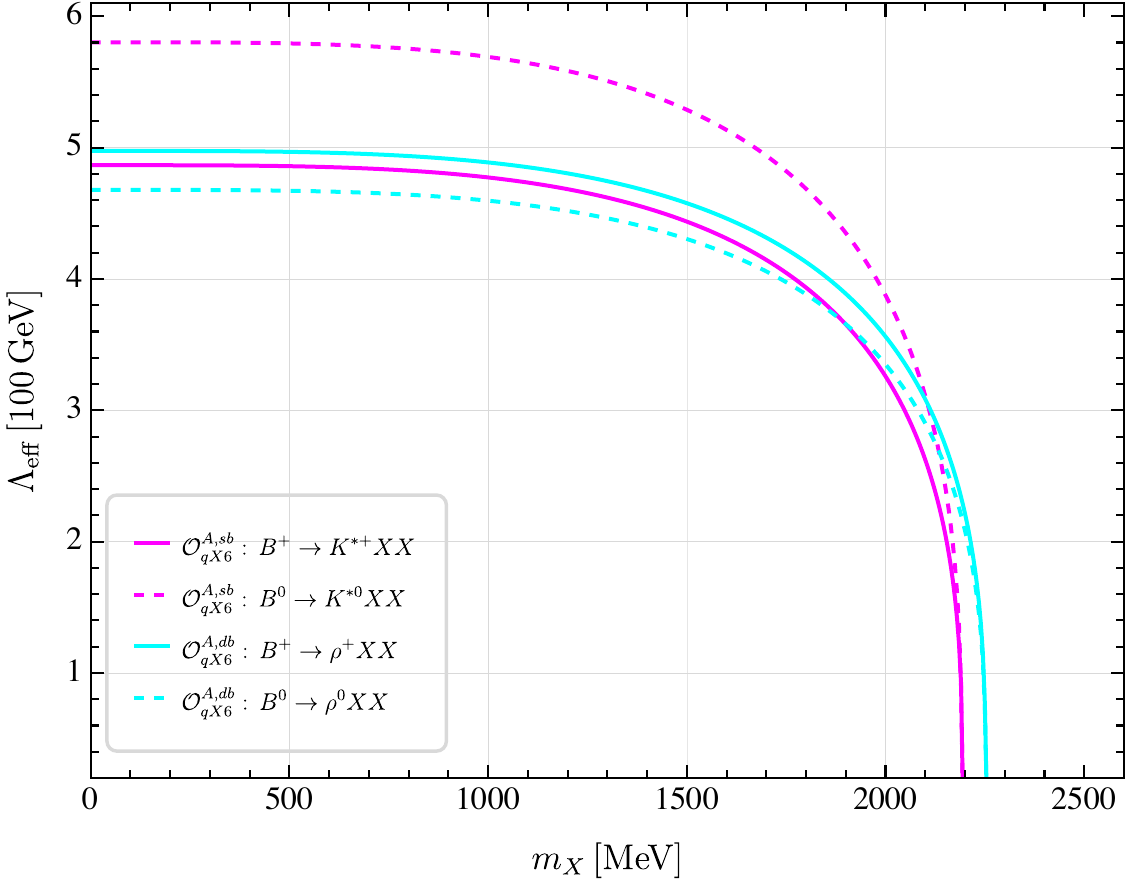}

\caption{Constraints on the effective new physics scale for the 4 dim-6 operators $\calO_{qX3,6}^{V,A}$ as a function of the DM mass $m$ from $B\to K(\pi) \slashed{E}$ channels. }
\label{fig:OqX36VA}
\end{figure}

\begin{figure}
\centering
\includegraphics[width=7cm]{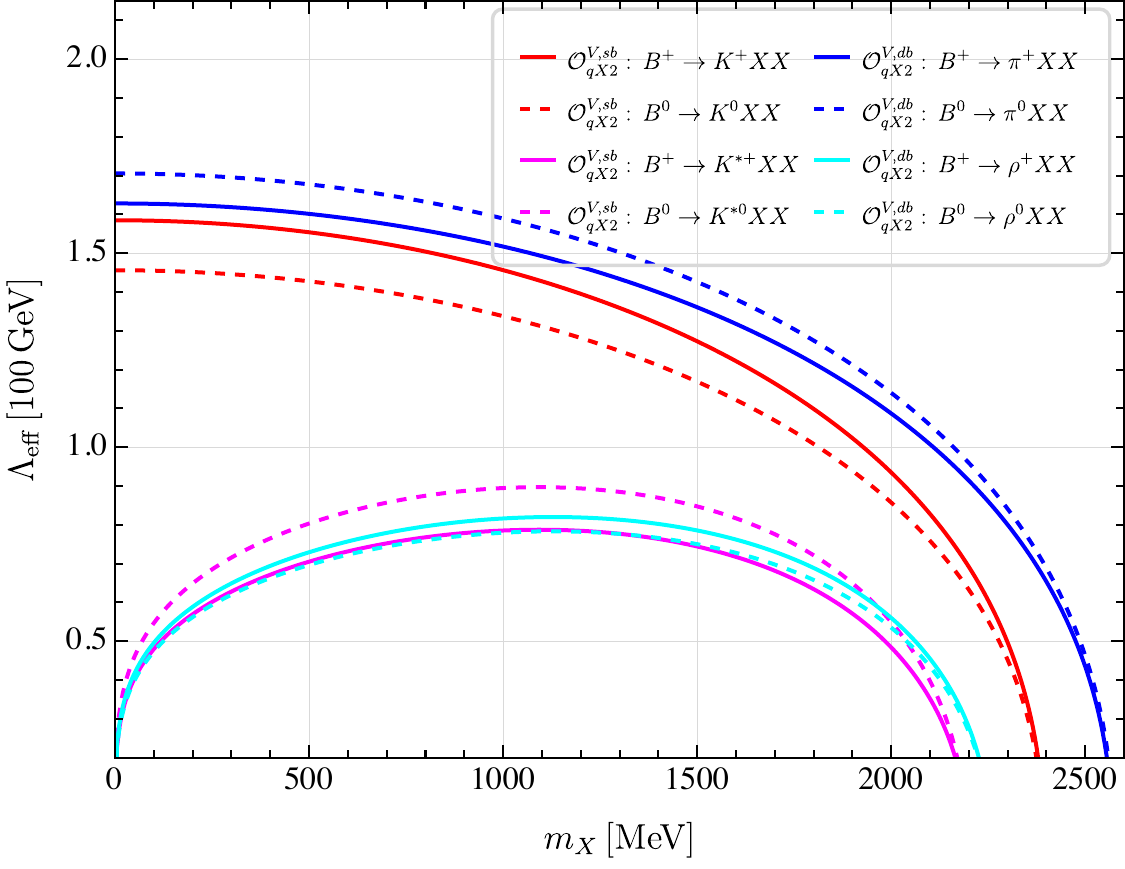}\qquad\quad
\includegraphics[width=7cm]{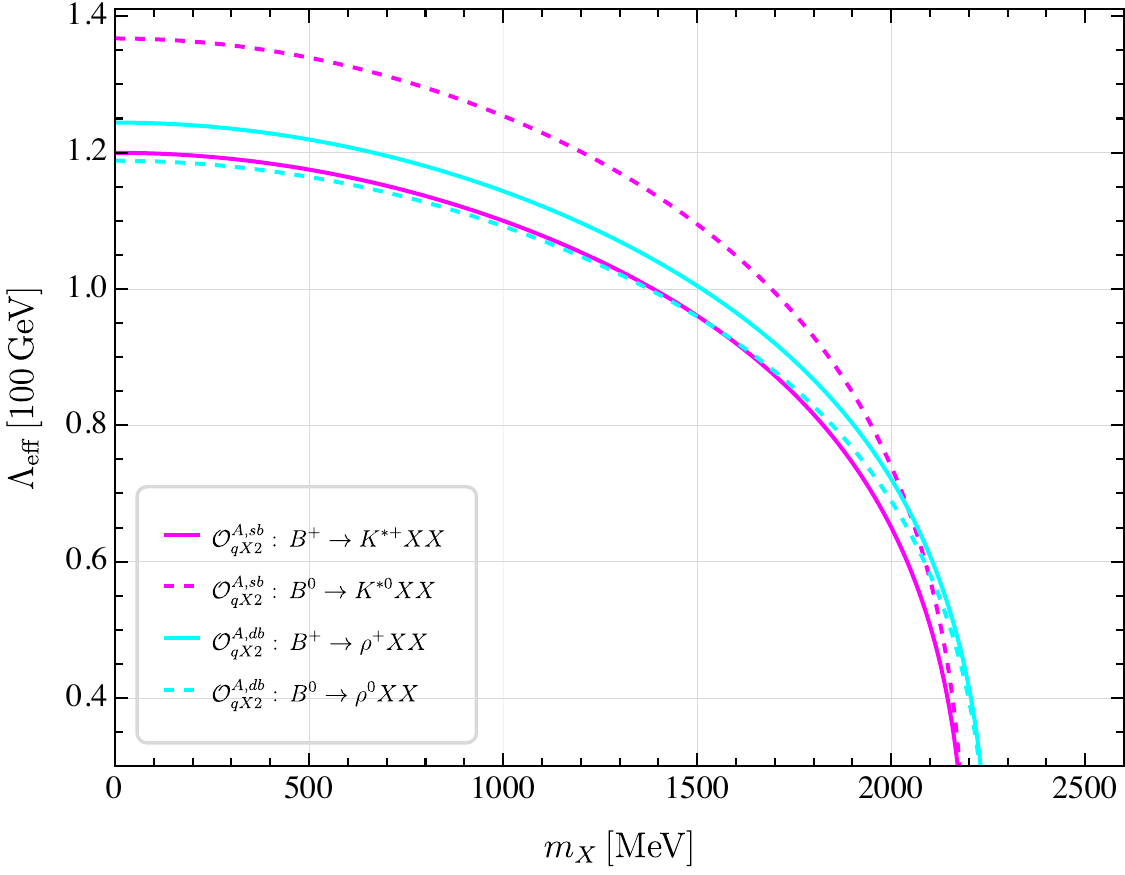}

 \bigskip 
\includegraphics[width=7cm]{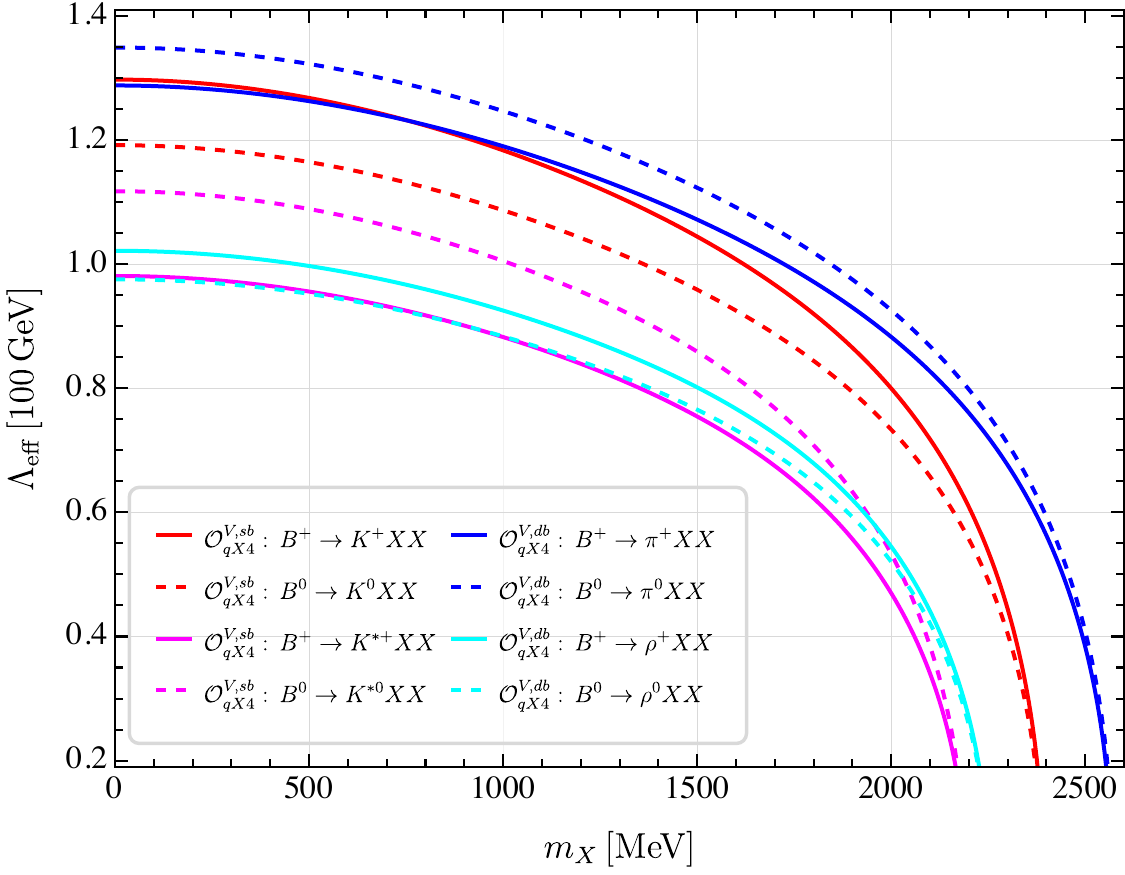}\qquad\quad
\includegraphics[width=7cm]{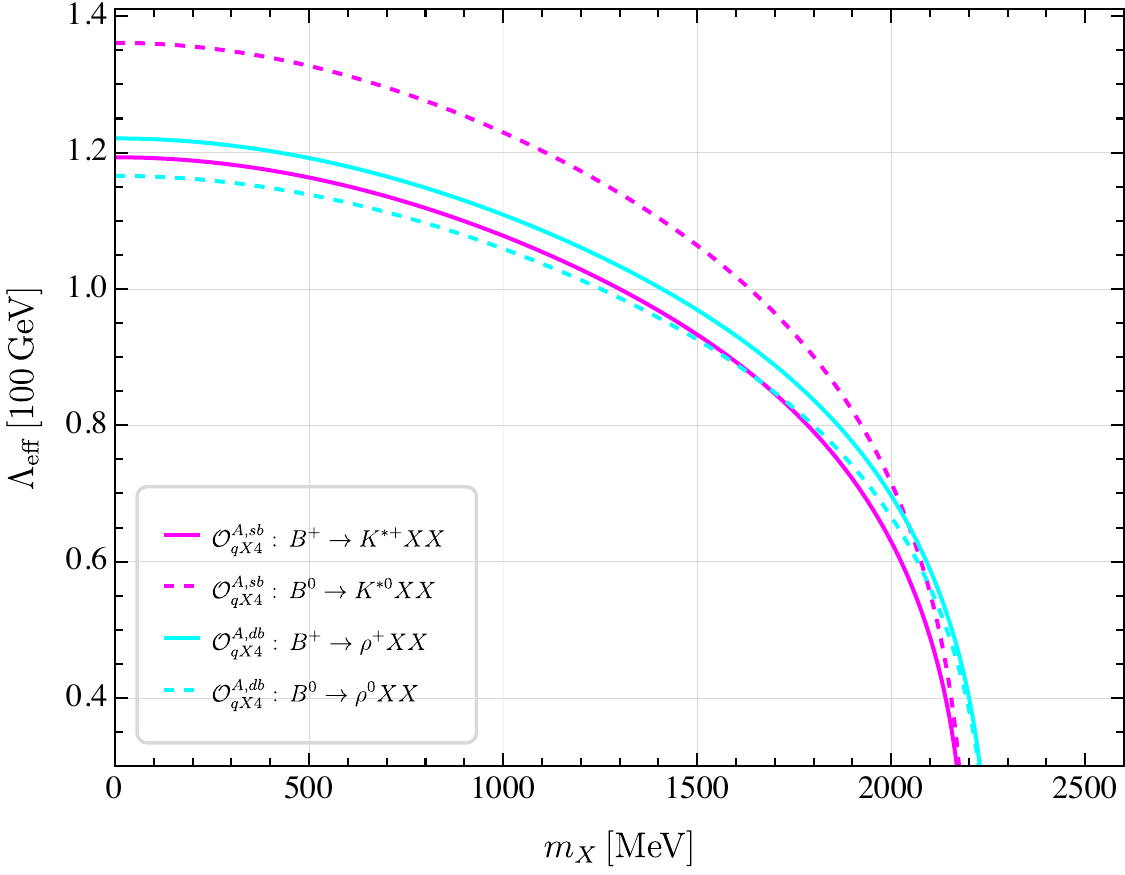}

 \bigskip
 \includegraphics[width=7cm]{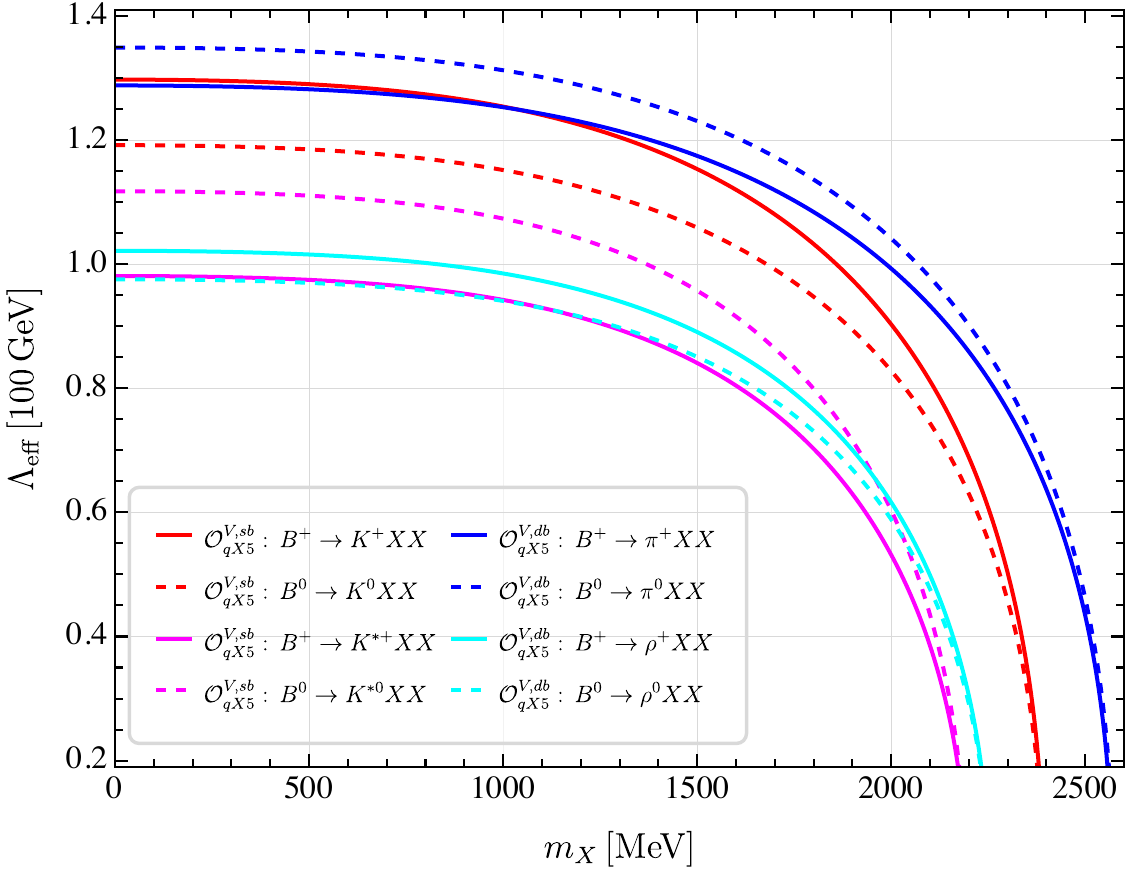}\qquad\quad
 \includegraphics[width=7cm]{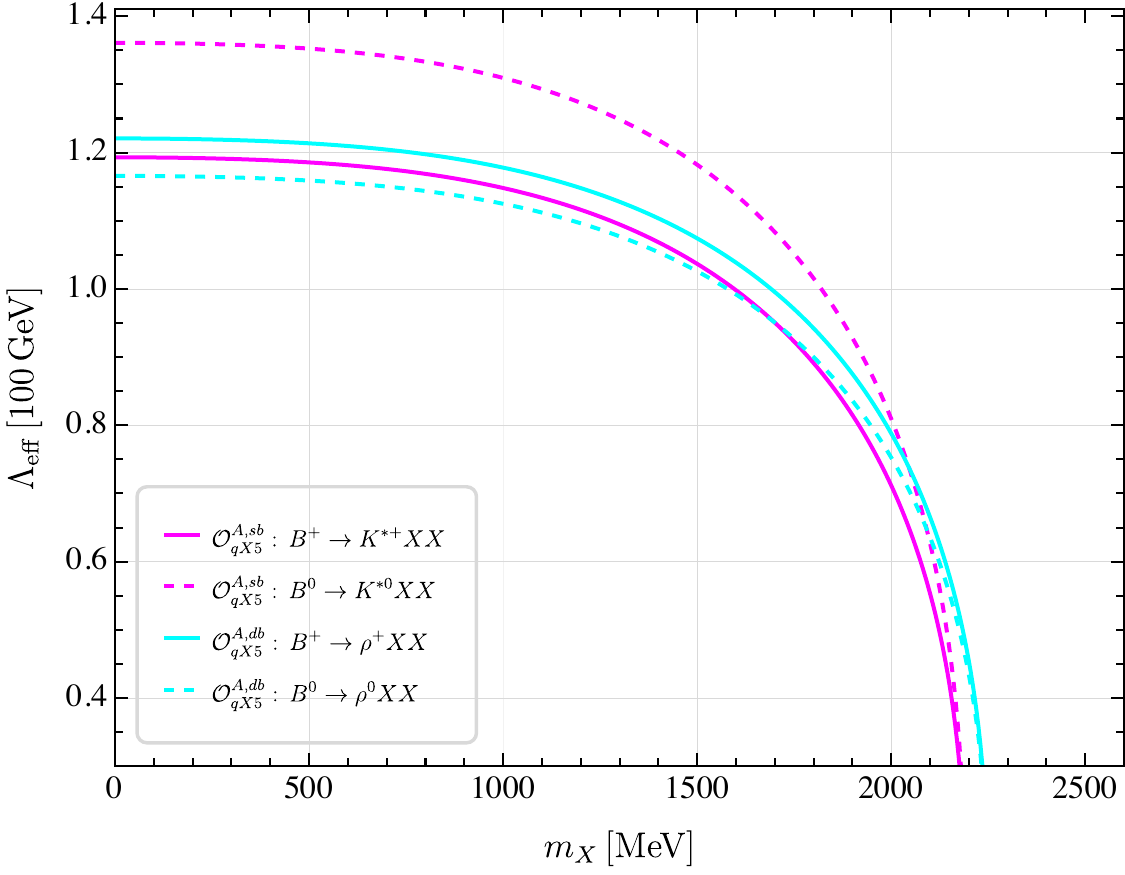} 
\caption{Constraints on the effective new physics scale for the 6 dim-6 operators $\calO_{qX2,4,5}^{V,A}$ as a function of the DM mass $m$ from $B\to K(\pi) \slashed{E}$ channels.}
\label{fig:OqX245VA}
\end{figure}

For the tensor operator $\calO_{qX1}^{T} (\calO_{qX2}^{T})$ (lower two panels), the neutral mode $B^0\to K^{*0} XX$ gives a stronger bound for DM mass $m \lesssim 2.2 (1.9)\, \rm GeV$ and the charged mode $B^+\to K^+XX$ for $m \gtrsim 2.2 (1.9)\, \rm GeV$ for $(sb)$ flavor indices. For $(db)$ flavor indices instead, the neutral mode $B^0\to \pi^0 XX$ gives a stronger bound in the full DM mass range for $\calO_{qX1}^{T}$. For $\calO_{qX2}^{T}$ with $(db)$ flavor indices, it is the charged mode $B^+\to \rho^+XX$ that gives a stronger bound for $m \lesssim 1.5\, \rm GeV$ and the neutral mode $B^0\to\pi^0XX$  for $m \gtrsim 1.5\, \rm GeV$. 
The different behavior of the operator $\calO_{qX2}^{T}$ from the $B\to K(\pi)XX$ modes is due to a quadratic (rather than quartic) inverse dependence on $m$ in the decay widths. 
In all the four cases, the effective scale is constrained to be above a few hundreds of GeV, validating our use of an EFT framework for this discussion.

Fig.\,\ref{fig:OqX36VA} shows the constraints for the 4 dim-6 operators with quark (axial-)vector currents, $\calO_{qX3,6}^{V,A}$, and following Eq.\,\eqref{eq:CqX2mLam}. It can be seen that these constraints observe a similar behavior to  those for the quark (axial-)vector current operators $\calO_{q\phi}^{V,A}$ in the scalar DM case (Fig.\,\ref{fig:scalar_constraint}). The limit on $\Lambda_{\rm eff}$ is weaker by roughly an order of magnitude,  due to the different dimensionality of the operators.  
Fig.\,\ref{fig:OqX245VA} shows the  results for the remaining dim-6 operators, $\calO_{qX2,4,5}^{V,A}$. 
The constraints for $\calO_{qX2}^V$ from the $B\to K^*(\rho)XX$ exhibit similar behavior to those for $\calO_{qX2}^T$ in Fig.\,\ref{fig:OqXSPT}. They are much weaker due to the higher dimensionality and result in limits on $\Lambda_{\rm eff}$ of order a few tens of GeV. These are still much larger than the $B$ meson mass implying that the LEFT framework is valid.  On the other hand, it may be difficult to interpret them within SMEFT.
It would also be difficult to UV complete these operators, as UV completions would very likely predict new states with collider accessible masses.

\begin{figure}
\centering
\includegraphics[width=7cm]{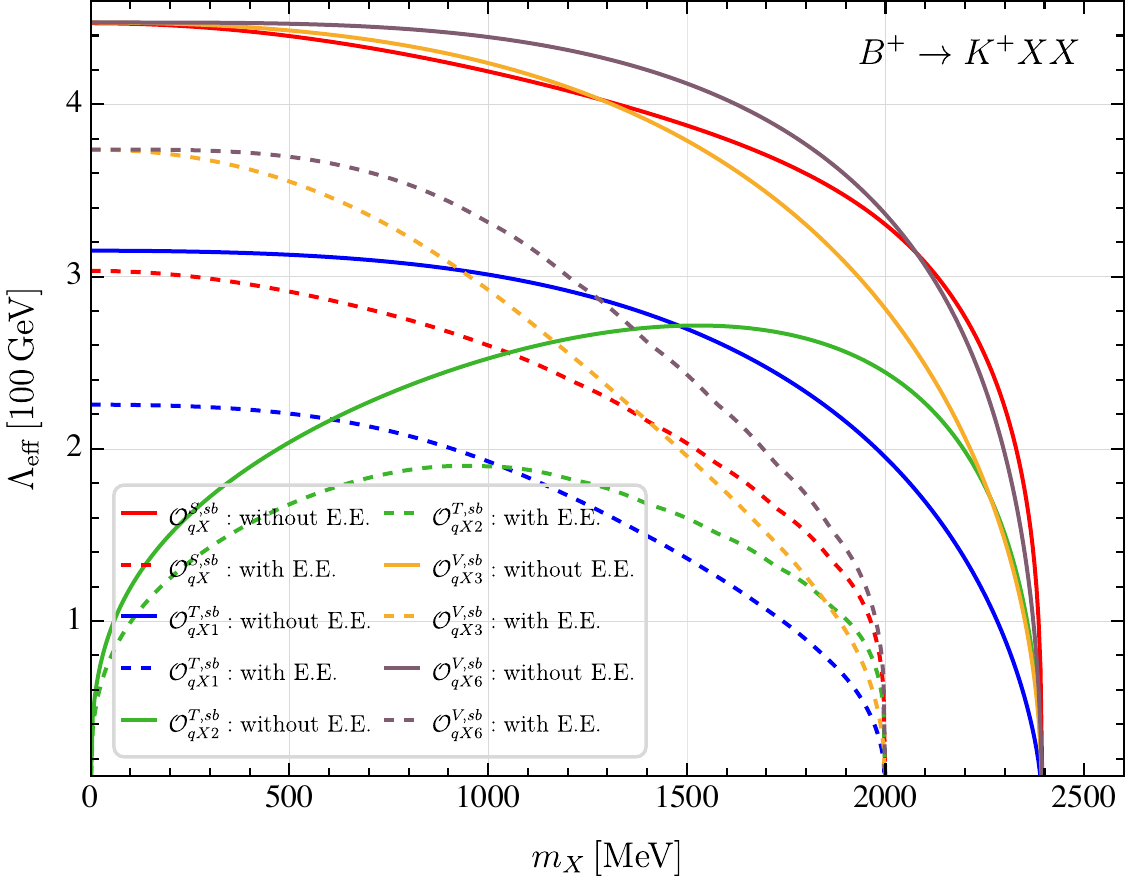}\qquad\quad
\includegraphics[width=7.15cm]{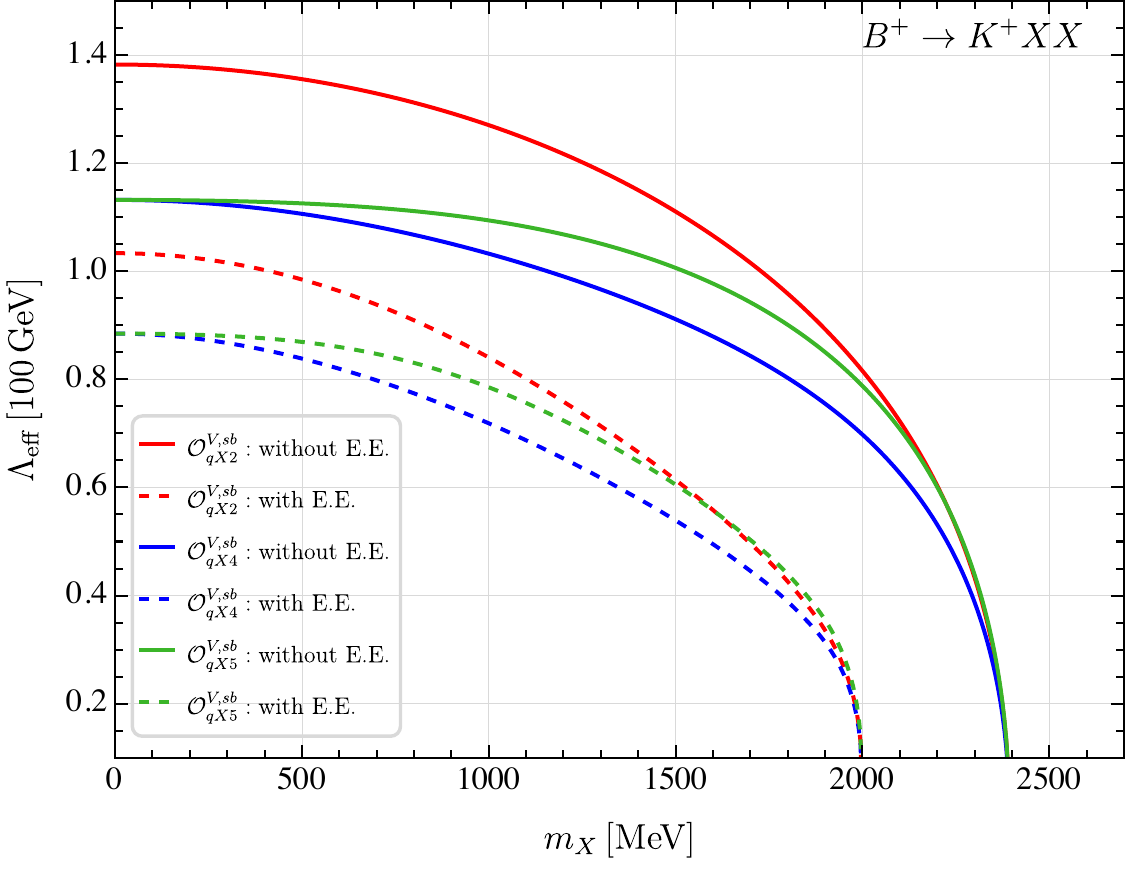}
\caption{Constraints on the effective new physics scale as a function of the DM mass $m$ from the inclusive tag Belle II $B^+\to K^+ \nu\bar\nu$ for the vector DM of scenario A.  }
\label{fig:vectorABelleeff}
\end{figure}

Following the discussion on the scalar DM case with the inclusion of the Belle II experimental efficiency in Fig.\,\ref{fig:scalarBelleeff}, we show the similar plots for the vector DM of scenario A from $B^+\to K^+XX$ mode in Fig.\,\ref{fig:vectorABelleeff}. 
It can be seen that the sensitivity on $\Lambda_{\rm eff}$ for $m =0$ is weaker by a factor of about 1.2-1.5 (with the specific value depending on the operator) when the efficiency is included, and then gradually decreases as $m$ increases. Furthermore, the sensitivity in this case is limited to $m \lesssim 2\,\si{GeV}$ by $4 m^2 \lesssim q^2_{\rm max}$, with the maximum $q^2_{\rm max} \approx 16\,\si{GeV}^2 $ corresponding to the non-vanishing signal efficiency region in Belle II.

\subsection{$B \to M + XX$ with vector DM $X$: scenario B }

For these operators there is no issue with the $m\to 0$ limit, and in addition all the form factors needed have been estimated in the literature before.  
The non-vanishing amplitudes for the two processes $B(p) \to P(k) X( k_1) X^*(k_2)$ and $B(p) \to V(k) X( k_1) X^*(k_2)$ take the following  form, 
\begin{subequations}
\begin{eqnarray}
i\calM_{B\to PXX}^B  &= & \epsilon^{*}_\rho (k_1) \epsilon^{*}_\sigma (k_2)
\left\{ \left[ 
 2(  k_1^\sigma k_2^\rho - k_1\cdot k_2 g^{\rho\sigma} )\tilde C_{qX1}^{S,xb} 
+ 2\epsilon^{\mu\nu\rho\sigma}k_{1\mu} k_{2\nu}  \tilde C_{qX2}^{S,xb}  \right]  
 \langle P(k)|\bar q_x b| B(p)\rangle 
 \right.
 \nonumber
\\
& + & 
{i \over 2}\left[ (  k_1^\alpha  k_2^\beta g^{\rho\sigma}+k_1\cdot k_2 g^{\alpha\rho} g^{\beta\sigma} -  k_1^\alpha k_2^\rho g^{\beta\sigma}- k_1^\sigma k_2^\beta g^{\alpha \rho})
\left(2 g_{\mu\alpha} g_{\nu \beta} \tilde C_{qX1}^{T,xb}
+ \epsilon_{ \mu\nu \alpha\beta}  \tilde C_{qX2}^{T,xb}\right) \right]
\nonumber
\\
& \times & \left. 
\langle P(k)| \bar q_x \sigma^{\mu\nu} b |B(p)\rangle
\right\},
\\%
i\calM_{B\to VXX}^B &= &  \epsilon^{*}_\rho (k_1) \epsilon^{*}_\sigma (k_2)
\left\{ \left[ 
 2(  k_1^\sigma k_2^\rho - k_1\cdot k_2 g^{\rho\sigma} )\tilde C_{qX1}^{P,xb} 
+ 2\epsilon^{\mu\nu\rho\sigma}k_{1\mu} k_{2\nu}  \tilde C_{qX2}^{P,xb}  \right]  
 \langle V(k)|\bar q_x i\gamma_5 b| B(p)\rangle 
 \right.
 \nonumber
\\
& + & 
{i \over 2} \left[ (  k_1^\alpha  k_2^\beta g^{\rho\sigma}+k_1\cdot k_2 g^{\alpha\rho} g^{\beta\sigma} -  k_1^\alpha k_2^\rho g^{\beta\sigma}- k_1^\sigma k_2^\beta g^{\alpha \rho})
\left( 2 g_{\mu\alpha} g_{\nu \beta} \tilde C_{qX1}^{T,xb}
+\epsilon_{ \mu\nu \alpha\beta}  \tilde C_{qX2}^{T,xb} \right) \right]
\nonumber
\\
&\times &
\left. \langle V(k)| \bar q_x \sigma^{\mu\nu} b |B(p)\rangle
\right\}.
\end{eqnarray}
\end{subequations}
From these amplitudes, the differential decay widths result in the following compact forms,
\begin{eqnarray}
{d\Gamma_{B\to PXX}^B \over d q^2}  &= & 
{(m_B^2 - m_P^2)^2(s^2 - 4 m^2  s +6 m^4) \over 128 \pi^3 m_B^3(m_b - m_{q_x})^2 } 
\lambda^{1\over2}(m_B^2, m_P^2, s)  \kappa^{1\over2}(m^2,s)  f_0^2 \left| \tilde C_{qX1}^{S,xb}\right|^2
\nonumber
\\
& + & 
{(m_B^2 - m_P^2)^2 s^2 \over 128 \pi^3 m_B^3(m_b - m_{q_x})^2 } 
\lambda^{1\over2}(m_B^2, m_P^2, s)  \kappa^{3\over2}(m^2,s)  f_0^2 \left| \tilde C_{qX2}^{S,xb}\right|^2
\nonumber
\\
& + & 
{s(s+2m^2) \over 1536 \pi^3 m_B^3(m_B+m_P)^2 } 
\lambda^{3\over2}(m_B^2, m_P^2, s)  \kappa^{3\over2}(m^2,s)  f_T^2 \left| \tilde C_{qX1}^{T,xb}\right|^2
\nonumber
\\
& + & 
{s^2 - 2m^2 s +4m^4 \over 1536 \pi^3 m_B^3(m_B+m_P)^2 } 
\lambda^{3\over2}(m_B^2, m_P^2, s)  \kappa^{1\over2}(m^2,s)  f_T^2 \left| \tilde C_{qX2}^{T,xb}\right|^2,
\\%
{d\Gamma_{B\to VXX}^B \over d q^2}  &= &
{s^2 - 4 m^2  s +6 m^4 \over 128 \pi^3 m_B^3(m_b + m_{q_x})^2 } 
\lambda^{3\over2}(m_B^2, m_V^2, s)  \kappa^{1\over2}(m^2,s) A_0^2 \left| \tilde C_{qX1}^{P,xb}\right|^2
\nonumber
\\
& + & {s^2 \over 128 \pi^3 m_B^3(m_b + m_{q_x})^2 } 
\lambda^{3\over2}(m_B^2, m_V^2, s)  \kappa^{3\over2}(m^2,s) A_0^2 \left| \tilde C_{qX2}^{P,xb}\right|^2
\nonumber
\\
& + &
{1 \over 768 \pi^3 m_B^3 s } 
 \lambda^{1\over2}(m_B^2, m_V^2, s)  \kappa^{1\over2}(m^2,s) 
 \left\{
(s+2m^2)(s-4m^2) \lambda(m_B^2, m_V^2, s) T_1^2 
\right.
 \nonumber
 \\
 & + &
\left.
 (s^2 - 2m^2 s +4 m^4) \left[ 
 (m_B^2 - m_V^2)^2  T_2^2 
+ { 8 m_B^2 m_V^2 s \over (m_B+m_V)^2 }T_{23}^2  \right]
\right\}
\left|\tilde  C_{qX1}^{T,xb}\right|^2
\nonumber
\\
& + &
{1 \over 768 \pi^3 m_B^3  s } 
 \lambda^{1\over2}(m_B^2, m_V^2, s)  \kappa^{1\over2}(m^2,s) 
 \left\{
 (s^2 - 2m^2 s +4 m^4)  \lambda(m_B^2, m_V^2, s) T_1^2 
\right.
 \nonumber
 \\
 & + &
\left.
(s+2m^2)(s-4m^2)\left[ 
 (m_B^2 - m_V^2)^2  T_2^2 
+ { 8 m_B^2 m_V^2 s \over (m_B+m_V)^2 }T_{23}^2  \right]
\right\}
\left|\tilde  C_{qX2}^{T,xb}\right|^2. 
\end{eqnarray}
Similarly to the case of scalar DM, there is no interference between the different operators due to their different parity and/or  charge conjugation.  Interestingly, in the above expressions, the contributions from each pair of operators with similar quark Lorentz structure (characterized by the same superscript ``S/P/T'') 
\footnote{Note that the two tensor currents are ``similar'' to each other because
$\bar q \sigma^{\mu\nu}\gamma_5 q= {i \over 2} \epsilon^{\mu\nu\rho\sigma} \bar q \sigma_{\rho\sigma} q$.}
become the same in the limit of $m\to 0$.  

\begin{figure}
\centering
\includegraphics[width=6.5cm]{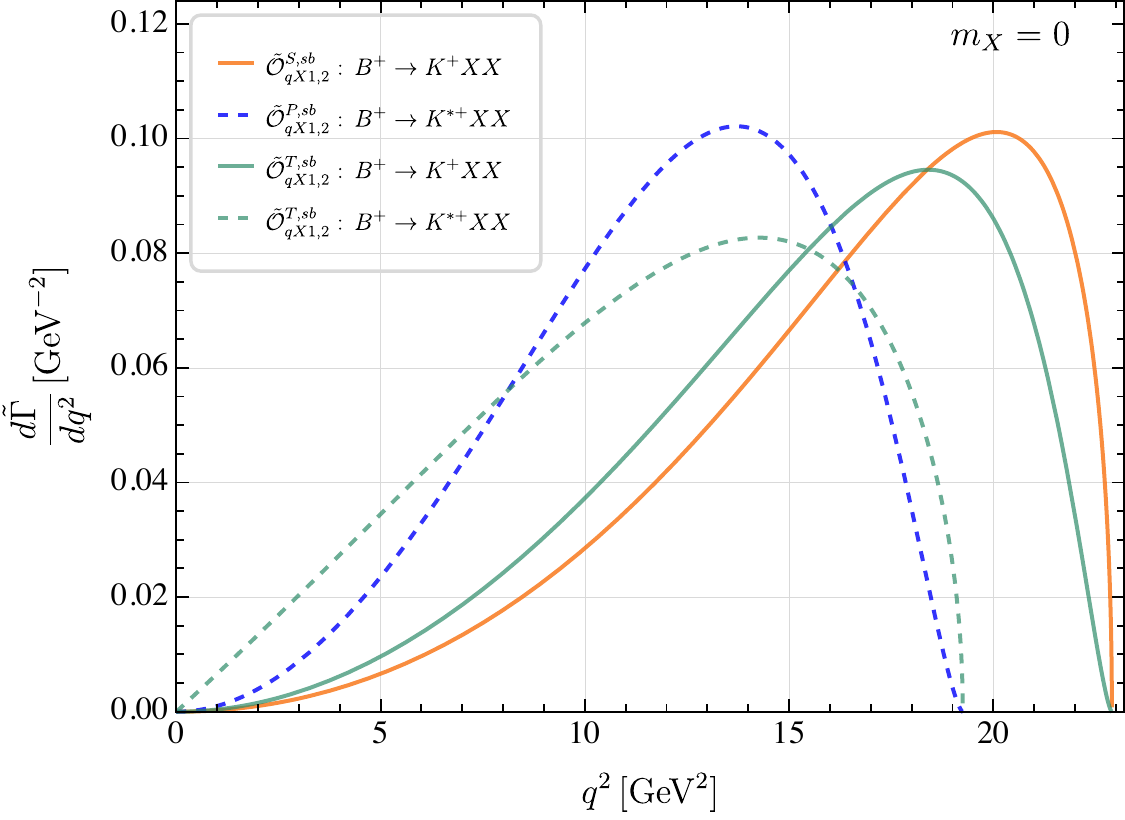}\qquad\quad
\includegraphics[width=6.5cm]{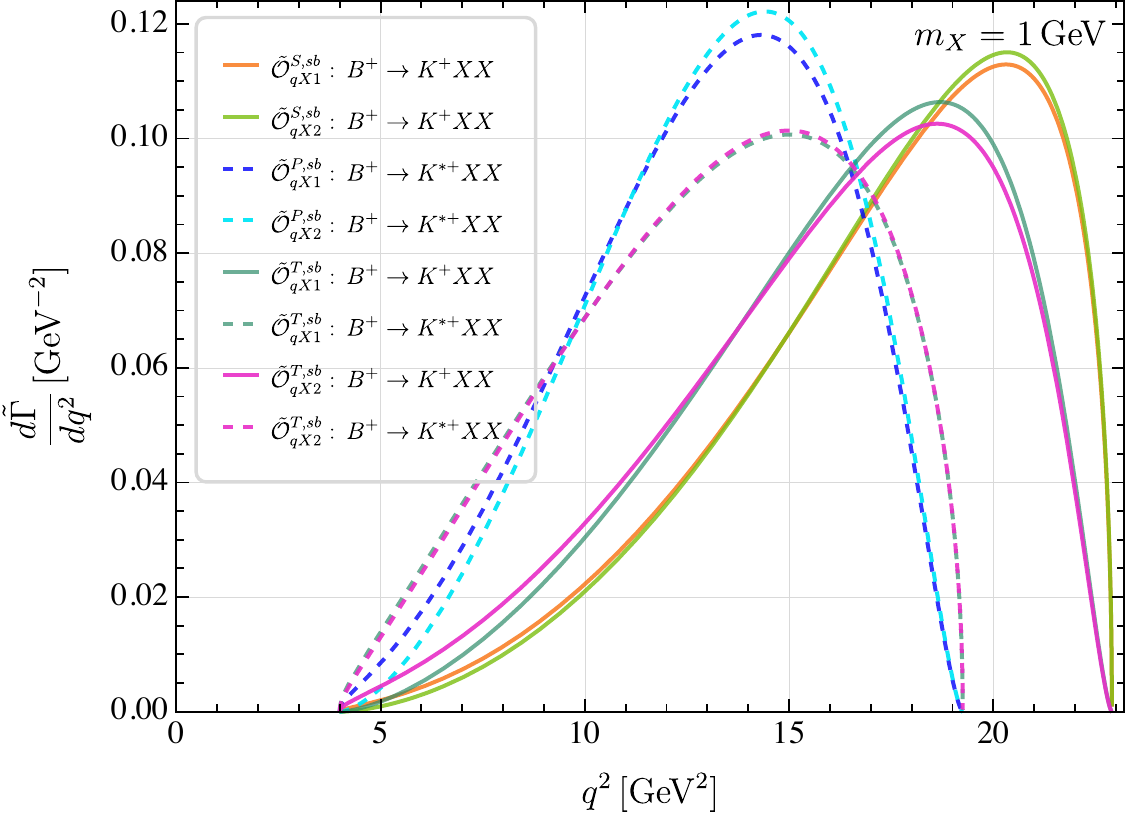}
\caption{Normalized differential decay width for $B\to K^{(*)+}XX$ from different operators in the second scenario for vector DM. {\it Left panel}: $m=0$; {\it Right panel}: $m=1\,\rm GeV$. }
\label{fig:dBdq2svector}
\end{figure}

\begin{figure}
\centering
\includegraphics[width=7cm]{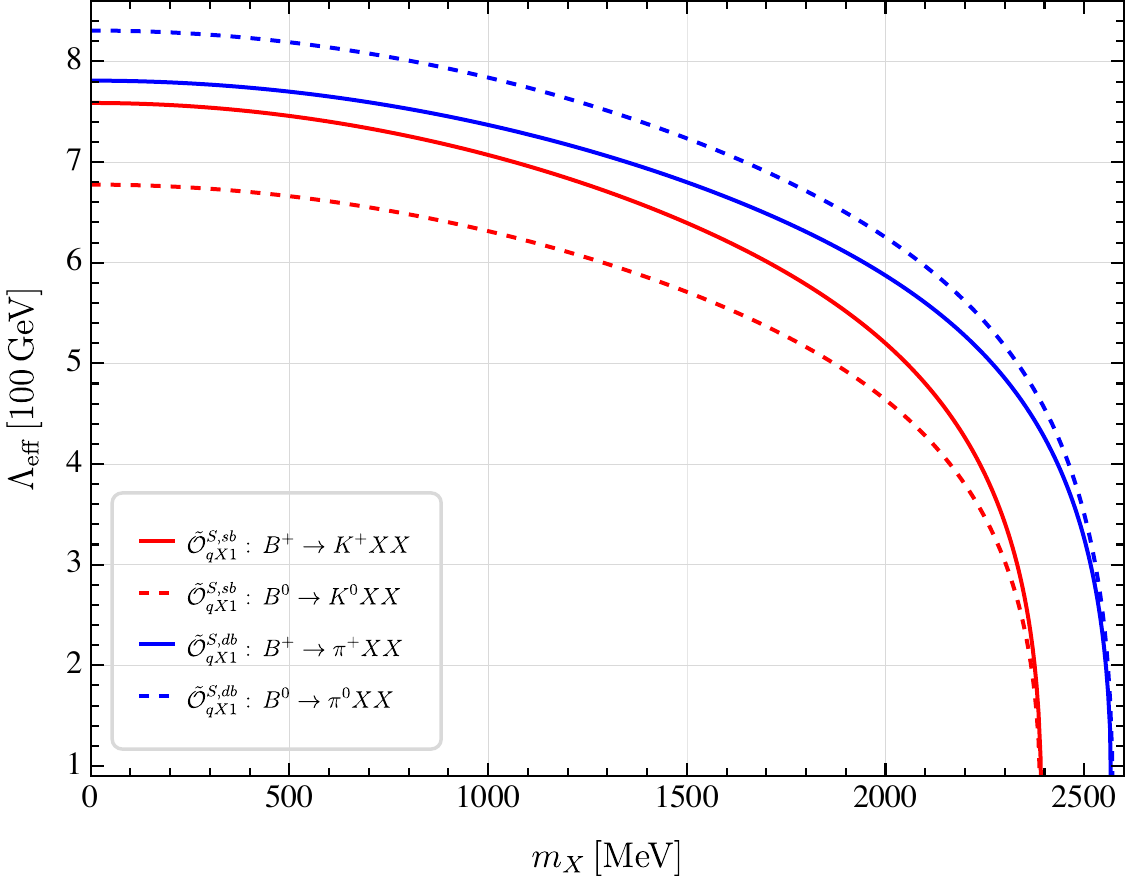}\qquad\quad
\includegraphics[width=7cm]{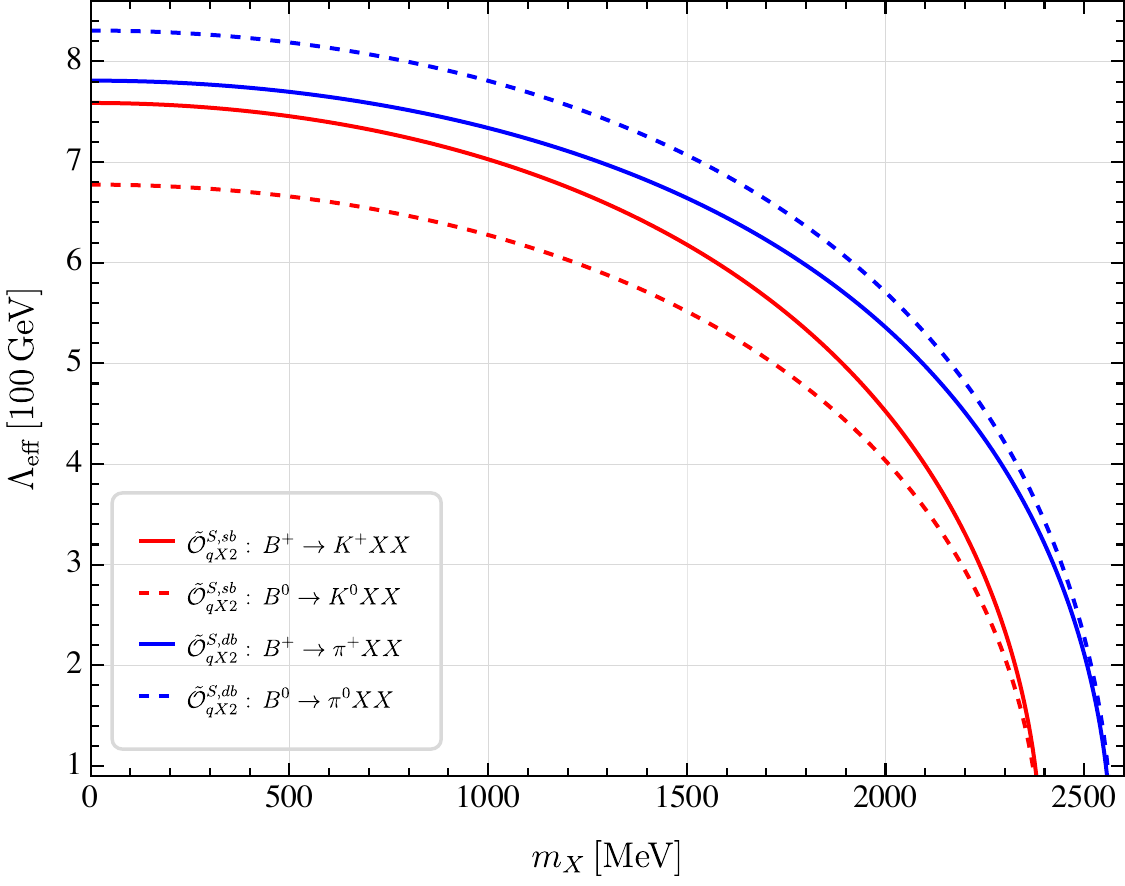}

 \bigskip 
\includegraphics[width=7cm]{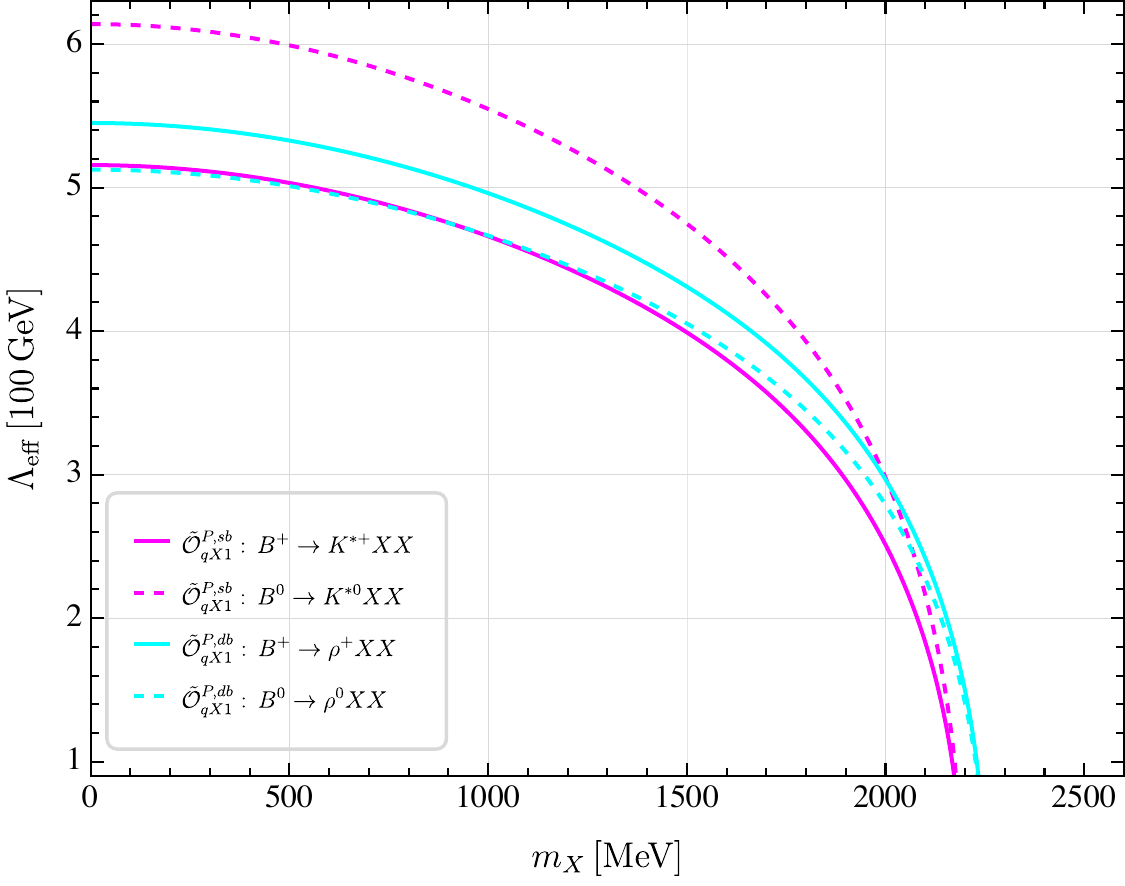}\qquad\quad
\includegraphics[width=7cm]{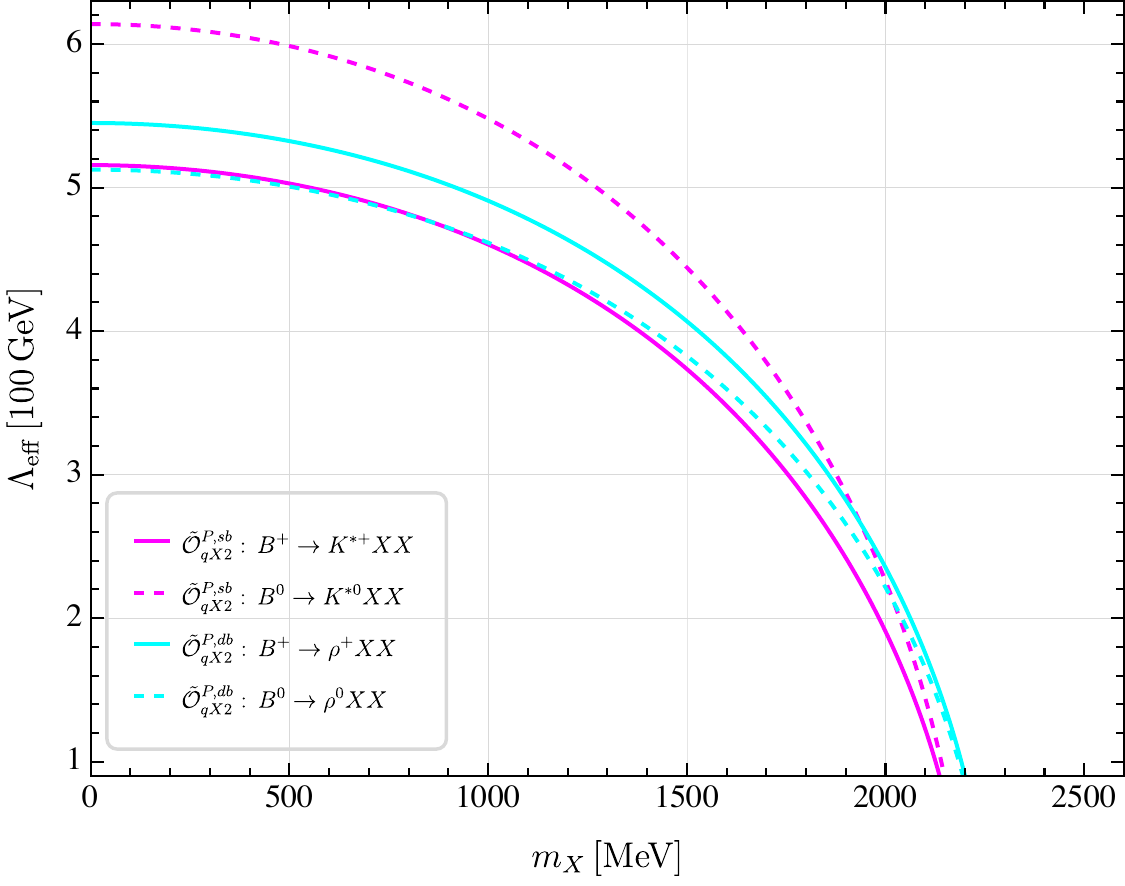} 

 \bigskip 
\includegraphics[width=7cm]{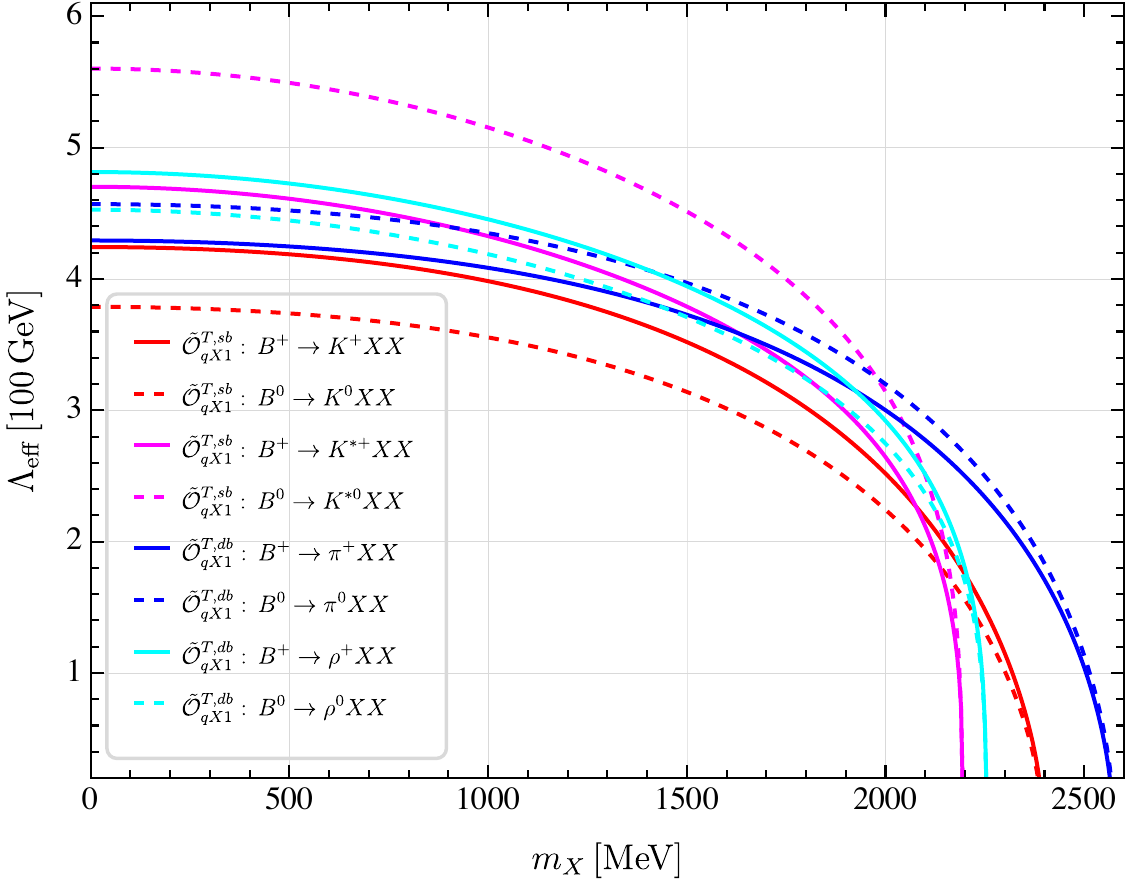}\qquad\quad
\includegraphics[width=7cm]{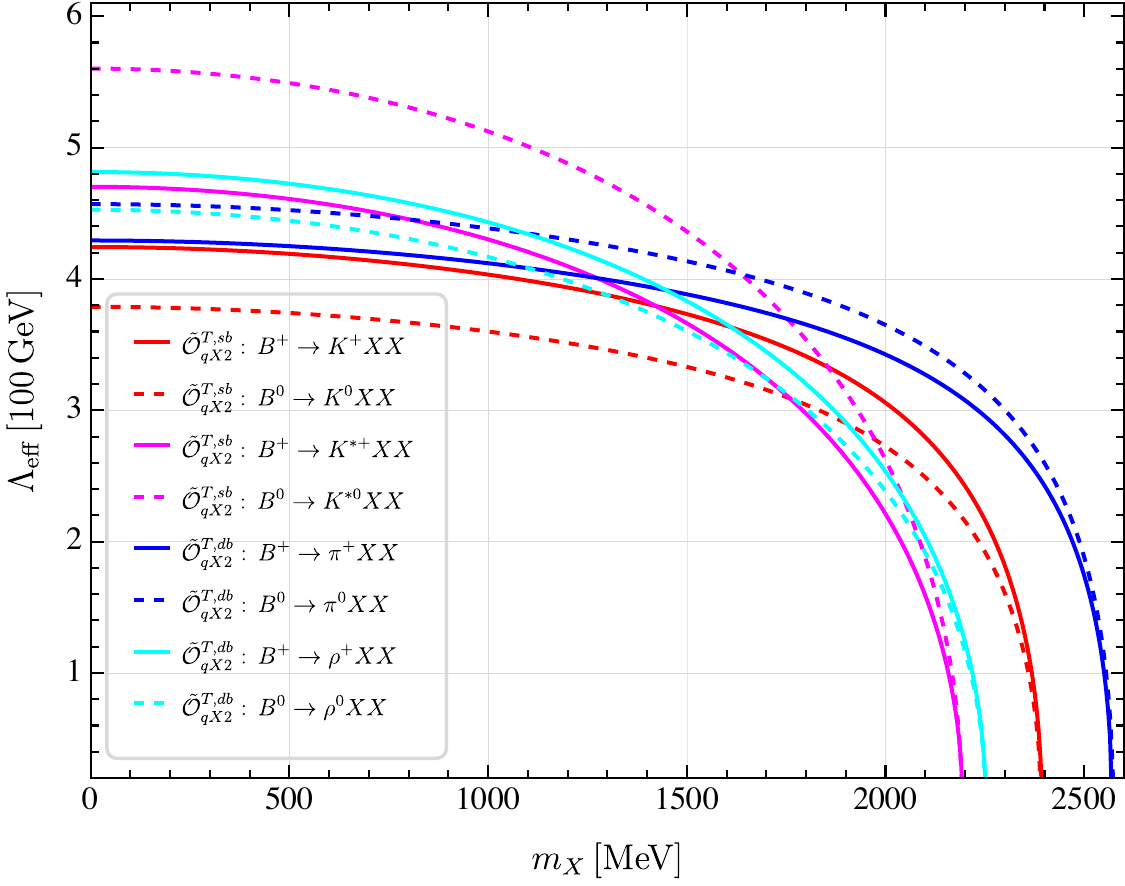}
\caption{Constraints on the effective new physics scale for each operator as a function of the DM mass $m$ from $B\to K(\pi) \slashed{E}$ channels. }
\label{fig:OqXB}
\end{figure}
\begin{figure}
\centering
\includegraphics[width=7cm]{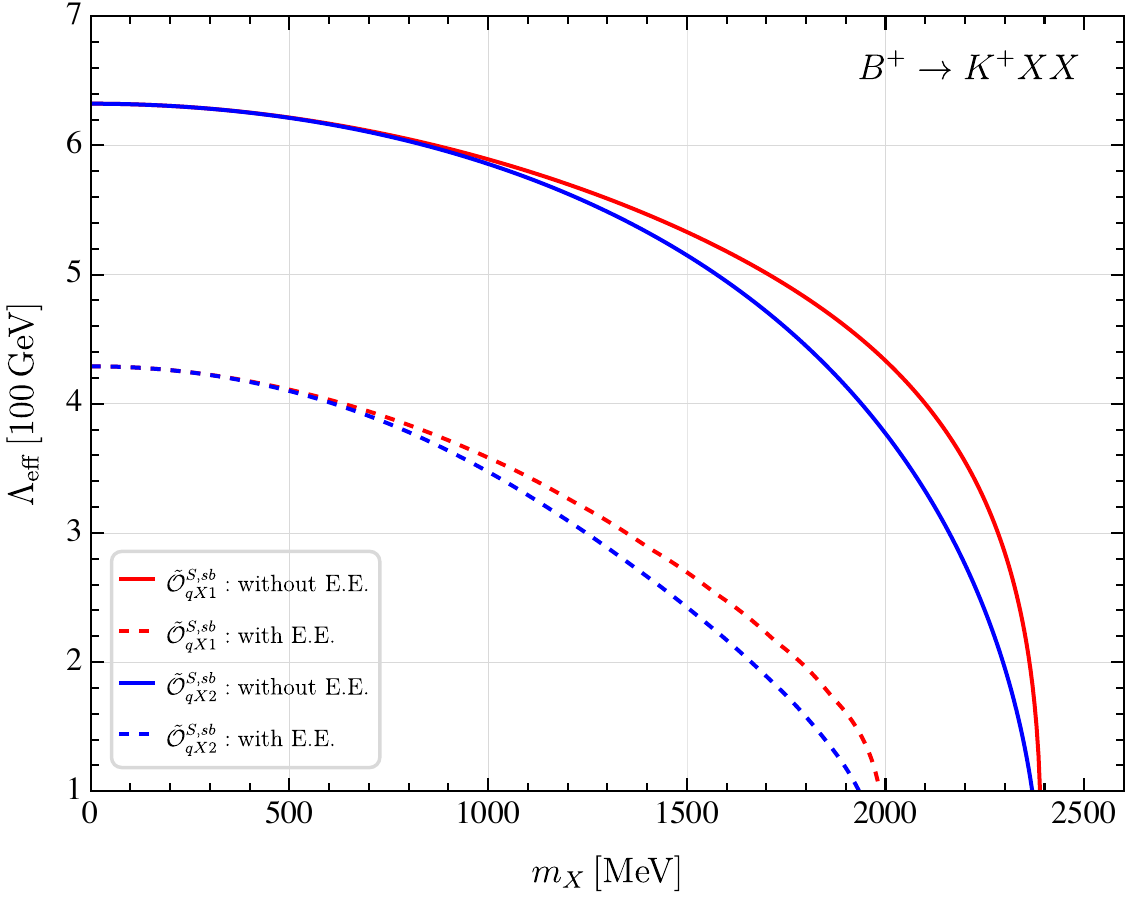}\qquad\quad
\includegraphics[width=7cm]{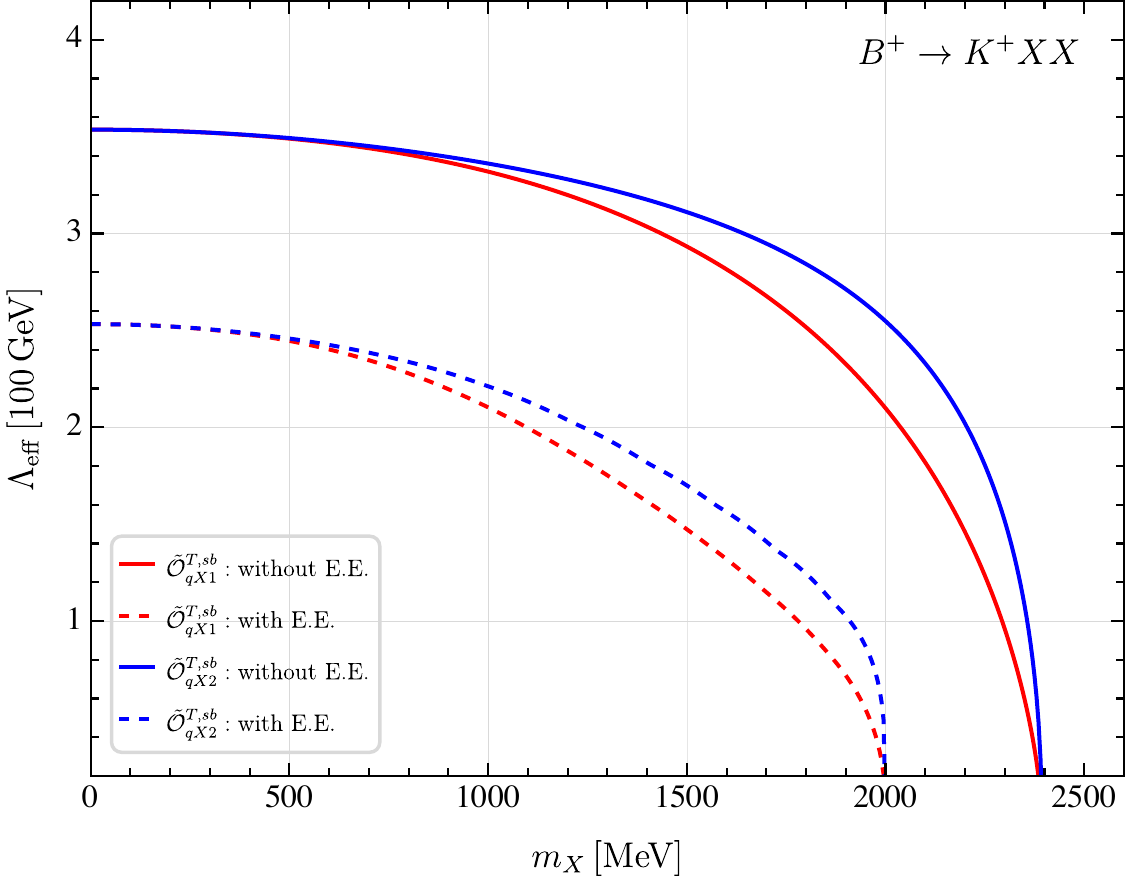}
\caption{ Similar to Fig.\,\ref{fig:scalarBelleeff} and Fig.\,\ref{fig:vectorABelleeff}, constraints on the effective new physics scale  as a function of the DM mass $m$ from the inclusive tag Belle II $B^+\to K^+ \nu\bar\nu$ for the vector DM of scenario B. }
\label{fig:vectorBBelleeff}
\end{figure}

Fig.\,\ref{fig:dBdq2svector} shows the normalized distributions  for $B\to K^{+}XX$ (solid lines) and $B\to K^{*+}XX$ (dashed lines) for different operators with $(sb)$  flavor indices. In the left (right) panel, we set the vector DM mass to $m=0$ and  $1$\,GeV respectively. 
One can see in the left panel that the distributions are the same for the two operators 
with the same superscript ``S/P/T''. 
Even for $m=1\, \rm GeV$ as shown in the right panel, this degeneracy still holds except for a rather narrow range of $q^2$ value, which implies the $d\tilde \Gamma /dq^2$ observable alone cannot be used to distinguish these interactions.  

To set numerical constraints we include the new physics scale in the Wilson coefficients as $\tilde C_{qX1,2}^{S,P,T} \equiv \Lambda_{\rm eff}^{-3}$.  Fig.\,\ref{fig:OqXB} shows the bounds on the $m$-$\Lambda_{\rm eff}$ plane for each operator. These bounds are comparable to those shown in Fig.\,\ref{fig:OqXSPT} for operators with the same quark current as they have the same dependence on $\Lambda_{\rm eff}$. The effective scale $\Lambda_{\rm eff}$ which can be probed depends on the value of DM mass,  and is above a few hundreds of GeV for $m \lesssim 2\,\si{GeV}$. 
Finally, Fig.\,\ref{fig:vectorBBelleeff} shows the sensitivity after including the experimental efficiency for $B^+\to K^+ XX$ from Belle II. Similar to the other two DM cases, the constraint is weaker by a factor of 1.5 (1.4) for $\tilde{\cal O}_{qX1,2}^{S,sb}$ ($\tilde{\cal O}_{qX1,2}^{T,sb}$) at $m=0$ when the efficiency is included, with a sensitivity limit for $m$ around 2\,GeV.

\section{$K\to \pi$+DM+DM}
\label{sec:kaondecay}

Since the kaon decay process $K\rightarrow\pi \slashed{E}$ involves only the light quarks $q=u,~d,~s$, its transition matrix element due to effective interactions in LEFT can be evaluated by matching onto chiral perturbation theory ($\chi$PT), which is the low energy effective field theory of QCD. $\chi$PT is based on the fact that the QCD Lagrangian has the approximate chiral symmetry $SU(3)_L\times SU(3)_R$ for the three light quarks which is spontaneously broken by the quark condensate $\langle\bar qq\rangle$ to the diagonal flavor $SU(3)_V$. The symmetry breakdown results in eight pseudo-Nambu-Goldstone bosons (pNGBs), which are identified with the octet of the lowest-lying pseudoscalars 
$\pi^\pm,\,\pi^0,\,K^\pm,\,K^0,\,\bar{K}^0,\,\eta$. In the $\chi$PT formalism they are represented by the element in the coset space $SU(3)_L\times SU(3)_R/SU(3)_V$ and take the matrix form,
\begin{eqnarray}
U(x) = \exp\left(\frac{i\sqrt{2}\Pi(x)}{F_0}\right), 
\quad
\Pi = 
\begin{pmatrix}
\frac{\pi^0}{\sqrt{2}}+\frac{\eta}{\sqrt{6}} & \pi^+ & K^+
\\
\pi^- & -\frac{\pi^0}{\sqrt{2}}+\frac{\eta}{\sqrt{6}} & K^0
\\
K^- & \bar{K}^0 & -\sqrt{\frac{2}{3}}\eta
\end{pmatrix},
\end{eqnarray}
where $F_0$ is the pion decay constant in the chiral limit. Corresponding to the chiral transformations of quarks $q_L\to Lq_L$ and $q_R\to Rq_R$, $U$ transforms as $U\rightarrow  LUR^\dagger$ with $L\in SU(3)_L$ and $R\in SU(3)_R$.

The interactions of pNGBs with DM due to the effective operators in Eqs.\,(\ref{eq:Oqphi}-\ref{eq:OtildeqX}) can be realized through the external source method in which the global chiral symmetry is promoted to a local one \cite{Gasser:1983yg,Gasser:1984gg,Cata:2007ns}.
At the quark-gluon level, the QCD Lagrangian with all possible external sources associated with quark bilinear currents is parameterized as follows,
\begin{subequations}
\begin{eqnarray}
\mathcal{L}&=&
\mathcal{L}_\textrm{QCD}
+\overline{q}\hat v_\mu \gamma^\mu q
+\overline{q} \hat  a_\mu \gamma^\mu \gamma_5 q
- \overline{q}\hat  s q 
+  \overline{q} \hat p i\gamma_5 q 
+\overline{q} \hat t_{\mu\nu} \sigma^{\mu\nu} q 
\\
& = &
\mathcal{L}_\textrm{QCD}
+\overline{q_L}l_\mu \gamma^\mu q_L+\overline{q_R}r_\mu \gamma^\mu q_R
- \left[\overline{q_R}(s + ip)q_L
- \overline{q_R}t_l^{\mu\nu}\sigma_{\mu\nu}q_L+\hc\right],
\label{eq:QCDL}
\end{eqnarray}
\end{subequations}
where $\mathcal{L}_\textrm{QCD}$ is the QCD Lagrangian for $u,\,d,\,s$ quarks in the massless limit. The external sources,  
$\hat v_\mu=\hat v_\mu^\dagger$,
 $\hat  a_\mu=\hat  a_\mu^\dagger$, 
 $\hat s=\hat s^\dagger$, 
 $\hat p=\hat p^\dagger$, 
 $\hat t_{\mu\nu}=(\hat t_{\mu\nu})^{\dagger}$, are $3\times 3$ Hermitian traceless matrices in flavor space. In the second line, we rewrite the Lagrangian in terms of chiral quark fields to make the chiral symmetry manifest.  The  external sources in the second line are related to those in the first line by the relations, 
 \begin{eqnarray}
 l_\mu = \hat v_\mu - \hat a_\mu, 
\quad
r_\mu = \hat v_\mu + \hat a_\mu, 
\quad
\chi=2B(\hat s - i\hat p),
\quad
t_l^{\mu\nu} = P_L^{\mu\nu\alpha\beta} \hat t_{\alpha\beta},
\quad
t_r^{\mu\nu} = P_R^{\mu\nu\alpha\beta} \hat t_{\alpha\beta} = t_l^{\mu\nu \dagger}, 
 \end{eqnarray}
where the chiral projection for the tensor currents is defined by $P_{R,L}^{\mu\nu\alpha\beta} = {1 \over 4}(g^{\mu\alpha} g^{\nu\beta} - g^{\mu\beta} g^{\nu \alpha} \pm i \epsilon^{\mu\nu\alpha\beta} )$ \cite{Cata:2007ns}, with the property $\hat t^{\mu\nu}= t_l^{\mu\nu}+t_r^{\mu\nu}$.   
Under chiral transformations, 
$l_{\mu}\to L l_{\mu} L^{\dagger}+i L \partial_{\mu} L^{\dagger}$, 
 $r_{\mu}\to R r_{\mu} R^{\dagger}+i R \partial_{\mu} R^{\dagger}$, 
 $\chi\to L \chi R^{\dagger}$, and $t_l^{\mu\nu}\to R t_l^{\mu\nu} L^{\dagger}$, respectively. 
 One should note that the tensor external sources have mass dimension one in our convention.\footnote{Our convention for $U$ and $\chi$ are equivalent to $U^\dagger $ and $\chi^\dagger$ in \cite{Cata:2007ns}.}
The constant $B$ is related to the quark condensate and $F_0$ by the relation $B=-\langle\bar{q}q\rangle /(3F_0^2)$. For  numerical estimates, we use
$F_0=87\,\rm MeV$ \cite{Colangelo:2003hf} and $B\approx 2.8\, \rm GeV$. 

By comparing the external sources in Eq.\,\eqref{eq:QCDL} with the effective interactions in LEFT  in Eqs.\,(\ref{eq:Oqphi}-\ref{eq:OtildeqX}) we see that the correspondence needed to calculate the $K \to\pi $ transition with scalar DM is,
\begin{eqnarray}
(\hat v^\mu)_{ds} = C_{q\phi}^{V,ds} \phi^\dagger i \overleftrightarrow{\partial_\mu} \phi, 
\quad
(\hat a^\mu)_{ds} =  C_{q\phi}^{A,ds} \phi^\dagger i \overleftrightarrow{\partial_\mu} \phi, 
\quad
(\hat s)_{ds} =  -  C_{q\phi}^{S,ds} \phi^\dagger \phi,
\quad
(\hat p)_{ds} = C_{q\phi}^{P,ds}  \phi^\dagger \phi, 
\end{eqnarray}
and the corresponding Hermitian conjugates interchanging labels $s$ and $d$. 
For the vector DM case, it is also easily to identify the relevant external sources from the operators in
Eqs.\,(\ref{eq:OqX}-\ref{eq:OtildeqX}). For the second scenario of Eq.\,\eqref{eq:OtildeqX}, the non-vanishing sources are
\begin{subequations}
\begin{eqnarray} 
(\hat s)_{ds} & = & - \tilde C_{qX1}^{S,ds} X_{\mu\nu}^\dagger X^{\mu\nu} - \tilde C_{qX2}^{S,ds} X_{\mu\nu}^\dagger \tilde X^{\mu\nu},
\\
(\hat p)_{ds} & = &  \tilde C_{qX1}^{P,ds} X_{\mu\nu}^\dagger X^{\mu\nu} + \tilde C_{qX2}^{P,ds} X_{\mu\nu}^\dagger \tilde X^{\mu\nu},
\\
 (\hat t^{\mu\nu})_{ds} & = &{i\over 2}\tilde C_{qX1}^{T,ds} ( X^{\dagger \mu}_{\,\,\,\,\rho} X^{\rho\nu}-X^{\dagger \nu}_{\,\,\,\,\rho} X^{\rho\mu})
+ {i \over 2}\tilde C_{qX2}^{T,ds}\epsilon^{\mu\nu\alpha\beta} X^{\dagger }_{\alpha\rho} X^{\rho}_{\,\beta},  
\end{eqnarray}
\end{subequations}
and the corresponding Hermitian conjugates exchanging the $s$ and $d$ labels. 

In $\chi$PT the vector and scalar sources first appear at order $\calO(p^2)$ in the chiral power counting scheme \cite{Gasser:1983yg,Gasser:1984gg}
\begin{eqnarray}
\mathcal{L}^{(2)}_{\chi\rm PT}
=\frac{F_0^2}{4}{\Tr}\left[ D_\mu U (D^\mu U)^\dagger \right]+\frac{F_0^2}{4}{\Tr} \left[\chi U^\dagger +U\chi^\dagger \right],
\quad
D_\mu U\equiv \partial_\mu U-i l_\mu U+i U r_\mu,
\label{eq:chptp2}
\end{eqnarray}
whereas the tensor sources first appear at $\calO(p^4)$~\cite{Cata:2007ns}
\begin{eqnarray}
\mathcal{L}^{(4)}_{\chi\rm PT}\ni i\Lambda_2 {\Tr}\left[ t_r^{\mu\nu}(D_\mu U)^\dagger U (D_\nu U)^\dagger+t_l^{\mu\nu}D_\mu UU^\dagger D_\nu U\right],
\label{eq:chptp4}
\end{eqnarray}
with $\Lambda_2$ being a new low energy constant with mass dimension one. For our numerical  estimates we follow \cite{Jiang:2012ir} and use $|\Lambda_2| \approx 0.018\,\rm GeV$, noting this is comparable with naive dimensional analysis estimates which find $\Lambda_2\sim {\Lambda_\chi \over 16\pi^2 }\sim 0.008\, \rm GeV$ with the chiral symmetry breaking scale $\Lambda_\chi\sim 1.2\, \rm GeV$. To find the relevant local interactions mediating $K\to \pi\slashed{E}$ in question from the $\chi$PT formalism, it suffices to expand the above Lagrangian in Eq.\,\eqref{eq:chptp2} and Eq.\,\eqref{eq:chptp4} to linear order in the kaon and pion fields as well as each external source. This leads to the following local interactions mediating $K\to \pi\slashed{E}$, 
\begin{subequations}
\begin{eqnarray}
\mathcal{L}_{K\to \pi} & = & 
- B \hat s_{sd}  \pi^- K^+ 
+ {B \over \sqrt{2} } \hat s_{sd} \pi^0 K^0 
 \nonumber
\\
 & + & i (\hat v_\mu)_{sd} (\partial^\mu \pi^- K^+ - \pi^- \partial^\mu K^+ )
- {i \over \sqrt{2} } (\hat v_\mu)_{sd} (\partial^\mu \pi^0 K^0 - \pi^0 \partial^\mu K^0 ) 
\nonumber
 \\
 & - &
 i { \Lambda_2 \over F_0^2 } \hat t^{\mu\nu}_{sd}(\partial_\mu \pi^- \partial_\nu K^+ - \partial_\nu \pi^- \partial_\mu K^+ )
+ i {\Lambda_2 \over \sqrt{ 2 }  F_0^2 } \hat t^{\mu\nu}_{sd}(\partial_\mu \pi^0 \partial_\nu K^0 -\partial_\nu \pi^0 \partial_\mu K^0 )
+ \hc
\\
 & \ni &- B \hat s_{sd}  \pi^- K^+ 
+ {B \over 2} (\hat s_{sd}+ \hat s_{ds})\pi^0 K_L 
 \nonumber
\\
 & + & i (\hat v_\mu)_{sd} (\partial^\mu \pi^- K^+ - \pi^- \partial^\mu K^+ )
- {i \over 2 } [(\hat v_\mu)_{sd}-(\hat v_\mu)_{ds}] (\partial^\mu \pi^0 K_L - \pi^0 \partial^\mu K_L ) 
\nonumber
 \\
 & - &
 i { \Lambda_2 \over F_0^2 } \hat t^{\mu\nu}_{sd}(\partial_\mu \pi^- \partial_\nu K^+ - \partial_\nu \pi^- \partial_\mu K^+ )
+ i {\Lambda_2 \over 2  F_0^2 }( \hat t^{\mu\nu}_{sd} - \hat t^{\mu\nu}_{ds})(\partial_\mu \pi^0 \partial_\nu K_L -\partial_\nu \pi^0 \partial_\mu K_L ).
\end{eqnarray}
\end{subequations}
In the second equation we ignore CP violation in kaon mixing to write $K^0 (\bar K^0) \approx {1\over \sqrt{2}}(K_L \pm K_S)$. It can be seen in the above effective Lagrangian that the neutral mode $K_L \to \pi^0\slashed{E}$ can be obtained from the charged mode by replacing the relevant Wilson coefficients $C_i^{sd}$ with $(C_i^{sd}\pm C_i^{ds})/2 \sim \Re[C_i^{ds}] (\Im[C_i^{ds}])$. The plus sign (real) applies to the scalar current while the minus sign (imaginary) applies to the vector and tensor currents. The above Lagrangian leads to the following form factors
\begin{subequations}
\label{eq:formfacK2pi}
\begin{eqnarray}
\langle \pi^+(k) | {\bar s} d | K^+(p) \rangle & \simeq &  B, 
\\
\langle  \pi^+(k) | {\bar s} \gamma^\mu d | K^+(p) \rangle & \simeq & \left( p+k \right)^\mu 
,
\\
\langle  \pi^+(k) | {\bar s} \sigma^{\mu \nu} d | K^+(p) \rangle & \simeq &
i {\Lambda_2 \over F_0^2} ( p^\mu k^\nu - p^\nu k^\mu ).
\end{eqnarray}
\end{subequations}
These matrix elements have exactly the same Lorentz structure as those for $B\to P$ transitions given in Eq.\,\eqref{eq:formfacB2P} in the limit of $q^2\to0$ and noticing that $f_+(0) = f_0(0)$.  
Thus, the differential decay width for $K\to \pi\slashed{E}$ can be obtained directly from the result for $B$ decay with suitable replacements of variables. The differential decay width for the charged mode and scalar DM is then 
\begin{eqnarray}
{d\Gamma_{K^+\to \pi^+\phi\phi} \over d q^2} 
& = & 
{B^2 \over 256 \pi^3 m_K^3 }
\lambda^{1\over 2}(m_K^2, m_\pi^2, s) \kappa^{1\over2}(m^2,s) \left|C_{q\phi}^{S,ds}\right|^2 
\nonumber
\\
& + &  {1 \over 768 \pi^3 m_K^3 } \lambda^{3\over2}(m_K^2, m_\pi^2, s)  \kappa^{3\over2}(m^2,s) \left|C_{q\phi}^{V,ds}\right|^2. 
\end{eqnarray}
For the two vector DM scenarios, using Eqs.\,(\ref{eq:Oqphi}-\ref{eq:OtildeqX}) and Eq.\,\eqref{eq:formfacK2pi}, the differential decay widths are 
\begin{eqnarray}
{d\Gamma_{K^+\to \pi^+XX}^A \over d q^2} 
& = & 
{B^2(s^2 - 4 m^2  s +12 m^4) \over 1024 \pi^3 m_K^3 m^4 } 
\lambda^{1\over 2}(m_K^2, m_\pi^2, s)  \kappa^{1\over 2}(m^2,s) \left|C_{qX}^{S,ds}\right|^2
\nonumber
\\
& + & 
{\Lambda_2^2 s(s + 4 m^2) \over 12288 \pi^3 F_0^4 m_K^3 m^4} 
\lambda^{3\over2}(m_K^2, m_\pi^2, s)  \kappa^{3\over2}(m^2,s)  \left|C_{qX1}^{T,ds}\right|^2
\nonumber
\\
& + & 
{\Lambda_2^2 (s + 2 m^2) \over 3072 \pi^3 F_0^4 m_K^3  m^2 } 
\lambda^{3\over2}(m_K^2, m_\pi^2, s)  \kappa^{1\over2}(m^2,s)  \left|C_{qX2}^{T,ds}\right|^2
\nonumber
\\
& + & 
{s \over 3072 \pi^3 m_K^3 m^4}   \lambda^{1\over2}(m_K^2, m_\pi^2, s)  \kappa^{3\over2}(m^2,s)
\nonumber
\\
& \times & 
\left[ 3(s- 4 m^2) (m_K^2 - m_\pi^2)^2 +  4 m^2 \lambda(m_K^2, m_\pi^2, s) \right]
\left|C_{qX2}^{V,ds}\right|^2
\nonumber
\\
& + & 
{1 \over 768 \pi^3 m_K^3 m^2} \lambda^{1\over2}(m_K^2, m_\pi^2, s)   \kappa^{3\over2}(m^2,s)
\nonumber
\\
& \times & 
\left[ 6 m^2 (m_K^2 - m_\pi^2)^2 + (s - 4 m^2) \lambda(m_K^2, m_\pi^2, s) \right]
\left|C_{qX3}^{V,ds}\right|^2
\nonumber
\\
& + & 
{s^2 - 4m^2 s +12 m^4  \over 3072 \pi^3 m_K^3 m^4 } 
\lambda^{3\over2}(m_K^2, m_\pi^2, s)  \kappa^{3\over2}(m^2,s) \left|C_{qX4}^{V,ds}\right|^2
\nonumber
\\
& + & 
{s(s + 4 m^2) \over 3072 \pi^3 m_K^3 m^4} 
\lambda^{3\over2}(m_K^2, m_\pi^2, s)  \kappa^{3\over2}(m^2,s) \left|C_{qX5}^{V,ds}\right|^2
\nonumber
\\
& + & 
{s + 2 m^2 \over 768 \pi^3 m_K^3 m^2} 
\lambda^{3\over2}(m_K^2, m_\pi^2, s)  \kappa^{1\over2}(m^2,s)  \left|C_{qX6}^{V,ds}\right|^2
+\cdots, 
\\%
{d\Gamma_{K^+\to \pi^+XX}^B \over d q^2}  &= & 
{B^2 (s^2 - 4 m^2  s +6 m^4) \over 128 \pi^3 m_K^3} 
\lambda^{1\over2}(m_K^2, m_\pi^2, s)   \kappa^{1\over2}(m^2,s)\left| \tilde C_{qX1}^{S,ds}\right|^2
\nonumber
\\
& + & 
{ B^2 s^2 \over 128 \pi^3 m_K^3} 
\lambda^{1\over2}(m_K^2, m_\pi^2, s)   \kappa^{3\over2}(m^2,s)\left| \tilde C_{qX2}^{S,ds}\right|^2
\nonumber
\\
& + & 
{\Lambda_2^2 s(s+2m^2) \over 6144 \pi^3 F_0^4 m_K^3} 
\lambda^{3\over2}(m_K^2, m_\pi^2, s)   \kappa^{3\over2}(m^2,s) \left| \tilde C_{qX1}^{T,ds}\right|^2
\nonumber
\\
& + & 
{\Lambda_2^2 (s^2 - 2m^2 s +4m^4) \over 6144 \pi^3 F_0^4 m_K^3} 
\lambda^{3\over2}(m_K^2, m_\pi^2, s)  \kappa^{1\over2}(m^2,s)\left| \tilde C_{qX2}^{T,ds}\right|^2. 
\end{eqnarray}
The corresponding results for the neutral kaon mode $K_L\to \pi^0\slashed{E}$ are obtained from these ones  replacing $C_i^{S,ds}$ by their real parts $\Re[C_i^{S,ds}]$ for  scalar quark currents and $C_i^{V(T),ds}$ by their imaginary parts  $\Im[C_i^{V(T),ds}]$ for  vector and tensor quark currents, respectively. 

\begin{figure}
\centering
\includegraphics[width=7.13cm]{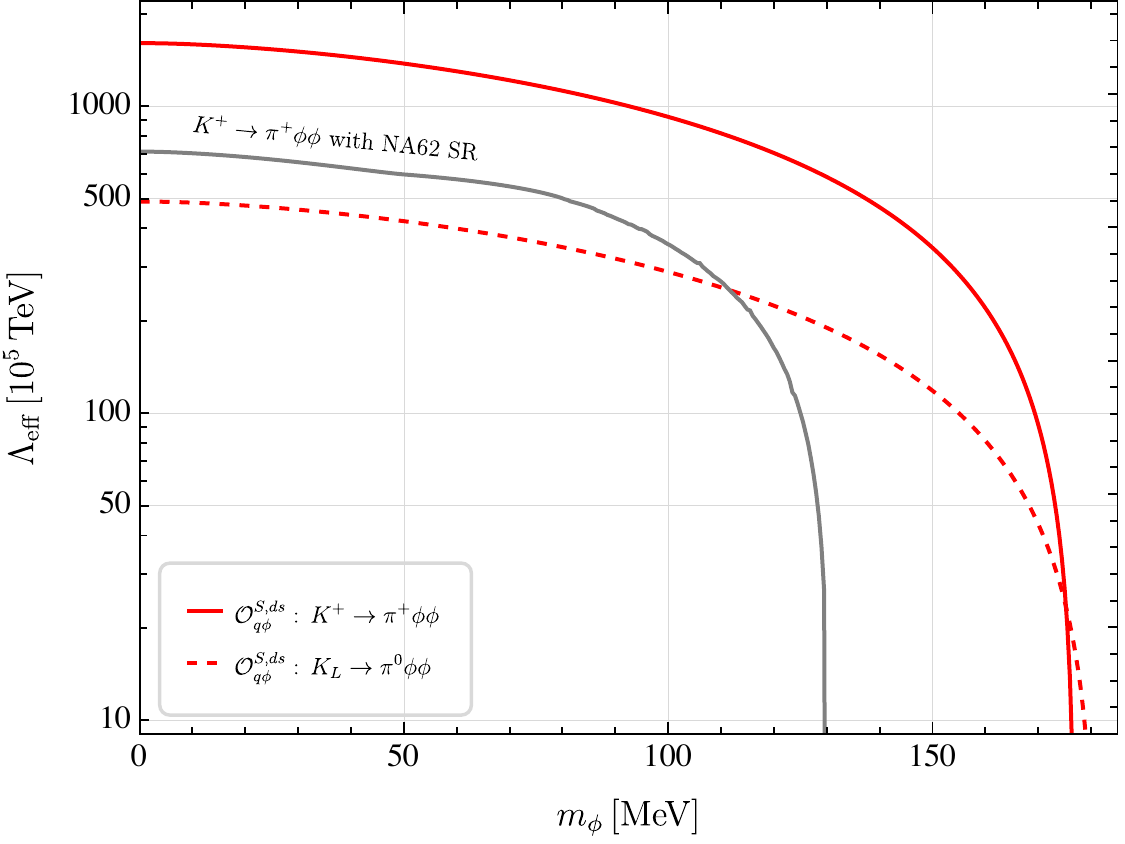}\qquad\quad
\includegraphics[width=7.cm]{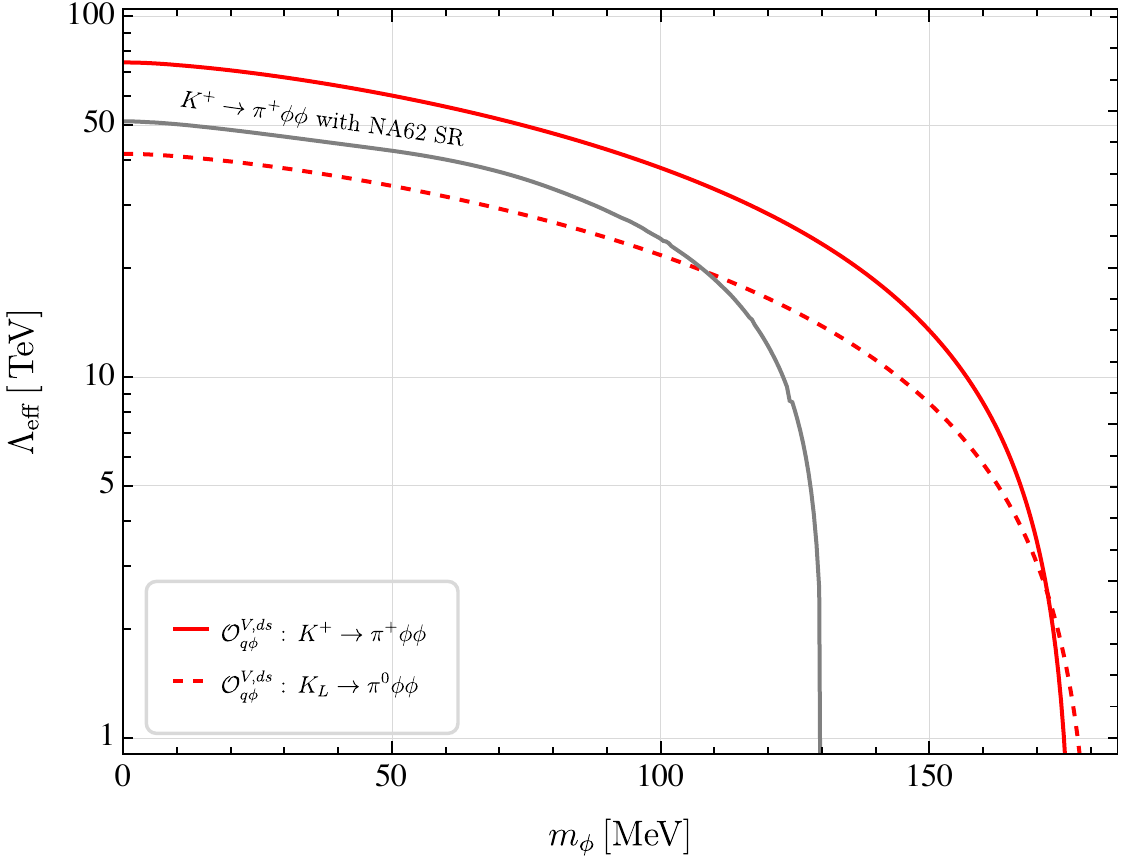}
\caption{Constraints on the effective new physics scale for the two operators $\calO_{a\phi}^{S,ds}$ and $\calO_{a\phi}^{V,ds}$ involving $d,s$ quarks as a function of the DM mass $m$ from $K\to \pi \slashed{E}$ channel.  }
\label{fig:Oqphids}
\end{figure}

Fig.\,\ref{fig:Oqphids} shows the bounds obtained on $\Lambda_{\rm eff}$  from $K\to\pi \phi\phi$ for scalar DM. 
Kaon decays only cover the low mass region, $m \lesssim 180$\,MeV, but the constraints on $\Lambda_{\rm eff}$ in this region are much stronger than the corresponding ones in $B$ meson decay shown in Fig.\,\ref{fig:scalar_constraint}. The difference,  a factor of $\calO(10^4)$ ($\calO(10)$) for scalar (vector) 
current operators, is due both to the  much longer kaon lifetime and the much stronger experimental bounds on kaon modes.

\begin{figure}
\centering
\includegraphics[width=7.13cm]{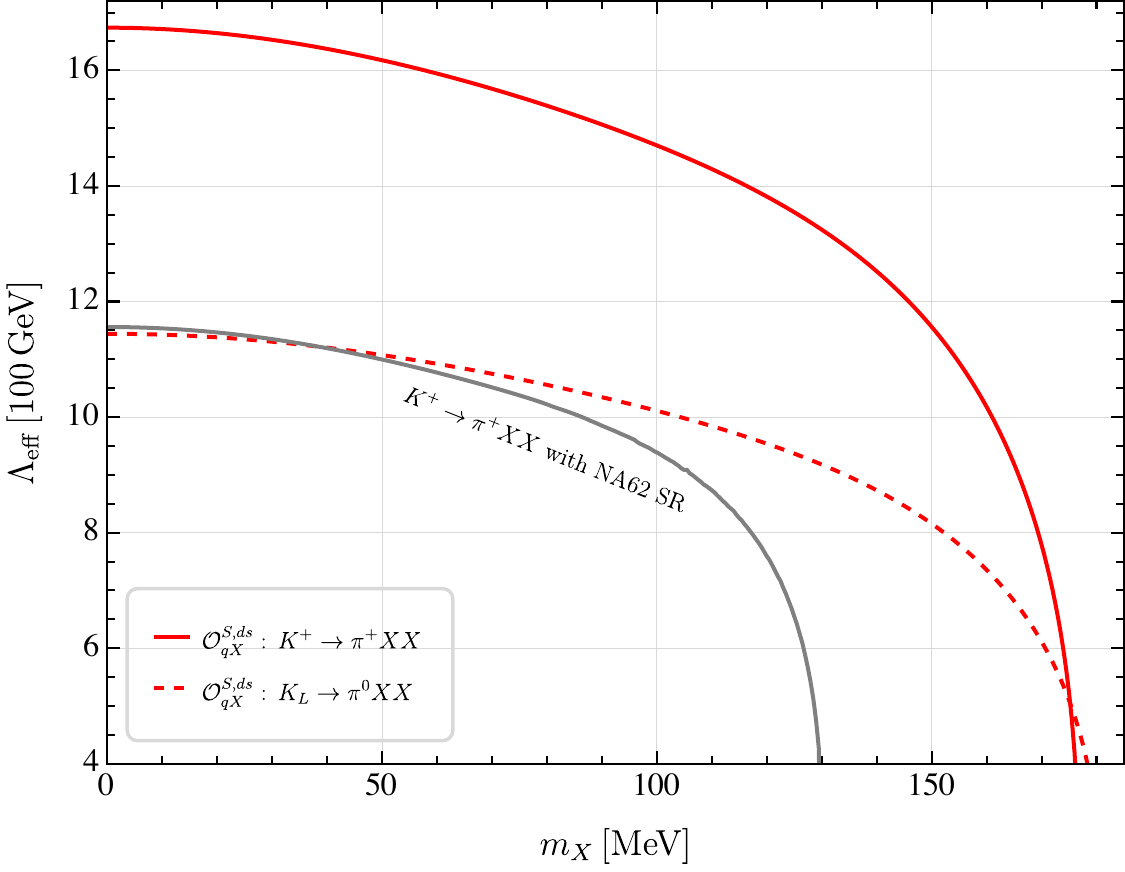}\qquad\quad
\includegraphics[width=7cm]{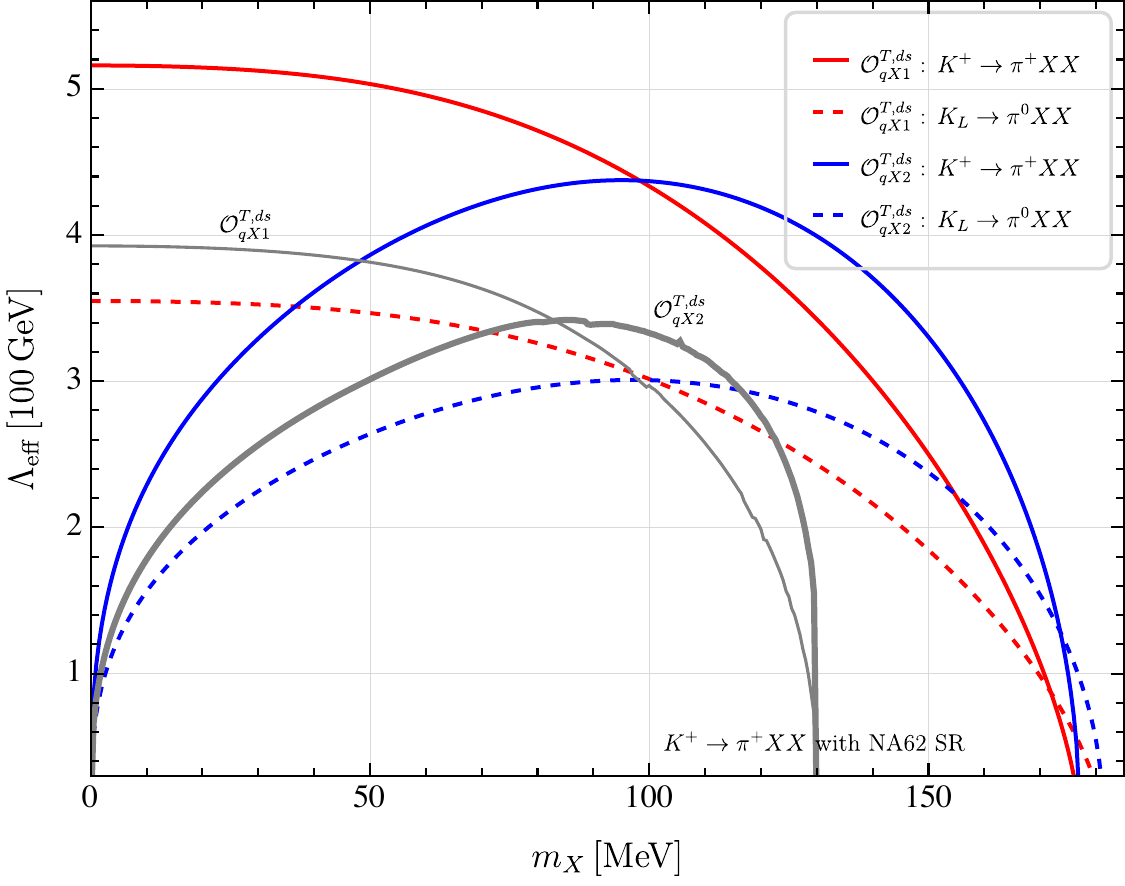}

\bigskip 

\includegraphics[width=7.13cm]{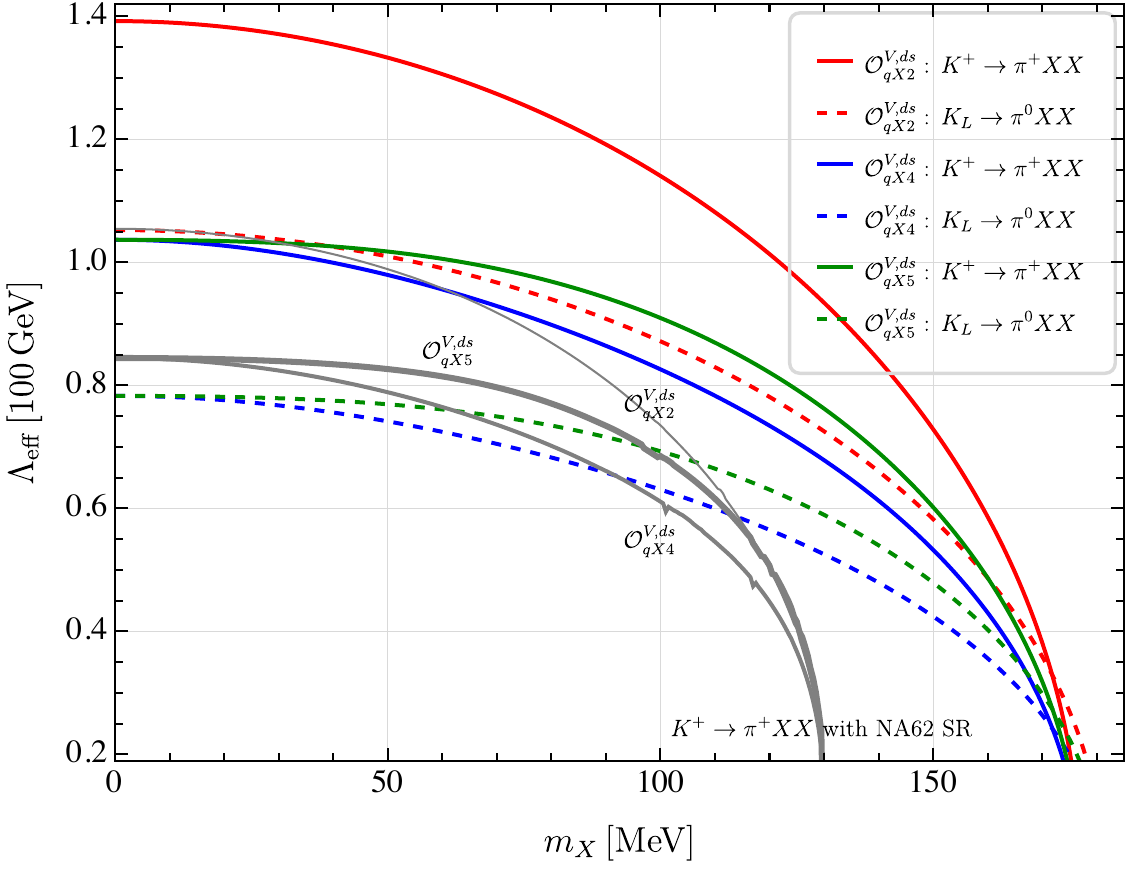}\qquad\quad
\includegraphics[width=7cm]{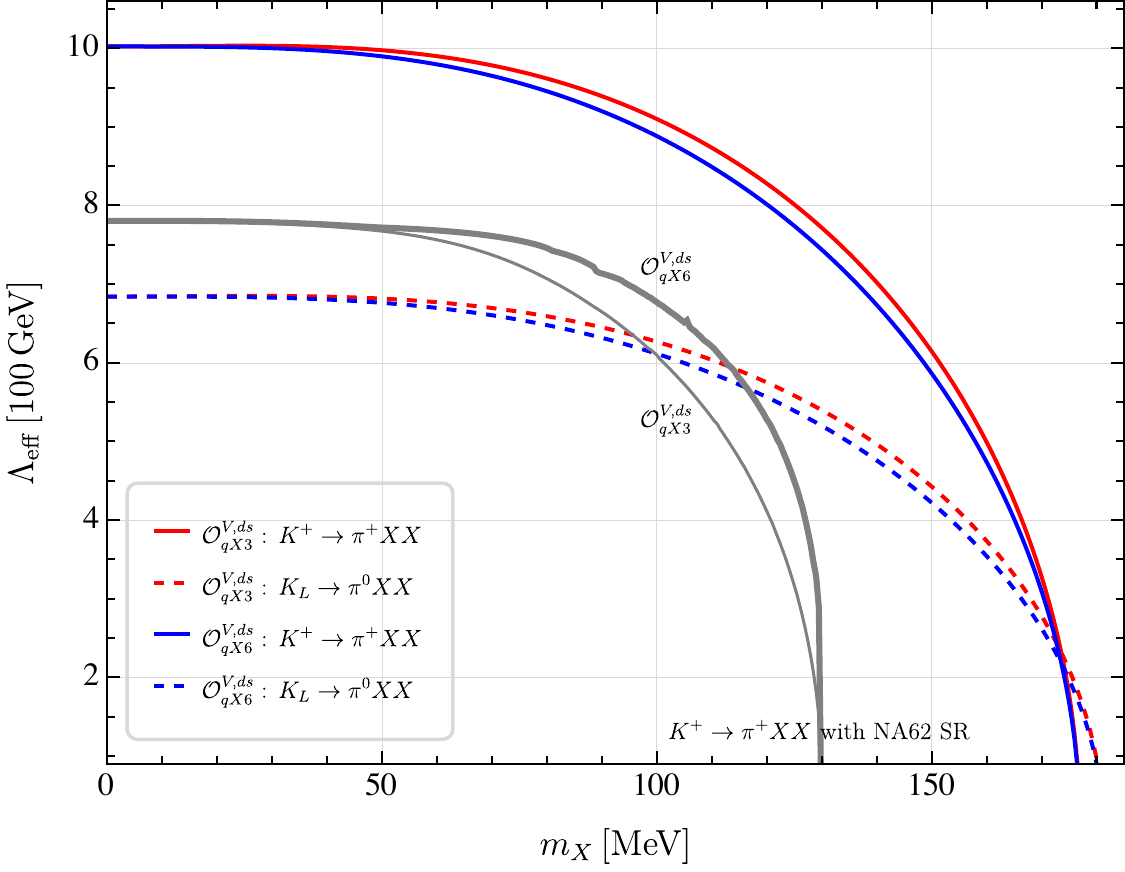}

\bigskip 
\includegraphics[width=7.13cm]{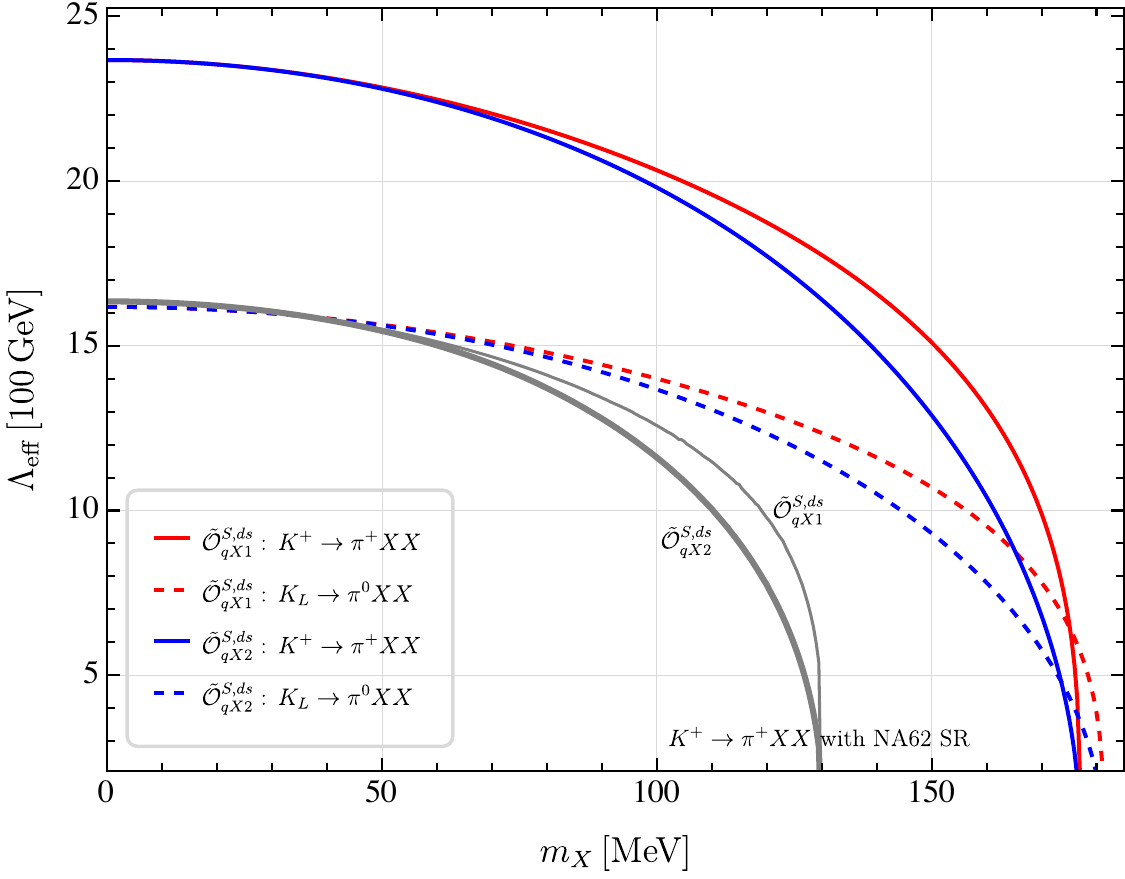}\qquad\quad
\includegraphics[width=7cm]{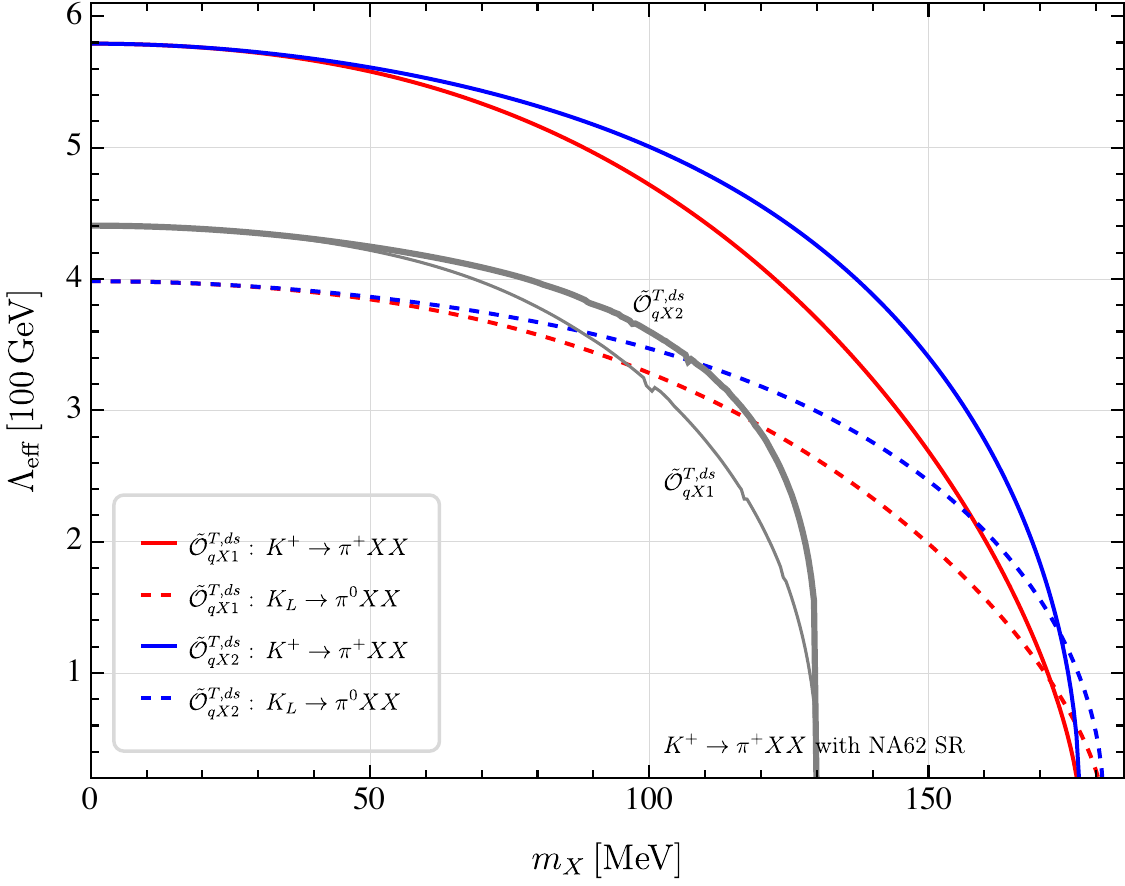}
\caption{Constraints on the effective new physics scale for the vector DM operators with $d,s$ quarks as a function of the DM mass $m$ from $K\to \pi \slashed{E}$ channel. }
\label{fig:OqXds}
\end{figure}

Fig.\,\ref{fig:OqXds} shows the constraints for the vector DM cases: the upper four panels for scenario A and the lower two panels for scenario B. Unlike the case of scalar DM, the constraints for vector DM case from kaon and $B$ meson decays are similar as seen in Figs.\,\ref{fig:OqXSPT}, \ref{fig:OqX36VA}, \ref{fig:OqX245VA}, and \ref{fig:OqXB}. This can be understood from dimensional arguments: the decay width from vector DM operators scales approximately as $\Gamma \sim {m_{K,B}^7 /\Lambda_{\rm eff}^6} ({m_{K,B}^9 /\Lambda_{\rm eff}^8})$ for operators $\calO^{S}_{qX}, \calO^{T}_{qX1,2}, \calO^{V}_{qX3,6},\tilde \calO^{S}_{qX1,2}, \tilde\calO^{T}_{qX1,2},$ ($\calO^{V}_{qX2,4,5}$). 
This large enhancement of $(m_B/m_K)^{7(9)}$ for  $B$ mesons compensates for its shorter lifetime and weaker experimental bounds, resulting in comparable constraints on $\Lambda_{\rm eff}$ (of course, for different flavor indices).  

\begin{table}
\centering
\resizebox{\linewidth}{!}{
\renewcommand{\arraystretch}{1.1}
\begin{tabular}{|c | l |c |c |c | c |c |c |}
\hline
Cases & Operators  & \multicolumn{6}{|c|}{Lower bound on $\Lambda_{\rm eff}\, \rm [GeV]$ }
\\\hline
& & \multicolumn{2}{|c|}{ $(sb)$-flavor } &\multicolumn{2}{|c|}{ $(db)$-flavor} & \multicolumn{2}{|c|}{ $(ds)$-flavor }
\\\hline
&  & $m=0$ & $m=2\, \rm GeV$  & $m=0$ & $m=2\, \rm GeV$  & $m=0$ & $m=150\, \rm MeV$
\\\hline
\multirow{4}*{scalar}   & $\calO_{q\phi}^{S} \,[\Lambda_{\rm eff}^{-1}]$ 
& $2.4\times 10^7$ &   $7.8\times 10^6$
& $2.6\times 10^7$ &  $1.1\times 10^7$
&  $1.6\times 10^{11} $ & $3.5\times 10^{10} $
\\\cline{2-8}
& $\calO_{q\phi}^{P} \,[\Lambda_{\rm eff}^{-1}]$ 
& $1.9\times 10^7 $ &  $1.7\times 10^6 $
& $1.2\times 10^7  $ & $1.6\times 10^6 $
& --- & ---
\\\cline{2-8}
& $\calO_{q\phi}^{V} \,[\Lambda_{\rm eff}^{-2}] $ 
& $7.0\times10^3$ & $2.2\times10^3$
& $7.0\times10^3$ & $2.9\times10^3$
&   $7.4\times 10^4$ & $1.3\times 10^4$
\\\cline{2-8}
& $\calO_{q\phi}^{A}\,[\Lambda_{\rm eff}^{-2}] $ 
& $8.0\times10^3$ &  $1.9\times10^3$
& $6.2\times10^3$ &  $1.7\times10^3$
& ---& ---
\\\hline
\multirow{14}*{vector A}   & $\calO_{qX}^{S} \,[m^2 \Lambda_{\rm eff}^{-3}]$
& $540$ & $400$
& $590$ & $470$
& $ 1.7 \times 10^3 $ & $ 1.2 \times 10^3 $
\\\cline{2-8}
& $\calO_{qX}^{P} \,[m^2 \Lambda_{\rm eff}^{-3}]$
& $430$ &  $230$
& $390$ &  $230$
& --- & ---
\\\cline{2-8}
& $\calO_{qX1}^{T} \,[m^2 \Lambda_{\rm eff}^{-3}] $
& $410$ & $310$
& $410$ & $300$
& $520 $ & $250 $
\\\cline{2-8}
& $\calO_{qX2}^{T}\,[m^2 \Lambda_{\rm eff}^{-3}]$
& $470$ & $290$
& $400$ & $340$ 
& --- &  $330 $
\\\cline{2-8}
& $\calO_{qX2}^{V}\,[m^2 \Lambda_{\rm eff}^{-4}]$
& $160$ & $94$
& $170$ & $110$
& $140 $ & $73$
\\\cline{2-8}
& $\calO_{qX3}^{V}\,[m \Lambda_{\rm eff}^{-3}]$
& $540$ & $340$
& $550$ & $410$
& $1.0\times10^3$ & $610$
\\\cline{2-8}
& $\calO_{qX4}^{V}\,[m^2 \Lambda_{\rm eff}^{-4}]$
& $130$ & $80$
& $130$ & $93$
&  $ 100 $ & $53$
\\\cline{2-8}
& $\calO_{qX5}^{V}\,[m^2 \Lambda_{\rm eff}^{-4}]$
& $130$ & $90$
& $130$ & $100$
& $ 100 $ &  $60$
\\\cline{2-8}
& $\calO_{qX6}^{V}\,[m \Lambda_{\rm eff}^{-3}]$
& $540$ & $400$
& $550$ & $450$  
& $1.0\times10^3$ & $590$
\\\cline{2-8}%
& $\calO_{qX2}^{A}\,[m^2 \Lambda_{\rm eff}^{-4}]$
& $140$ &  $74$ 
& $120$ &  $72$ 
& --- & ---
\\\cline{2-8}
& $\calO_{qX3}^{A}\,[m \Lambda_{\rm eff}^{-3}]$
& $580$ & $200$
& $500$ & $200$  
& --- & ---
\\\cline{2-8}
& $\calO_{qX4}^{A}\,[m^2 \Lambda_{\rm eff}^{-4}]$
& $140$ & $72$ 
& $120$ & $70$   
& --- & ---
\\\cline{2-8}
& $\calO_{qX5}^{A}\,[m^2 \Lambda_{\rm eff}^{-4}]$
& $140$ & $81$
& $120$ & $79$
& ---& ---
\\\cline{2-8}
& $\calO_{qX6}^{A}\,[m \Lambda_{\rm eff}^{-3}]$
& $580$ & $390$
& $500$ & $360$
& ---& ---
\\\hline
\multirow{6}*{vector B}   & $\tilde \calO_{qX1}^{S}\,[\Lambda_{\rm eff}^{-3}]$
& $760$ & $520$
& $830$ & $630$
& $2.4\times10^3$ & $1.5\times10^3$
\\\cline{2-8}
& $\tilde \calO_{qX2}^{S}\,[\Lambda_{\rm eff}^{-3}]$
& $760$ & $450$
& $830$ & $570$
&  $2.4\times10^3$ & $1.3\times10^3$
\\\cline{2-8}
& $\tilde \calO_{qX1}^{P}\,[\Lambda_{\rm eff}^{-3}]$
& $610$ &  $300$
& $550$ &  $300$
& --- & ---
\\\cline{2-8}
& $\tilde \calO_{qX2}^{P}\,[\Lambda_{\rm eff}^{-3}]$
& $610$ & $230$
& $550$ & $240$
& --- & ---
\\\cline{2-8}
& $\tilde \calO_{qX1}^{T}\,[\Lambda_{\rm eff}^{-3}]$
& $560$ & $310$ 
& $480$ & $320$
& $580$ & $270$
\\\cline{2-8}
& $\tilde \calO_{qX2}^{T}\,[\Lambda_{\rm eff}^{-3}]$
& $560$ & $310$
& $480$ & $360$
& $580$ & $340$
\\\hline
\end{tabular}
}
\caption{The strongest bounds on the effective scale $\Lambda_{\rm eff}$ associated with all FCNC operators with two representative DM masses: $m=0$ and $m=2\, \rm GeV$ (or $m=150\,\rm MeV$ for $(ds)$-flavor). For the $(ds)$-flavor, all the constraints in the table come from the charged channel $K^+\to\pi^+\slashed{E}$. }
\label{tab:Lambda}
\end{table}

In practice, the decay-in-flight search of $K^+\to \pi^+ +\slashed{E}$ by NA62 experiment \cite{NA62:2020fhy, NA62:2018ctf} has only two signal regions corresponding to $0 < q^2 < 0.01\,\si{GeV}^2$ (region 1) and $0.026\,\si{GeV}^2 < q^2 < 0.068\,\si{GeV}^2$ (region 2), with both signal regions also being constrained in the pion momentum by $15\,\si{GeV} < |\pmb{p}_\pi| < 35\,\si{GeV}$ (equivalently, the pion energy by $15\,\si{GeV} < E_\pi < 35\,\si{GeV}$). When restricting the phase space to these kinematic windows, the corresponding sensitivity bounds on the parameter space shift from the solid color lines into the solid gray lines in Figs.\,\ref{fig:Oqphids} and \ref{fig:OqXds}. The bound weakens by factors of a few relative to what could be obtained from the entire phase space.  The sensitivity drops quickly for heavier DM masses and vanishes around $m \approx 130\,\si{MeV}$, the cutoff value being determined by  the largest $q^2$, $q^2_{\rm max} \approx 0.068\,\si{GeV}^2$, in the NA62 signal region.  

Finally, in Tab.\,\ref{tab:Lambda}, we summarize the strongest constraints on $\Lambda_{\rm eff}$ for all FCNC interactions with two representative DM masses: $m=0$ and $m=2\, \rm GeV$ (or $m=150\,\rm MeV$ for kaon decays).  In the second column,  we show the scaling behavior of the Wilson coefficients employed to obtain the bounds. One should note that all results we present correspond to the case of complex DM fields. For the case of real DM fields, the operators $\calO_{q\phi}^{V,A}$ in Eq.\,\eqref{eq:Oqphi}, 
 $\calO_{qX1,2}^{T},\calO_{qX4,5,6}^{V,A}$ in Eq.\,\eqref{eq:OqX},  and $\tilde \calO_{qX1,2}^{T}$ in Eq.\,\eqref{eq:OtildeqX} do not exist. For the remaining ones, the bounds on $\Lambda_{\rm eff}$ will be enhanced by a factor of $2^{1/2n}$ with $n$ being  the power of $\Lambda_{\rm eff}^{-1}$ in the corresponding Wilson coefficient, shown in the second column of Tab.\,\ref{tab:Lambda}.  

\section{Summary and conclusions}
\label{sec:conclusion}

In this paper we have carried out a systematic study of possible flavor changing neutral current $B$ and $K$ meson decays with a pair of light scalar or vector invisible particles in the final state using the effective field theory approach. This completes the existing studies of $B(K)\to M\slashed{E}$ transitions where the missing energy is attributed to a pair of new invisible particles in the context of effective field theory. The case of two invisible fermions was studied before, and we have now addressed the cases of two invisible scalars or vectors. This study is particularly relevant when new symmetries forbid the appearance of single DM particles. 

We first constructed the local quark-DM interactions relevant to these processes in the low energy effective field theory framework at leading order. We tabulated results for invisible scalar, fermion, or vector particles completing  and correcting existing lists in the literature.   
We then used the effective interactions to consider $B \to (K, \pi, K^*,\rho)$+DM+DM transitions using the form factor formalism, followed by $K\to \pi$+DM+DM transitions using chiral perturbation theory.  
We describe the different characteristic features of each operator in the differential decay rate,  which could be exploited in future detailed experimental searches to differentiate between various possibilities of new physics.  
Finally, with the help of the most recent experimental results on these modes, we set constraints on all the relevant effective operators. The sensitivity to the new physics scale strongly depends on the operator structure as well as the DM mass. In the massless limit,  for the $B$ meson decay with scalar (vector) DM, we find that the current experimental data can probe the effective new physics scale up to  $\calO(10^{7}) (\calO(10^3))$\,GeV for some operators, while for the kaon decays $\calO(10^{11}) (\calO(10^3))$\,GeV can be reached. 

In two cases, $B^+\to K^+ \slashed E$ and $K^+\to \pi^+ \slashed E$, we considered the effect that experimental efficiency affects the theoretical constraints. For the former we relied on the current Belle II sensitivity to bins of different $q^2$ and for the latter on the signal window in NA62.

\acknowledgements

We would like to thank Shao-Zhou Jiang for clarifying their result for the LEC $\Lambda_2$ associated with one $p^4$ chiral Lagrangian term for the tensor external source in \cite{Jiang:2012ir}, and Ulrik Egede for the useful discussions.
This work was supported in part by the National Natural Science Foundation of China (Nos. 12090064, 11975149, 11735010),  the Shanghai Pujiang Program (20PJ1407800), 
Chinese Academy of Sciences Center for Excellence in Particle Physics (CCEPP), and Key Laboratory for Particle Physics, Astrophysics and Cosmology, Ministry of Education, and Shanghai Key Laboratory for Particle Physics and Cosmology. XGH was also supported in part by the MOST (Grant No. MOST 106- 2112-M-002-003-MY3 ). 
GV and XGH  were supported in part by the Australian Government through the Australian Research Council.
\appendix

\section{Differential decay rate}
\label{sec:phasespace}

For reference we detail here the kinematics relevant for the modes discussed in this paper. 
For the three-body decay, $B(p) \to M(k) + {\rm DM}(k_1) + {\rm DM}(k_2)$, there are two independent Mandelstam variables describing the kinematics that are denoted as $s\equiv (p- k)^2 =(k_1 + k_2)^2$ and $t\equiv (p -k_1)^2 = (k_2 + k)^2$.  
Then the scalar product of any pair of four vectors can be expressed in terms of the two Mandelstam variables  and masses as follows, 
\begin{eqnarray}
&&p \cdot k_1 = {m_B^2 + m_1^2 - t \over 2}, \quad
 p \cdot k_2 = {s+ t - m_1^2 - m_M^2 \over 2}, \quad
  p \cdot k={m_B^2 + m_M^2 - s \over 2},
  \\
 && k_1\cdot k_2 = { s - m_1^2 - m_2^2 \over 2 }, \quad
 k_1\cdot k = { m_B^2 +m_2^2 - s - t \over 2 }, \quad 
  k_2\cdot k = { t - m_2^2 - m_M^2 \over 2 },
\end{eqnarray}
where $m_B$ is the initial $B$ meson mass while $m_M$ for the mass of final state meson; $m_1 =m_2 \equiv m$ is the mass of the DM particle. For the process to happen, the largest possible DM mass is restricted by kinematics to be  $m \leq (M_B - m_M)/2$. 

The differential decay width can be expressed as
\begin{eqnarray}
{d\Gamma \over d q^2} ={1 \over S}{1 \over 256\pi^3 m_B^3}  \int_{t_-}^{t_+} d t \overline{|\calM|^2},
\end{eqnarray}
where $S$ is a possible symmetry factor for identical DM particles at the final state, and $\overline{|\calM|^2}$ is the spin-averaged squared matrix element for the relevant decay process. The integration domain for $t$ is 
\begin{eqnarray}
&&t_\pm = (E_2^* + E_3^*)^2 - \left(\sqrt{E_2^{*2} - m_2^2} \mp \sqrt{E_3^{*2} - m_M^2}   \right)^2, 
\nonumber
\\
&& E_2^* = {s - m_1^2 + m_2^2 \over 2\sqrt{s} }, \quad
E_3^* = { m_B^2 - s - m_M^2 \over 2\sqrt{s}}.
\end{eqnarray}

To extract the constraints we take one operator at a time, therefore ignoring interference between different operators. As the DM  are not SM particles, there is never interference with the SM. If these interactions exist, they thus contribute additively to the SM and can be directly constrained by the ``room for new physics'' of the last column in Tab.\,\ref{tab:B2KmisE2}. The branching ratio can be written as  a sum of numerical coefficients times the squares of the couplings of the new operators, schematically
\begin{eqnarray}
{\cal B}_{\rm NP} = {1\over \Gamma_B^{\rm tot}} \int d q^2 {d\Gamma \over d q^2} 
 \equiv \sum_i \hat B_i(m) |C_{i}|^2, 
\end{eqnarray}
where the generic range of $q^2$ goes from $(m_1 +m_2)^2$ to $(m_B- m_M)^2$. 

We allow for two modifications to the integration range to accommodate reported details of existing experiments. 
For $B^+\to K^+\slashed E$ we take into account the experimental efficiency as a function of $q^2$ reported in Fig.\,3 (supplementary material) of Belle II \cite{Belle-II:2021rof}, with the details of our analysis being given at the end of subsection \ref{subsec:scalarDM}.
For $K^+\to \pi^+\slashed E$, we limit the integration region to the two signal regions of NA62 \cite{NA62:2020fhy,NA62:2018ctf}
\begin{eqnarray}
&& 15\leq |\pmb{p}_\pi| \leq 35\,{\rm GeV} 
\text{ for the pion momentum in the NA62 rest  frame,}
\nonumber \\
&& 0 \leq q^2 \leq 0.1\,{\rm GeV}^2~{\rm or}~ 0.026\leq q^2 \leq 0.068\,{\rm GeV}^2  
\text{ for the two signal regions.}
\end{eqnarray}
In this case we work on the NA62 lab frame, with the kaon momentum $|\pmb{p}_K| = 75\,\si{GeV}$.

Requiring that the NP contribution of each operator to the branching ratio does not exceed the  value $ {\cal B}^{\rm UL}$ given in Tab.\,\ref{tab:B2KmisE2}, we set the constraints
\begin{eqnarray}
|C_{i}|^2 \leq  { {\cal B}^{\rm UL} \over \hat B_i(m) } .
\end{eqnarray}
To interpret this as a bound on new physics we then write $C_i\equiv \Lambda_{\rm eff}^{-n}$ (with the power $n$ depending on the dimension of the corresponding operator), leading to
\begin{eqnarray}
\Lambda_{\rm eff}(m) \geq \left( {\hat B_i(m) \over {\cal B}^{\rm UL} } \right)^{1\over 2n}. 
\end{eqnarray}
%

\section{Detailed analysis for the operators with a vector DM}
\label{sec:reductionofVope}

Here we describe in detail how  to obtain the operators with a vector DM in Eq.\,\eqref{eq:OqX} for scenario A. First, we can always choose the operators to be self-conjugate for the flavor diagonal case as given in Eq.\,\eqref{eq:OqX}, and we 
denote the Hermitian and anti-Hermitian combination of vector DM fields as, 
$S^{\mu\nu} \equiv X^{\mu\dagger}X^\nu +X^{\nu\dagger}X^\mu$ and 
$A^{\mu\nu} \equiv X^{\mu\dagger}X^\nu - X^{\nu\dagger}X^\mu$. 
It can be seen that they are automatically symmetric and anti-symmetric in their two Lorentz indices, respectively. For a combination of the vector quark current $\overline{q}\gamma_\mu q$ with two DM fields $X^\dagger_\alpha X_\beta$ to form dim-6 operators by attaching an   additional derivative, we have the following possibilities 
\begin{subequations}
\label{eq:opeall}
\begin{eqnarray}
 (\overline{q}\gamma_\mu q) X^{\dagger}_\nu X^\nu 
 & \overset{i\partial_\mu}{\Rightarrow} &
  (\overline{q}\gamma^\mu q) X^{\dagger}_\nu  i \overleftrightarrow{\partial_\mu} X^\nu ,  
 \\%
 (\overline{q}\gamma_\mu q) S^{\mu \nu} 
& \overset{i\partial_\nu}{\Rightarrow} &
(\overline{q}\gamma_\mu q) \partial_\nu S^{\mu \nu}, \,
(\overline{q}\gamma_\mu i\overleftrightarrow{D_\nu} q) S^{\mu \nu}, \,
\\%
 (\overline{q}\gamma_\mu q) A^{\mu \nu} 
& \overset{i\partial_\nu}{\Rightarrow} &
(\overline{q}\gamma_\mu q) i \partial_\nu A^{\mu \nu}, \, 
\boxed{  (\overline{q}\gamma_\mu \overleftrightarrow{D_\nu} q) A^{\mu \nu}}, 
\\%
 (\overline{q}\gamma_\mu q)X^\dagger_\rho X_\sigma \epsilon^{\mu\nu\rho\sigma} 
& \overset{\partial_\nu}{\Rightarrow} &
(\overline{q}\gamma_\mu q)(X^\dagger_\rho  \overleftrightarrow{\partial_\nu}X_\sigma)\epsilon^{\mu\nu\rho\sigma},\,
(\overline{q}\gamma_\mu q) \partial_\nu A_{\rho \sigma}  \epsilon^{\mu\nu\rho\sigma}, \,
\boxed{ (\overline{q}\gamma_\mu \overleftrightarrow{D_\nu}  q) A_{\rho \sigma}  \epsilon^{\mu\nu\rho\sigma}}, 
\end{eqnarray}
\end{subequations}
where  we have used IBP and the on-shell condition $\partial_\mu X^\mu=0$.  
For the quark axial-vector current $\overline{q}\gamma_\mu \gamma_5 q$, a similar operator realization can be obtained by replacing the vector gamma matrix $\gamma_\mu$ in the quark current by $\gamma_\mu \gamma_5$. 

Above, we have written 8 possible operators. However, the two operators in a ``$\Box $'' are redundant and they can be transformed into others appearing in our basis given in Eq.\,\eqref{eq:OqX}.  
Using the Dirac gamma matrix identities (DIs), 
\begin{eqnarray}
\gamma^\mu\gamma^\nu= g^{\mu\nu} - i \sigma^{\mu\nu},
\quad
\gamma^\mu \gamma^\nu \gamma^\rho=
g^{\mu\nu} \gamma^\rho + g^{\nu\rho} \gamma^\mu - g^{\mu\rho} \gamma^\nu + 
i \epsilon^{\mu\nu\rho\sigma} \gamma_\sigma \gamma_{5},
\quad
\sigma^{\mu\nu}\gamma_5 = {i \over 2} \epsilon^{\mu\nu\rho\sigma} \sigma_{\rho\sigma},  
\end{eqnarray}
the two operators can be manipulated as follows,
\begin{subequations}
\begin{eqnarray}
2(\overline{q_a}\gamma_\mu \overleftrightarrow{D_\nu} q_b) A^{\mu \nu} & = &
\left[ \overline{q_a} ( \gamma_{\mu}\gamma_\nu \slashed{D}
-\overleftarrow{\slashed{D}}\gamma_\nu\gamma_{\mu}) q_b
+\overline{q_a} (\gamma_{\mu} \slashed{D} \gamma_\nu
- \gamma_\nu \overleftarrow{\slashed{D}}\gamma_{\mu}) q_b \right] A^{\mu \nu}
\nonumber
\\
& \overset{\rm DI}{=} &
- \left[ \overline{q_a} ( \sigma_{\mu\nu} i\slashed{D}
+ i \overleftarrow{\slashed{D}}\sigma_{\mu\nu}) q_b
+\overline{q_a} (\gamma_{\mu} \slashed{D} \gamma_\nu
+ \gamma_\mu \overleftarrow{\slashed{D}}\gamma_{\nu}) q_b \right] A^{\mu \nu}
\nonumber
\\
& \overset{\rm EoM}{=} &
(m_a - m_b)(\overline{q_a}\sigma_{\mu\nu} q_b)A^{\mu\nu}
- \partial^\alpha (\overline{q_a} i \epsilon_{\mu\alpha\nu \beta}\gamma^\beta\gamma_5 q_b)  A^{\mu \nu} 
\nonumber
\\
& \overset{\rm IBP}{=} &
 (m_a - m_b) (\overline{q_a} \sigma_{\mu\nu} q_b)A^{\mu \nu} 
+ (\overline{q_a} \gamma^{\mu}\gamma_5 q_b) i\partial^\nu A^{\rho \sigma} \epsilon_{\mu\nu\rho\sigma}, 
\\%
 2 (\overline{q_a}\gamma^\mu \overleftrightarrow{D^\nu} q_b) A^{\rho\sigma}\epsilon_{\mu\nu\rho\sigma} 
& \overset{\rm DI}{=} &
- \left[ \overline{q_a} ( \sigma^{\mu\nu} i\slashed{D}
+ i \overleftarrow{\slashed{D}}\sigma^{\mu\nu}) q_b
+\overline{q_a} (\gamma^{\mu} \slashed{D} \gamma^\nu
+ \gamma^\mu \overleftarrow{\slashed{D}}\gamma^{\nu}) q_b \right]A^{\rho\sigma}\epsilon_{\mu\nu\rho\sigma} 
\nonumber
\\
& \overset{\rm EoM}{=} &(m_a - m_b)(\overline{q_a}\sigma^{\mu\nu} q_b)A^{\rho\sigma}\epsilon_{\mu\nu\rho\sigma} 
-  \partial_\alpha (\overline{q_a} i \epsilon^{\mu\alpha\nu \beta}\gamma_\beta\gamma_5 q_b)  A^{\rho\sigma}\epsilon_{\mu\nu\rho\sigma}
\nonumber
\\
& \overset{\rm IBP}{=} &2i (m_b -m_a)  (\overline{q_a} \sigma^{\mu\nu}\gamma_5 q_b)A_{\mu\nu}
+ (\overline{q_a} \gamma_\beta \gamma_5 q_b) i \partial_\alpha  A^{\rho\sigma}
2(\delta^\alpha_\rho \delta^\beta_\sigma - \delta^\beta_\rho \delta^\alpha_\sigma )
\nonumber
\\
& = & 2 i (m_b -m_a)  (\overline{q_a} \sigma_{\mu\nu}\gamma_5 q_b)A^{\mu \nu}
- 4 (\overline{q_a}  \gamma_\mu \gamma_5 q_b) i \partial_\nu  A^{\mu\nu}. 
\end{eqnarray}
\end{subequations}
In the reduction of the second operator, we used the identity  $\epsilon^{\mu\nu\alpha \beta}\epsilon_{\mu\nu\rho\sigma} =-2(\delta^\alpha_\rho \delta^\beta_\sigma - \delta^\beta_\rho \delta^\alpha_\sigma )$. Since the final 4 operators  are already in our basis, we conclude that the two operators in a ``$\Box $'' are redundant.  
Similarly, for the axial-vector quark current, the two corresponding operators are redundant and  can be transformed into those in our basis as follows, 
\begin{subequations}
\begin{eqnarray}
2(\overline{q_a}\gamma_\mu \overleftrightarrow{D_\nu}\gamma_5 q_b) A^{\mu \nu} & = &
(m_a + m_b) (\overline{q_a} \sigma_{\mu\nu}\gamma_5 q_b) A^{\mu \nu}
+ (\overline{q_a} \gamma^{\mu} q_b) i\partial^\nu A^{\rho \sigma} \epsilon_{\mu\nu\rho\sigma},
\\%
(\overline{q_a}\gamma^\mu \overleftrightarrow{D^\nu}\gamma_5 q_b) 
A^{\rho\sigma}\epsilon_{\mu\nu\rho\sigma} 
& = & 
  -  i (m_a + m_b)  (\overline{q_a} \sigma_{\mu\nu}q_b)A^{\mu \nu}
- 2 (\overline{q_a}  \gamma_\mu q_b) i \partial_\nu  A^{\mu\nu}. 
\end{eqnarray}
\end{subequations}
In conclusion, there are 6 independent operators for each quark current and they are listed in Eq.\,\eqref{eq:OqX} after normalization and (anti-)symmetrization. 

When using the {\tt Basisgen} package \cite{Criado:2019ugp}, we only find 4 dim-6 operators containing a derivative, the remaining 8 operators are missing. The reason for this undercounting is an oversimplified treatment of the equation of motion of the vector field in that package.\footnote{We thank J. C. Criado for confirming this.} 
Since the partial derivative $\partial_\mu$ and the vector field $X_\nu$ both belong to the vector representation $\left({1\over 2}, {1\over 2}\right)$ of the Lorentz algebra $(\mathfrak{su}(2),\mathfrak{su}(2))$, the general irreducible representation decomposition of their product $\partial_\mu X_\nu$ under Lorentz algebra is (for example, see eq. (34.31) in \cite{Srednicki:2007qs})
\begin{eqnarray}
\left({1\over 2}, {1\over 2}\right)\otimes \left({1\over 2}, {1\over 2}\right) 
={\color{cyan} (0,0)_S} \oplus {\color{purple} (1,1)_S} \oplus {\color{orange} (1,0)_A \oplus (0,1)_A}, 
\end{eqnarray}
or in terms of fields and derivatives, 
\begin{eqnarray}
\partial_\mu X_\nu ={\color{cyan} {1\over 4} g_{\mu\nu} \partial_\alpha X^\alpha }
+{\color{purple}{1\over 2}\left(\partial_\mu X_\nu  +\partial_\nu X_\mu - {1\over 2}g_{\mu\nu} \partial_\alpha X^\alpha  \right)}
+{\color{orange} {1\over 4}X_{\mu\nu}^+ +{1\over 4}X_{\mu\nu}^- },
\end{eqnarray}
where $X_{\mu\nu}^\pm =X_{\mu\nu} \mp i \tilde X_{\mu\nu}$ with $X_{\mu\nu} = \partial_\mu X_\nu  -\partial_\nu X_\mu$ and $\tilde X_{\mu\nu}=(1/2)\epsilon_{\mu\nu\rho\sigma}X^{\rho\sigma}$. 
The scalar component $\color{cyan} (0,0)_S$ vanishes for on-shell vector field as we described above.
The traceless symmetric component $\color{purple} (1,1)_S$ is the only one retained in the {\tt Basisgen} package as implemented in its python script. The last two components $\color{orange}(1,0)_A$ (self-dual 2-form field) and $\color{orange}(0,1)_A$ (anti-self-dual 2-form field),  are missing in the package and lead to the difference. After including these two components, using the method outlined in section 2.2 in \cite{Criado:2019ugp}, we indeed obtain the same total number of operators, i.e., 12.

\section{Lepton-DM interaction in LEFT}
\label{sec:leptonDMope}

For completeness we list here the independent operators involving a lepton current and two dark sector particles. Denoting the charged leptons as $\ell\in\{e,\mu,\tau\}$, the charged lepton-DM interactions can be directly obtained from the quark-DM interactions given in section \ref{sec:LEFT_ope} by exchanging the quark flavor label $q$ by the lepton label $\ell$. Following the conventions for quark-DM interactions in section \ref{sec:LEFT_ope}, they are:

\noindent
{\bf Fermion case}: 
\begin{subequations}
\begin{align}
\calO_{\ell\chi1}^{S} &= (\overline{\ell } \ell )(\overline{\chi}\chi),
&
\calO_{\ell\chi2}^{S}  &= (\overline{\ell } \ell )(\overline{\chi}i \gamma_5\chi), 
\\
\calO_{\ell\chi1}^{P} &=  (\overline{\ell } i \gamma_5 \ell )(\overline{\chi}\chi),
&
\calO_{\ell\chi2}^{P} & = (\overline{\ell } \gamma_5 \ell )(\overline{\chi} \gamma_5\chi), 
\\
\calO_{\ell\chi1}^{V} &=  (\overline{\ell }\gamma^\mu  \ell )(\overline{\chi}\gamma_\mu  \chi),
\, (\times)
&
\calO_{\ell\chi2}^{V} &= (\overline{\ell }\gamma^\mu \ell )(\overline{\chi}\gamma_\mu  \gamma_5\chi), 
\\
\calO_{\ell\chi1}^{A} &=  (\overline{\ell }\gamma^\mu\gamma_5  \ell )(\overline{\chi}\gamma_\mu  \chi),
\, (\times)
&
\calO_{\ell\chi2}^{A} & = (\overline{\ell }\gamma^\mu\gamma_5 \ell )(\overline{\chi}\gamma_\mu  \gamma_5\chi), 
\\
\calO_{\ell\chi1}^{T} &=  (\overline{\ell }\sigma^{\mu\nu}  \ell )(\overline{\chi}\sigma_{\mu\nu}   \chi),
\, (\times)
&
\calO_{\ell\chi2}^{T} &= (\overline{\ell }\sigma^{\mu\nu} \ell )(\overline{\chi}\sigma_{\mu\nu}  \gamma_5\chi), 
\, (\times)
\end{align}
\end{subequations}
\noindent
{\bf Scalar case}: 
\begin{subequations}
\begin{eqnarray}
\calO_{\ell\phi}^S &= & (\overline{\ell } \ell )(\phi^\dagger \phi), 
\\
\calO_{\ell\phi}^P &= & (\overline{\ell } i \gamma_5 \ell )(\phi^\dagger \phi), 
\\
\calO_{\ell\phi}^V &= & (\overline{\ell }\gamma^\mu \ell ) (\phi^\dagger i \overleftrightarrow{\partial_\mu} \phi), \, (\times) 
\\
\calO_{\ell\phi}^A &= & (\overline{\ell }\gamma^\mu\gamma_5 \ell ) (\phi^\dagger i \overleftrightarrow{\partial_\mu} \phi),  \, (\times).  
\end{eqnarray}
\end{subequations}
\noindent 
{\bf Vector case A}:
\begin{subequations}
\begin{eqnarray}
\calO_{\ell X}^S &= & (\overline{\ell } \ell )(X_\mu^\dagger X^\mu), 
\\
\calO_{\ell X}^P &= & (\overline{\ell }i \gamma_5 \ell )(X_\mu^\dagger X^\mu), 
\\
\calO_{\ell X1}^T &= &  {i \over 2} (\overline{\ell }  \sigma^{\mu\nu} \ell ) (X_\mu^\dagger X_\nu - X_\nu^\dagger X_\mu),  \, (\times) 
\\
\calO_{\ell X2}^T &= & {1\over 2} (\overline{\ell }\sigma^{\mu\nu}\gamma_5 \ell ) (X_\mu^\dagger X_\nu - X_\nu^\dagger X_\mu),  \, (\times) 
 \\
 \calO_{\ell X1}^V &= &{1\over 2} [ \overline{\ell }\gamma_{(\mu} i \overleftrightarrow{D_{\nu)} } \ell ] (X^{\mu \dagger} X^\nu + X^{\nu \dagger} X^\mu  ), 
\\
\calO_{\ell X2}^V &= & (\overline{\ell }\gamma_\mu \ell )\partial_\nu (X^{\mu \dagger} X^\nu + X^{\nu \dagger} X^\mu  ), 
\\
\calO_{\ell X3}^V &= & (\overline{\ell }\gamma_\mu \ell )( X_\rho^\dagger \overleftrightarrow{\partial_\nu} X_\sigma )\epsilon^{\mu\nu\rho\sigma}, 
\\
\calO_{\ell X4}^V &= & (\overline{\ell }\gamma^\mu \ell )(X_\nu^\dagger  i \overleftrightarrow{\partial_\mu} X^\nu), 
 \, (\times)
 \\
\calO_{\ell X5}^V &= & (\overline{\ell }\gamma_\mu \ell )i\partial_\nu (X^{\mu \dagger} X^\nu - X^{\nu \dagger} X^\mu  ),  \, (\times)
 \\
\calO_{\ell X6}^V &= & (\overline{\ell }\gamma_\mu \ell ) i \partial_\nu ( X^\dagger_\rho X_\sigma )\epsilon^{\mu\nu\rho\sigma},
\, (\times)  
 \\
\calO_{\ell X1}^A &= &{1\over 2} [\overline{\ell }\gamma_{(\mu} \gamma_5 i \overleftrightarrow{D_{\nu)} }  \ell ](X^{\mu \dagger} X^\nu + X^{\nu \dagger} X^\mu  ), 
\\
\calO_{\ell X2}^A &= & (\overline{\ell }\gamma_\mu \gamma_5 \ell )\partial_\nu (X^{\mu \dagger} X^\nu + X^{\nu \dagger} X^\mu  ), 
\\ 
\calO_{\ell X3}^A &= & (\overline{\ell }\gamma_\mu\gamma_5 \ell ) (X_\rho^\dagger \overleftrightarrow{ \partial_\nu} X_\sigma )\epsilon^{\mu\nu\rho\sigma}, 
\\
\calO_{\ell X4}^A &= & (\overline{\ell }\gamma^\mu\gamma_5 \ell )(X_\nu^\dagger  i \overleftrightarrow{\partial_\mu} X^\nu), 
 \, (\times)
  \\
\calO_{\ell X5}^A &= &  (\overline{\ell }\gamma_\mu \gamma_5 \ell )i \partial_\nu (X^{\mu \dagger} X^\nu - X^{\nu \dagger} X^\mu  ),  \, (\times)
 \\
\calO_{\ell X 6}^A &= & (\overline{\ell }\gamma_\mu\gamma_5 \ell )i \partial_\nu (  X^\dagger_\rho X_\sigma)\epsilon^{\mu\nu\rho\sigma},
\, (\times)  
\end{eqnarray}
\end{subequations}
\noindent
{\bf Vector case B}: 
\begin{subequations}
\begin{eqnarray}
\tilde \calO_{\ell X1}^S& = & (\overline{\ell }\ell )X_{\mu\nu}^\dagger  X^{\mu\nu},
\\
\tilde \calO_{\ell X2}^S& = & (\overline{\ell }\ell )X_{\mu\nu}^\dagger \tilde X^{ \mu\nu},
\\
\tilde \calO_{\ell X1}^P& = & (\overline{\ell }i \gamma_5\ell )X_{\mu\nu}^\dagger X^{ \mu\nu},
\\
\tilde \calO_{\ell X2}^P& = & (\overline{\ell }i \gamma_5\ell )X_{\mu\nu}^\dagger \tilde X^{ \mu\nu},
\\
\tilde \calO_{q\ell X1}^T& = &{i \over 2} (\overline{q}\sigma^{\mu\nu} \ell )(X^{\dagger}_{ \mu\rho} X^{\rho}_{\,\nu}-X^{\dagger}_{ \nu\rho} X^{\rho}_{\,\mu}), \, (\times) 
\\
\tilde \calO_{\ell X2}^T& = & {1\over 2}  (\overline{\ell } \sigma^{\mu\nu}\gamma_5 \ell )(X^{\dagger}_{ \mu\rho} X^{\rho}_{\,\nu}-X^{\dagger}_{ \nu\rho} X^{\rho}_{\,\mu}), \,  (\times) 
\end{eqnarray}
\end{subequations}
In addition to the usually considered electron-DM scattering process for direct DM detection, the above lepton-DM interactions can induce lepton flavor violating transitions, $\ell_j\to \ell_i$+DM+DM.  The charged leptons can be replaced by SM neutrinos leading to exotic interactions between neutrinos and DM particles. We defer a study of these possibilities to a future publication.

\section{Form factors for the $B\to P(V)$ transitions }
\label{sec:formfactor}

In Tabs.\,\ref{tab:formfactorP} and  \ref{tab:formfactorV} we collect the fitted parameters for the form factor parameterizations with $q^2\neq 0$ as given in Eq.\,\eqref{eq:FFpara1}  and Eq.\,\eqref{eq:FFpara2} relevant to $B\to P$ and $B\to V$ transitions respectively.  
\begin{table}[!h]
\setlength{\tabcolsep}{0.5cm}
\centering
\begin{tabular*}{\textwidth}{@{}@{\extracolsep{\fill}}ccccc}
\hline
Form factor & $r_1$ & $r_2$ & $m_{\rm fit}^2 (\si{GeV^2})$  & $m_R (\si{GeV})$
\\\hline 
$f_+^\pi$ & 0.744  & -0.486 & 40.73      & 5.32
\\
$f_0^\pi$ & 0         & 0.258   & 33.81      & ---
\\
$f_T^\pi$ & 1.387  & -1.134  & 32.22      & 5.32
\\\hline 
$f_+^K$  & 0.162  & 0.173   & ---           & 5.41
\\
$f_0^K$  & 0         & 0.330   & 37.46     & ---
\\
$f_T^K$  & 0.161  & 0.198   & ---           & 5.41
\\\hline
\end{tabular*}
\caption{Parameters appearing in the form factors of the $B\to K(\pi)$ transitions \cite{Ball:2004ye}. $f_T$ is a scale-dependent quantity and the value is given at $\mu=4.8 \, \si{GeV}$.}
\label{tab:formfactorP}
\end{table}

\begin{table}[htb]
\setlength{\tabcolsep}{0.5cm}
\centering
\begin{tabular*}{\textwidth}{@{}@{\extracolsep{\fill}}ccccc}
\hline
$F_i$& $B\to K^*$ &$m_{R,i}^{b \to s}/$GeV& $B\to\rho$&$m_{R,i}^{b \to d}/$GeV \\
\hline
$\alpha_0^{A_0}$ & $0.36 \pm 0.05$ && $0.36 \pm 0.04$ &\\
$\alpha_1^{A_0}$ & $-1.04 \pm 0.27$ &$5.366$& $-0.83 \pm 0.20$&$5.279$  \\
$\alpha_2^{A_0}$ & $1.12 \pm 1.35$ && $1.33 \pm 1.05$ &\\
\hline
$\alpha_0^{A_1}$ & $0.27 \pm 0.03$ && $0.26 \pm 0.03$ &\\
$\alpha_1^{A_1}$ & $0.30 \pm 0.19$ &$5.829$& $0.39 \pm 0.14$ &$5.724$\\
$\alpha_2^{A_1}$ & $-0.11 \pm 0.48$ && $0.16 \pm 0.41$ &\\
\hline
$\alpha_0^{A_{12}}$ & $0.26 \pm 0.03$ && $0.30 \pm 0.03$ &\\
$\alpha_1^{A_{12}}$ & $0.60 \pm 0.20$ &$5.829$& $0.76 \pm 0.20$  &$5.724$\\
$\alpha_2^{A_{12}}$ & $0.12 \pm 0.84$ && $0.46 \pm 0.76$ &\\
\hline
$\alpha_0^{V_0}$ & $0.34 \pm 0.04$ && $0.33 \pm 0.03$ &\\
$\alpha_1^{V_0}$ & $-1.05 \pm 0.24$ &$5.415$& $-0.86 \pm 0.18$ &$5.325$\\
$\alpha_2^{V_0}$ & $2.37 \pm 1.39$ && $1.80 \pm 0.97$ &\\
\hline
$\alpha_0^{T_1}$ & $0.28 \pm 0.03$ && $0.27 \pm 0.03$ &\\
$\alpha_1^{T_1}$ & $-0.89 \pm 0.19$ &$5.415$& $-0.74 \pm 0.14$ &$5.325$\\
$\alpha_2^{T_1}$ & $1.95 \pm 1.10$ && $1.45 \pm 0.77$ &\\
\hline
$\alpha_0^{T_2}$ & $0.28 \pm 0.03$ && $0.27 \pm 0.03$ &\\
$\alpha_1^{T_2}$ & $0.40 \pm 0.18$ &$5.829$& $0.47 \pm 0.13$ &$5.724$\\
$\alpha_2^{T_2}$ & $0.36 \pm 0.51$ && $0.58 \pm 0.46$ &\\
\hline
$\alpha_0^{T_{23}}$ & $0.67 \pm 0.08$ && $0.75 \pm 0.08$ &\\
$\alpha_1^{T_{23}}$ & $1.48 \pm 0.49$ &$5.829$& $1.90 \pm 0.43$ &$5.724$\\
$\alpha_2^{T_{23}}$ & $1.92 \pm 1.96$ && $2.93 \pm 1.81$ &\\
\hline\hline
\end{tabular*}
\caption{Parameters in the form factors of the $B\to \rho(K^*)$ processes with $k_{\rm max}=2$ \cite{Bharucha:2015bzk}.}
\label{tab:formfactorV}
\end{table}

\section{Specific renormalizable models to illustrate a possible origin of the LEFT operators}
\label{sec:models}

Here we illustrate with two simple renormalizable models a possible origin for the LEFT operators. First for the case of scalar DM generating the operators with a scalar mediator. Then 
for the case A of vector DM illustrating a possible origin for the additional mass factors that we argued  should accompany the operators in Eq.\,\eqref{eq:OqX} when $X_\mu$ is regarded as a dark sector gauge boson.

In these examples we will employ three dark sector fields: a light real scalar $\phi$ and a real vector gauge particle $X_\mu$ from a $U(1)_X$ gauge group, and a second scalar $\Delta$ that gives mass to $X$. These three particles are singlets under the SM gauge group. To generate the  FCNC in the quark sector we introduce two Higgs doublets $H_1$ and $H_2$.  Under the full gauge group $SU(3)_{\rm c}\times SU(2)_L \times U(1)_Y (U_X(1))$, these new scalars have the following charge assignments,   
\begin{eqnarray}
H_i (1,2, 1/2) (0) = \left ( \begin{array} {c} h^+_i\\ { v_i + h_i + i I_i \over \sqrt{2} } \end{array} \right ), \quad
\Delta (1,1,0)(1) = {v_\Delta + h_\Delta + i I_\Delta \over \sqrt{2}}. 
\end{eqnarray}
\noindent
{\bf Scalar case}: The relevant Lagrangian is given by  
\begin{eqnarray}
{\cal L}_{\rm scalar} \ni 
 \lambda_{ij}^{H\phi} H_i^\dagger H_j \phi \phi  
 +\left ( \bar Q_L Y_1 H_1 D_R + \bar Q_L Y_2 H_2 D_R + \hc \right), 
\end{eqnarray}
where $Q_L$ and $D_R$ are the SM left-handed quark doublet and right-handed down-type quark singlet, respectively. 
After spontaneous symmetry breaking and rotating back to the physical states,  $h_1$ and $h_2$ will mix giving rise to 
a SM Higgs ($h^m_1$) and a heavy Higgs ($h^m_2$), $h_i = \alpha_{ij} h_j^m$. Integrating out the physical Higgs bosons, we generate the operators $\calO_{q\phi}^{S}$ and $\calO_{q\phi}^{P}$. 

\noindent
{\bf Vector case}: The relevant interactions are given by
\begin{eqnarray}
{\cal L}_{\rm vector} \ni 
(D^\mu \Delta)^\dagger(D_\mu \Delta)  + \lambda^{H\Delta}_{ij}H^\dagger_i H_j \Delta^\dagger \Delta
+ \left( \bar Q_L Y_1 H_1 D_R + \bar Q_L Y_2 H_2 D_R    + \hc \right), 
\end{eqnarray}
where $D_\mu = \partial_\mu -i g_X X_\mu$. After $\Delta$ develops a vev, the $\Delta$ kinetic term will induce a mass for $X$ given by $m  = g_X v_\Delta$ and a $h_\Delta$-$X$-$X$ vertex with a coupling $g_X^2 v_\Delta $.  The Higgs quartic interaction leads to a mixing of $h_\Delta$ and $h_i$ with a mixing parameter ${1\over2 }\lambda_{ik}^{H\Delta} v_k v_\Delta$. Then integrating out $h_\Delta$ and $h_i$ we can obtain the operator $\calO_{qX}^{S}$  and $\calO_{qX}^{P}$, with  coefficients  proportional to $m^2$. 

One should note that if only one Higgs doublet is introduced, the diagonalization of the mass matrix also diagonalizes the Yukawa matrix $Y$ and there are no tree-level FCNC interactions. Introducing two Higgs doublets, where $Y_1$ and $Y_2$ cannot be diagonalized simultaneously, generates tree-level FCNC interactions. Since exchange of one of these Higgs bosons would generate undesirable FCNC interactions purely within the quark sector, this simple model should be considered as an existence proof. A realistic model would have to introduce many of the features that resolve this issue in two Higgs doublet models \cite{Foguel:2022unm}.


\end{document}